\documentclass[reprint,onecolumn,superscriptaddress,showpacs,floatfix]{revtex4-1}
\usepackage{amsmath}
\usepackage{mathrsfs}
\usepackage{txfonts}
\usepackage{amssymb}
\usepackage{graphicx}
\usepackage{hyperref}
\usepackage{ulem}
\usepackage{overpic}
\usepackage{psfrag}
\usepackage{tabularx}
\usepackage{multirow}
\usepackage{array}
\usepackage{placeins}
\usepackage[marginal]{footmisc}
\usepackage{color}
\newcommand{\PreserveBackslash}[1]{\let\temp=\\#1\let\\=\temp}

\newcommand{\ket} [1] {| #1 \rangle}

\makeatletter

\begin{document}

\title{Type-B Goldstone modes and a logarithmic spiral\\ in the staggered $\rm SU(4)$ ferromagnetic spin-orbital model}

\author{Qian-Qian Shi}
\affiliation{Centre for Modern Physics, Chongqing University, Chongqing 400044, The People's Republic of China}

\author{Huan-Qiang Zhou}
\affiliation{Centre for Modern Physics, Chongqing University, Chongqing 400044, The People's Republic of China}

\author{Ian P. McCulloch}
\affiliation{Department of Physics, National Tsing Hua University, Hsinchu 30013, Taiwan}
\affiliation{School of Mathematics and Physics, The University of Queensland, St. Lucia, QLD 4072, Australia}
\affiliation{Centre for Modern Physics, Chongqing University, Chongqing 400044, The People's Republic of China}

\author{Murray T. Batchelor}
\affiliation{Mathematical Sciences Institute, The Australian National University, Canberra ACT 2601, Australia}
\affiliation{Centre for Modern Physics, Chongqing University, Chongqing 400044, The People's Republic of China}

\begin{abstract}
It is found that the staggered $\rm SU(4)$ ferromagnetic spin-orbital model accommodates highly degenerate ground states arising from spontaneous symmetry breaking  with type-B Goldstone modes. The spontaneous symmetry breaking patterns are  ${\rm SU(4)} \rightarrow {\rm U(1)} \times {\rm U(1)} \times {\rm U(1)}$, with three type-B Goldstone modes or ${\rm SO(4)} \sim {\rm SU(2)} \times {\rm SU(2)}  \rightarrow {\rm U(1)} \times {\rm U(1)}$, with two type-B Goldstone modes, depending on the system size being even or odd. 
An abstract fractal constitutes the underlying structure of the ground-state subspace. For a sequence of atypical degenerate ground states the fractal dimension is identified with the number of type-B Goldstone modes. This connection is established by evaluating the entanglement entropy for these atypical degenerate ground states. The observed universal finite system-size scaling behavior of the entanglement entropy follows a logarithmic scaling relation with the block size in the thermodynamic limit. In addition, the ground state degeneracies, depending on the boundary conditions adopted, constitute the two Fibonacci-Lucas sequences. In the limit of large system size their  asymptotic forms become a self-similar logarithmic spiral. As a result, the model has a non-zero residual entropy $S_{\!r} = -2 \ln R $, where $R=(\! \sqrt{6}-\!\sqrt{2})/2$. 
\end{abstract}

\maketitle

\section{Introduction}

Much attention has been paid to exotic quantum states of matter associated with the orbital degrees of freedom in transition-metal oxides~\cite{metalbook,metaloxides}.
The underlying physics behind this fascinating but complicated quantum many-body problem is very rich~\cite{kugel}. 
For our purpose, we focus on the one-dimensional version, for which the physics is captured by the Hamiltonian
\begin{equation}
	\mathscr{H}=\sum_{j}(S_{\!j}\cdot S_{\!{j+1}}+\zeta)(T_j\cdot T_{j+1}+\eta).
	\label{hamist}
\end{equation}
Here the two coupling parameters $\zeta$ and $\eta$ describe the effective interactions between the spin and orbital degrees of freedom.
At the $j$-th site, $S_{\!j}$ denotes the spin-$1/2$ operator, and $T_{\!j}$ denotes the orbital pseudo spin-$1/2$ operator.  The sum over $j$ is from $1$ to $L-1$ for open boundary conditions (OBCs) or from $1$ to $L$ for periodic boundary conditions (PBCs), with 
$S_{\!L+1}= S_{\!1}$ and $T_{\!L+1}= T_{\!1}$, where $L$ denotes the system size.

\begin{figure}[ht]
	\centering
	\includegraphics[width=0.4\textwidth]{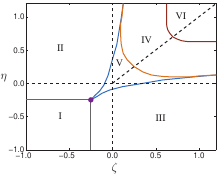}
	\hspace{5mm}
	\caption{The ground state phase diagram for the one-dimensional $\rm SO(4)$ spin-orbital model (\ref{hamist}) (adapted from Ref.~\cite{lundgren}).  
	The six distinct phases indicated are discussed in the text.
	The solid circle located at $(\zeta,\eta) = (-1/4, -1/4)$ is referred to as the staggered $\rm SU(4)$ ferromagnetic spin-orbital model.}
	\label{gspd}
\end{figure}

Extensive investigations~\cite{khomskii,lundgren}, both analytical and numerical, have revealed that the model (\ref{hamist}) exhibits distinct phases, as reflected in the ground state phase diagram shown in Fig.~\ref{gspd}.
The six distinct phases are labeled by I, II, III, IV, V and VI. 
Phase I is ferromagnetic, with the ground state being fully polarized in the spin and orbital sectors. 
Phase II is antiferromagnetic in the spin sector and ferromagnetic in the orbital sector.
Phase III is a variant of phase II, with the spin sector and orbital sector swapped, due to the fact that the Hamiltonian (\ref{hamist}) is invariant under  $S_{\!j} \leftrightarrow  T_{\!j}$ and $\zeta \leftrightarrow \eta$. 
Phases IV and VI are gapped dimerized phases, and phase V is a gapless phase, with central charge $c=3$.

Although the six distinct phases and the quantum phase transitions between them have been well understood, the precise nature of the particular point $\zeta=-1/4$ and $\eta=-1/4$ remains to be clarified, as far as the underlying physics of
 the spin-orbital model (\ref{hamist}) is concerned. 
This point is referred to as the staggered $\rm SU(4)$ ferromagnetic spin-orbital model, 
indicated by the solid circle in Fig.~\ref{gspd}, and surrounded by phases I, II, III and V.
The peculiarity of the point under investigation is that the model possesses the staggered ${\rm SU}(4)$ symmetry only if the system size $L$ is even. Indeed, the symmetry group reduces to the uniform ${\rm SU}(2) \times {\rm SU}(2)$ group if $L$ is odd under PBCs, as a result of an even-odd parity effect. 
This is in contrast to the ${\rm SU}(2)\times {\rm SU}(2)$ symmetry group for both even and odd $L$ at a generic point.
An interesting observation is that the Hamiltonian density (up to an additive constant) constitutes a realization of  the {\it same} representation of the Temperley-Lieb algebra~\cite{tla,baxterbook,martin}, 
together with other related but not necessarily unitarily equivalent realizations, such as the staggered ${\rm SU}(4)$ spin-$3/2$ ferromagnetic chain and the $16$-state Potts chain~\cite{barber}. 
In fact, the staggered $\rm SU(4)$ ferromagnetic spin-orbital model
is unitarily equivalent to the  staggered ${\rm SU}(4)$ spin-$3/2$ ferromagnetic model (cf. Appendix~A for the unitary transformation connecting the two models). 
It follows therefore that the staggered $\rm SU(4)$ ferromagnetic spin-orbital model is exactly solvable, falling into the realm of quantum Yang-Baxter integrability~\cite{baxterbook,sutherlandb,mccoy}.

In this article we investigate the staggered ${\rm SU}(4)$ ferromagnetic spin-orbital model in the context of spontaneous symmetry breaking (SSB) with type-B Goldstone modes (GMs), given that the counting rule of GMs~\cite{brauner-watanabe,watanabe,NG}, together with the redundancy principle~\cite{golden}, offers a powerful means to unlock mysteries behind various quantum many-body systems with distinct types of ferromagnetic interactions~\cite{FMGM,golden}.
As it turns out, the staggered $\rm SU(4)$ ferromagnetic spin-orbital model accommodates highly degenerate ground states arising from SSB  with type-B GMs. 
The SSB pattern is ${\rm SU(4)} \rightarrow {\rm U(1)} \times {\rm U(1)} \times {\rm U(1)}$, with the number of type-B GMs $N_B$ being $N_B=3$ or ${\rm SO(4)} \sim {\rm SU(2)} \times {\rm SU(2)}  \rightarrow {\rm U(1)} \times {\rm U(1)}$, with two type-B GMs, depending on the system size being even or odd.   For brevity, we mainly focus on even $L$, unless otherwise stated.

We thus reveal an abstract fractal underlying the ground state subspace. Its fractal dimension $d_f$, first exploited by Castro-Alvaredo and Doyon~\cite{doyon} to investigate scaling behavior of the entanglement entropy, is identified with the number of type-B GMs $N_B$.  
This connection is established by evaluating the entanglement entropy for  a set of orthonormal basis  states, which exhibits a universal finite system-size scaling behavior~\cite{finitesize}.  
The latter in turn implies that the entanglement entropy follows a logarithmic scaling relation with the block size in the thermodynamic limit.
In addition, the ground state degeneracies are seen to constitute the two Fibonacci-Lucas sequences, depending on either OBCs or PBCs. 
When the system size is large enough, their asymptotic forms become a logarithmic spiral -- a familiar self-similar geometric object. 
As a consequence, the residual entropy is non-zero, which we determine exactly.

\section{The SSB pattern for even $L$}
 
As already mentioned, the even-odd parity effect emerges in this model under PBCs. 
For odd $L$, the symmetry group is the uniform ${\rm SO(4)} \sim {\rm SU(2)} \times {\rm SU(2)}$ group generated by $S^x = \sum_j S^x_j$, $S^y = \sum_j S^y_j$, $S^z = \sum_j S^z_j$ and $T^x = \sum_j S^x_j$, $T^y = \sum_j S^y_j$, $T^z = \sum_j S^z_j$. The SSB pattern is from the ${\rm SU(2)} \times {\rm SU(2)}$ symmetry group to the residual symmetry group $\rm{U(1)} \times \rm{U(1)}$ generated by $S^z = \sum_j S^z_j$ and  $T^z = \sum_j T^z_j$, with the number of GMs being two.
For even $L$, the underlying symmetry group becomes the staggered $\rm{SU(4)}$ group, with a total number of 15 generators. In contrast, the symmetry group remains the staggered $\rm{SU(4)}$ group when OBCs are implemented. For this symmetry group, one may introduce the Cartan generators  $H_1$, $H_2$ and $H_3$,  along with three raising operators $E_1$, $E_2$ and $E_3$ and three lowering operators $F_1$, $F_2$ and $F_3$.
They are the generators of the three ${\rm SU(2)}$ subgroups, satisfying the following commutation relations: $[H_1,E_1]=2E_1$, $[H_1,F_1]=-2F_1$, $[E_1,F_1]=H_1$, $[H_2,E_2]=2E_2$, $[H_2,F_2]=-2F_2$, $[E_2,F_2]=H_2$,
$[H_3,E_3]=2E_3$, $[H_3,F_3]=-2F_3$, and $[E_3,F_3]=H_3$. 
In addition,  there are six more generators, denoted by $E_4$, $F_4$,  $E_5$, $F_5$, $E_6$ and $F_6$.
The detailed expressions for the Cartan generators  $H_1$, $H_2$ and $H_3$ and the others in terms of the spin-$1/2$ operators $S_{\!j}$ and $T_{\!j}$ at site $j$ are presented in Section A of the Supplementary Material (SM), along with the commutation relations between them.

The highest weight state $| \, \rm{hws}\rangle$ can be chosen to be $| \, {\rm hws}\rangle= \vert\otimes_{k=1}^{L}  \{\uparrow_s\uparrow_t\}_{\;k}\rangle$, where $|\uparrow_s\rangle$ and   $|\downarrow_s\rangle$ are the eigenvectors of $S_{j}^z$, with corresponding eigenvalues $1/2$ and $-1/2$. Similarly $|\uparrow_t\rangle$ and $|\downarrow_t\rangle$, are the eigenvectors of $T_j^z$, with  corresponding eigenvalues $1/2$ and $-1/2$. 
The expectation values of the local components $H_{1,j}$, $H_{2,j}$ and $H_{3,j}$ for the Cartan generators then follow as  
$\langle\uparrow_s\uparrow_t \!\! | \, H_{1,j} \, | \!\! \uparrow_s\uparrow_t\rangle=1$, $\langle\uparrow_s\uparrow_t  \!\! | \,  H_{2,j} \, | \!\! \uparrow_s\uparrow_t\rangle=1$, $\langle\uparrow_s\uparrow_t| \, H_{3,j} \, |\uparrow_s\uparrow_t\rangle=1$,
if $j$ is odd, and
$\langle\uparrow_s\uparrow_t| \, H_{1,j} \, |\uparrow_s\uparrow_t\rangle=0$, $\langle\uparrow_s\uparrow_t| \, H_{2,j} \, |\uparrow_s\uparrow_t\rangle=0$, $\langle\uparrow_s\uparrow_t|\, H_{3,j} \, |\uparrow_s\uparrow_t\rangle=1$, if $j$ is even.
For $E_\alpha$ and $F_\alpha$ ($\alpha=1, 2, \ldots, 6$),   $F_{\alpha,j}$ and $E_{\alpha,j}$ are  accordingly chosen as the interpolating fields~\cite{nielsen, inter}.
Given that $\langle[E_\alpha,F_{\alpha,j}]\rangle\propto \langle H_{\alpha,j}\rangle$, $\langle[E_{\alpha,j},F_\alpha]\rangle\propto \langle H_{\alpha,j}\rangle$ ($\alpha=1, 2, 3$),
$\langle[E_4,F_{4,j}]\rangle\propto \langle H_{2,j}-H_{1,j}\rangle$, $\langle[E_{4,j},F_4]\rangle\propto \langle H_{2,j}-H_{1,j}\rangle$,
$\langle[E_5,F_{5,j}]\rangle\propto \langle H_{3,j}-H_{2,j}\rangle$, $\langle[E_{5,j},F_5]\rangle\propto \langle H_{3,j}-H_{2,j}\rangle$,
$\langle[E_6,F_{6,j}]\rangle\propto \langle H_{3,j}-H_{1,j}\rangle$, and $\langle[E_{6,j},F_6]\rangle\propto \langle H_{3,j}-H_{1,j}\rangle$, and that $\langle H_{1,j} \rangle =\langle H_{2,j}\rangle$, but $\langle H_{3,j} \rangle \neq \langle H_{1,j}\rangle$ and $\langle H_{3,j} \rangle \neq \langle H_{2,j}\rangle$, we are led to conclude that
the generators $E_\alpha$ and $F_\alpha$ ($\alpha=$ 1, 2, 3, 5, 6) are spontaneously broken, with $\langle H_{\alpha,j}\rangle$ ($\alpha=1$, $2$, and $3$) being the local order parameters. 
It follows that the SSB pattern is from $\rm{SU(4)}$ to  ${\rm U(1)}\times{\rm U(1)}\times{\rm U(1)}$ via ${\rm U(1)}\times{\rm U(1)}\times{\rm SU(2)}$ successively. 
That is, there are ten broken generators. In contrast, there are only three (linearly) independent local order parameters, meaning that there are three type-B GMs: $N_B=3$.
However, the Mermin-Wagner-Coleman theorem~\cite{mwc} states that no type-A GM is permitted in one spatial dimension, from which it follows that the number of type-A GMs $N_A$ must be $N_A=0$.
As a consequence, there exists an apparent violation to the counting rule~\cite{watanabe} for the SSB pattern from $\rm{SU(4)}$ to $\rm{U(1)}\times\rm{U(1)}\times\rm{U(1)}$.

To maintain consistency with the GM counting rule~\cite{watanabe} we adopt the redundancy principle introduced in the discussion of the staggered $\rm{SU}(3)$ ferromagnetic spin-1 biquadratic model~\cite{golden}.
We state this principle here again as follows. 
Namely the number of GMs should not exceed the rank $r$ of the semisimple group $G$.
In our case, $r=3$  for the staggered symmetry group $\rm{SU(4)}$, the number of the type-B GMs must be equal to $N_B = 3$. 
Four broken generators are thus redundant for the SSB pattern from $\rm{SU(4)}$ to ${\rm U(1)}\times{\rm U(1)}\times{\rm U(1)}$.

A key point is that if only the highest weight state is taken into account, 
by itself the SSB pattern from $\rm{SU(4)}$ to  ${\rm U(1)}\times{\rm U(1)}\times{\rm U(1)}$ does not fully account for the exponential ground state degeneracies with $L$.
This is because the dimension of the irreducible representation, generated from the highest weight state $| \, \rm{hws}\rangle$, for the staggered $\rm{SU}(4)$ symmetry group, is $(L+1)(L+3)(L+5)/12$ for $L$ odd, and $(L+2)(L+4)(2L+3)/24$ for $L$ even.
Actually, for the staggered  $\rm{SU}(4)$ symmetry group, not all generators commute with an additional $Z_2$ symmetry operation, denoted as $\sigma$. Here $\sigma$ represents the one-site translation operation $T_1$ under PBCs or the bond-centred inversion operation $I_b$ for even system-size, and the site-centered inversion operation $I_s$ for odd system-size, under OBCs.
If PBCs are adopted, this is also reflected in the fact that the only degenerate factorized ground state, invariant under the one-site translation operation, is 
the highest weight state $\vert\otimes_{k=1}^{L}  \{\uparrow_s\uparrow_t\}_{\;k}\rangle$, modulo its time-reversed state, namely the lowest weight state $\vert\otimes_{k=1}^{L}  \{\downarrow_s\downarrow_t\}_{\;k}\rangle$.

Rather than being polynomial in $L$, the presence of the additional discrete $Z_2$ symmetry operation $\sigma$, together with  the time-reversal operation $K_s$ in the spin sector and the time-reversal operation $K_t$ in the orbital sector, lead to exponential ground state degeneracies ${\rm dim }(\Omega_L^{\rm OBC})$ and ${\rm dim }(\Omega_L^{\rm PBC})$ under PBCs and OBCs. They form the two Fibonacci-Lucas sequences.
More precisely, we find that the ground state degeneracies ${\rm dim }(\Omega_L^{\rm OBC})$ and ${\rm dim }(\Omega_L^{\rm PBC})$ are given by
\begin{align} \label{degens}
	{\rm dim }(\Omega_L^{\rm OBC}) &= \frac{(2+\sqrt 3)^{L+1} - (2-\sqrt 3)^{L+1}}{ 2 \sqrt{3}  },\nonumber \\
	{\rm dim }(\Omega_L^{\rm PBC}) &= (2+\sqrt 3)^L + (2-\sqrt 3)^L,
\end{align}
for $L \ge 2$ and for $L \ge 3$, respectively. Further details are given in Appendix~B. 
It follows that the residual entropy $S_{\!r}$ is non-zero, with $S_{\!r} = -2 \ln R $, where $R=(\!\sqrt{6}-\!\sqrt{2})/2$, an extension of the (inverse) golden ratio (cf.~Appendix~B).

In fact, there exist exponentially many generalized highest weight states that appear to be degenerate factorized ground states (cf. Sec.~B of the SM for degenerate factorized ground states when $L$ is small). The presence of such a generalized highest weight state is in fact a generic feature for a quantum many-body system undergoing SSB with type-B GMs, as long as the ground state degeneracies under OBCs and PBCs are exponential with size $L$, as also seen in the staggered $\rm{SU}(3)$ ferromagnetic spin-1 biquadratic model~\cite{golden}.
Exponential ground state degeneracies can be systematically understood in the context of emergent subsystem invertible symmetry operations specific to degenerate ground states~\cite{dimertrimer, emergentflatband}. In particular, Ref.~\cite{emergentflatband} explicitly demonstrates the construction of generalized highest weight states within the staggered ${\rm SU(4)}$ ferromagnetic spin-orbital model, using the first two sequences of factorized degenerate ground states as illustrative examples. A fundamental equivalence between three key phenomena is established: (1) the existence of emergent invertible subsystem symmetry operations tailored to degenerate ground states, (2) the exponential ground state degeneracies in system size, and (3) the emergence of Goldstone multi-magnon flat bands, arising in the context of SSB with type-B GMs.

\section{Degenerate ground states from the highest weight state and the generalized highest weight states}

In order to evaluate the entanglement entropy we need to further consider the construction of the highly degenerate ground states.
For instance, for the model under PBCs, the symmetry group is the staggered  $\rm{SU}(4)$ group when $L$ is even. In this case, the degenerate ground states can be constructed by the repeated action of the six lowering operators $F_1$, $F_2$, $F_3$, $F_4$, $F_5$ and $F_6$ on both the highest weight state and the generalized highest weight states (for a detailed description, cf.  Sec.~B of the SM). Actually, the lowering operators $F_4$, $F_5$ and $F_6$ may be expressed in terms of the lowering operators   $F_1$, $F_2$ and $F_3$,  in combination with the additional $Z_2$ symmetry operation $\sigma$,  the time-reversal operation $K_s$ in the spin sector and the time-reversal operation $K_t$ in the orbital sector.  Indeed, we have  $F_4=K_t F_3 K_t^{-1}$, $F_5=T_1F_1T_1$ and $F_6=T_1F_2T_1$, if PBCs are adopted, and  $F_4=K_t F_3 K_t^{-1}$, $F_5=I_bF_1I_b$ and $F_6=I_bF_2I_b$, if OBCs are adopted.

Here and hereafter, we call  periodic degenerate factorized ground states {\it atypical}~\cite{greenpara}, if they consist of local states $\vert \uparrow_s\uparrow_t \rangle$ and  $\vert \uparrow_s\downarrow_t \rangle$, in the sense that a {\it typical} degenerate factorized ground state is non-periodic~\cite{greenpara}, even if local states $ \vert \downarrow_s\uparrow_t \rangle$ and  $\vert \downarrow_s\downarrow_t \rangle$ are excluded.
An atypical generalized highest weight state, with a period $p$ an even integer, may be generically written as
an atypical (periodic) degenerate factorized ground state, namely $\vert\otimes_{k=1}^{L/p}  \{s_1t_1 \cdots s_pt_p\}_{\;k}\rangle$. 
A sequence of atypical degenerate ground states, with the system size $L$ a multiple of $p$, are then generated from the repeated action of  three lowering operators $F_1$, $F_2$ and $F_3$ on such an atypical generalized highest weight state $\vert\otimes_{k=1}^{L/p}  \{s_1t_1 \cdots s_pt_p\}_{\;k}\rangle$, combining with the one-site translation operation $T_1$ under PBCs and the bond-centered inversion operation $I_b$ under OBCs, together with the time-reversal operations $K_s$ in the spin sector and $K_t$ in the orbital sector.
In particular, the highest weight state $\vert\otimes_{k=1}^{L}  \{\uparrow_s\uparrow_t\}_{\;k}\rangle$ is itself exceptional, in the sense that it may be regarded as a special case with period $p =1$.
The degenerate ground states generated from the highest weight states are as a result generically periodic, with period $p=2$, due to the staggered structure of the lowering operators  $F_1$, $F_2$, $F_3$, $F_4$, $F_5$ and $F_6$.

It follows that atypical degenerate ground states with the period $p$, denoted as $| \, L,M_1,M_2,M_3\rangle_p$, are generated from the repeated action of the three lowering operators $F_1$, $F_2$ and $F_3$ on  an atypical generalized highest weight state $\vert\otimes_{k=1}^{L/p}  \{s_1t_1 \cdots s_pt_p\}_{\;k}\rangle$:
	\begin{equation}
		| \, L,M_1,M_2,M_3\rangle_p=\frac{1}{Z_p(L,M_1,M_2,M_3)}F_1^{M_1}F_2^{M_2}F_3^{M_3} \, \vert\otimes_{k=1}^{{L}/{p}}  \{s_1t_1 \cdots s_pt_p\}_{\;k}\rangle \,.
		\label{lm1m2m3p}
	\end{equation}
The term $Z_p(L,M_1,M_2,M_3)$ is a normalization factor, with  $M_1$, $M_2$ and $M_3$  subject to some constraints, which may be related to  $L$. Note that other atypical degenerate ground states may be constructed from the repeated action of $F_1$, $F_2$ and $F_3$, in combination with  the additional $Z_2$ symmetry operation $\sigma$ and the time-reversal operations $K_s$ and $K_t$, as long as they are organized periodically.

The occurrence of atypical degenerate ground states with period $p$ results physically from the fact that SSB occurs from the symmetry group $\mathscr{Z}_L$ under the one-site translation symmetry to a
discrete symmetry group $\mathscr{Z}_{L/p}$ under the $p$-site translation operation when PBCs are adopted, which always accompanies SSB from  ${\rm SU(4)}$ to ${\rm U(1)} \times {\rm U(1)} \times {\rm U(1)}$. As a result, the ground state subspace $\Omega_L^{\rm PBC}$ is decomposed into the direct sum of the sectors $\Omega_L^{\rm PBC}(p)$ invariant under the $p$-site translation operation, namely $\Omega_L^{\rm PBC} = \bigoplus_p \Omega_L^{\rm PBC}(p)$. Here the sum  $\bigoplus_p$ is taken over all possible $p$'s if $p$ divides $L$. 

Indeed, this offers a physical explanation for the fact that the $Z_2$ symmetry operation $\sigma$, together with the time-reversal operations $K_s$ and $K_t$, intertwine with the lowering operators $F_1$, $F_2$ and $F_3$,  which repeatedly act on the highest weight state 
$\vert\otimes_{k=1}^{L}  \{\uparrow_s\uparrow_t\}_{\;k}\rangle$ and the generalized highest weight states, thus yielding exponentially many degenerate ground states  (cf.~Sec.~B of the SM).
This is the reason underlying why the residual entropy is non-zero, since the residual entropy measures the degree of disorder when the $Z_2$ symmetry operation $\sigma$ and  the time-reversal operations $K_s$ and $K_t$ intertwine with the lowering operators $F_1$, $F_2$ and $F_3$, in addition to the degree of disorder in an atypical generalized highest weight state with a given period $p$ as well as in (non-periodic) typical generalized highest weight state (cf. Sec.~B of the SM). Stated differently, it is necessary to introduce both atypical and typical generalized highest weight states to account for the exponential ground state degeneracies with system size $L$.

This classification of degenerate factorized ground states is exactly the same as that for the staggered ${\rm SU}(3)$ ferromagnetic spin-1 biquadratic model~\cite{golden}. We are thus led to the conclusion that {\it atypical}  degenerate ground states are generated from acting lowering operator(s) on atypical factorized degenerate ground states, namely atypical generalized highest weight states.  Notably, the ensemble of highly degenerate atypical (periodic) ground states contains a monomerized ground state with $p=1$, dimerized ground states with $p=2$, trimerized ground states with $p=3$, tetramerized ground states with $p=4$, and so on,  when the thermodynamic limit is approached -- a characteristic feature for this novel type of quantum state of matter.

For our purpose, we shall mainly focus on atypical degenerate ground states. These states fall into two distinct classes: one class is generated from the highest weight state, while the other class is generated from special generalized highest weight states in a period $p$, which is greater than 2. Note that
the period $p$ is equal to 1 or 2 for an atypical degenerate ground state in the first class.
However, it is presumably a formidable task to evaluate the norm for exponentially many atypical degenerate ground states.
For this reason we restrict ourselves to two atypical (periodic) generalized highest weight states, with period $p=4$, in addition to the highest weight state.

\subsection{Atypical degenerate ground states generated from the highest weight state  $\vert\otimes_{k=1}^{L}  \{\uparrow_s\uparrow_t\}_{\;k}\rangle$}

Here we choose $| \, {\rm hws}\rangle= \vert\otimes_{k=1}^{L}  \{\uparrow_s\uparrow_t\}_{\;k}\rangle$ as the highest weight state, with $| \! \uparrow_s\rangle$ and $| \! \uparrow_t\rangle$ the eigenvectors of spin-$1/2$ operators $S_j^z$ and $T_j^z$ for the eigenvalue $1/2$.
Then a sequence of {orthonormal} degenerate ground states $|L,M_1,M_2,M_3\rangle_2$ ($M_1=0$, \ldots, $L/2$, $M_2=0$, \ldots, $L/2$, $M_3=0$, \ldots, $L$), with period 2, are generated from the repeated action of the lowering operators $F_1$, $F_2$ and $F_3$ on the highest weight state  $\vert\otimes_{k=1}^{L}  \{\uparrow_s\uparrow_t\}_{\;k}\rangle$:
	\begin{equation}
		| \, L,M_1,M_2,M_3\rangle_2=\frac{1}{Z_2(L,M_1,M_2,M_3)}F_1^{M_1}F_2^{M_2}F_3^{M_3} \, \vert\otimes_{k=1}^{L}  \{\uparrow_s\uparrow_t\}_{\;k}\rangle,
		\label{lm1m2m3}
	\end{equation}
where the normalization factor $Z_2(L,M_1,M_2,M_3)$ takes the form
	\begin{equation}
		Z_2(L,M_1,M_2,M_3)=\!M_1!M_2!M_3!\sqrt{C_{L/2}^{M_1}C_{L/2-M_1}^{M_2}C_{L-M_1-M_2}^{M_3}} \, .\label{zm1m2m3}
	\end{equation}
A detailed mathematical derivation of $Z_2(L,M_1,M_2,M_3)$ is given in Sec.~C of the SM.

\subsection{Atypical degenerate ground states generated from periodic generalized highest weight states}

A set of generalized highest weight states emerge in addition to the highest weight states, which are responsible for the exponential ground state degeneracies with $L$. Their presence also accounts for the difference between the ground state degeneracies under OBCs and PBCs (cf.~Sec.~B of the SM).
Here two atypical (periodic) generalized highest weight states are chosen as illustrative examples: 
$\vert\otimes_{k=1}^{L/4}  \{\downarrow_s\uparrow_t\uparrow_s\uparrow_t\uparrow_s\uparrow_t\uparrow_s\uparrow_t\}_{\;k}\rangle$ and 
$\vert\otimes_{k=1}^{L/4}  \{\downarrow_s\uparrow_t\downarrow_s\uparrow_t\uparrow_s\uparrow_t\uparrow_s\uparrow_t\}_{\;k}\rangle$, which exhibit a four-site periodic structure.

As an example, degenerate ground states with period $p=4$, denoted as  $|\, L,M_1,M_2,M_3\rangle_4$ ($M_1$, $M_2=0$, \ldots, $L/4$ and $M_3=0$, \ldots, $3L/4-M_1-M_2$), where $L$ is a multiple of four, are generated from the repeated action of the lowering operators $F_1$, $F_2$ and $F_3$ on an atypical generalized highest weight state $\vert\otimes_{k=1}^{L/4}  \{\downarrow_s\uparrow_t\uparrow_s\uparrow_t\uparrow_s\uparrow_t\uparrow_s\uparrow_t\}_{\;k}\rangle$. Mathematically, we have
\begin{equation}
 | \, L,M_1,M_2,M_3\rangle_4=\frac{1}{Z_4(L,M_1,M_2,M_3)}F_1^{M_1}F_2^{M_2}F_3^{M_3} \, \vert\otimes_{k=1}^{{L}/{4}}  \{\downarrow_s\uparrow_t\uparrow_s\uparrow_t\uparrow_s\uparrow_t\uparrow_s\uparrow_t\}_{\;k}\rangle,
		\label{lm1m2m3g}
\end{equation}
where the normalization factor $Z_4(L,M_1,M_2,M_3)$ takes the form
\begin{equation}
		Z_4(L,M_1,M_2,M_3)=M_1!M_2!M_3!\sqrt{C_{L/4}^{M_1}C_{L/4-M_1}^{M_2}C_{3L/4-M_1-M_2}^{M_3}} \, .
\end{equation}
A detailed mathematical derivation of $Z_4(L,M_1,M_2,M_3)$ is given in Sec.~C of the SM.
	
Meanwhile, degenerate ground states with period $p=4$, denoted as $| \, L,M_1,M_3,M_6\rangle_4$ ($M_1=0$, \ldots, $L/4$,  $M_3=0$, \ldots, $L/2$ and $M_6=0$, \ldots, $L/2$), where $L$ is a multiple of four, are generated from the repeated action of the lowering operators $F_1$, $F_3$ and $F_6$ on an atypical generalized highest weight state $\vert\otimes_{k=1}^{L/4}  \{\downarrow_s\uparrow_t\downarrow_s\uparrow_t\uparrow_s\uparrow_t\uparrow_s\uparrow_t\}_{\;k}\rangle$. Mathematically, we have
\begin{equation}
| \, L,M_1,M_3,M_6\rangle_4=\frac{1}{Z_4(L,M_1,M_3,M_6)}
F_1^{M_1}F_3^{M_3}F_6^{M_6} \, \vert\otimes_{k=1}^{{L}/{4}}  \{\downarrow_s\uparrow_t\downarrow_s\uparrow_t\uparrow_s\uparrow_t\uparrow_s\uparrow_t\}_{\;k}\rangle,
\label{lm1m3m6g}
\end{equation}
where the normalization factor $Z_4(L,M_1,M_3,M_6)$ takes the form
\begin{equation}
Z_4(L,M_1,M_3,M_6)=M_1!M_3!M_6!\sqrt{C_{L/4}^{M_1}\sum_{l=0}^{\min(M_6,L/4)}C_{L/4}^{l}C_{L/4}^{M_6-l}C_{L/2-M_1-M_6+l}^{M_3}} \,.
\end{equation}
A detailed mathematical derivation of $Z_4(L,M_1,M_2,M_3)$ is given in Sec.~C of the SM.

\section{Schmidt decomposition for atypical (periodic) degenerate ground states}

We restrict ourselves to the Schmidt decomposition for the two classes of atypical degenerate ground states. Mathematically, polynomially many degenerate ground states generated from the highest weight state constitute orthonormal basis states that span an irreducible representation space of the symmetry group in the ground state subspace. In contrast,  other degenerate ground states generated from  generalized highest weight states are exponential in system size, but it is sufficient to focus on atypical (periodic) degenerate ground states.  As we shall show, both of them are capable of sensing the presence of type-B GMs as a result of SSB. 
In particular,  the presence of atypical degenerate ground states implies the self-similarities underlying the ground state subspace, as already argued in Ref.~\cite{john}.

We choose the degenerate ground states $| \, L,M_1,M_2,M_3\rangle_2$ in Eq.~(\ref{lm1m2m3}), which admit an exact Schmidt decomposition:
\begin{equation}
		| \, L,M_1,M_2,M_3\rangle_2= \sum_{k_1=0}^{\min(M_1,n/2)}\sum_{k_2=0}^{\min(M_2,n/2)}\sum_{k_3=0}^{\min(M_3,n-k_1-k_2)}\lambda(L,k_1,k_2,k_3,M_1,M_2,M_3)
		\, | \, n,k_1,k_2,k_3\rangle_2 \, | \, L-n,M_1-k_1,M_2-k_2,M_3-k_3\rangle_2 \,. \label{svd}
\end{equation}
The Schmidt coefficients $\lambda(L,k_1,k_2,k_3,M_1,M_2,M_3)$ are given by
	\begin{equation*}
		\lambda(L,k_1,k_2,k_3,M_1,M_2,M_3)=C_{M_1}^{k_1}C_{M_2}^{k_2}C_{M_3}^{k_3}\frac{Z_2(n,k_1,k_2,k_3)Z_2(L-n,M_1-k_1,M_2-k_2,M_3-k_3)}{Z_2(L,M_1,M_2,M_3)} \,.
	\end{equation*}
This result in turn may be rewritten as
\begin{equation}
\lambda(L,k_1,k_2,k_3,M_1,M_2,M_3)=\sqrt{\frac{C_{n/2}^{k_1}C_{n/2-k_1}^{k_2}C_{n-k_1-k_2}^{k_3}C_{(L-n)/2}^{M_1-k_1}C_{(L-n)/2-M_1+k_1}^{M_2-k_2}C_{L-n-M_1+k_1-M_2+k_2}^{M_3-k_3}} {C_{L/2}^{M_1}C_{L/2-M_1}^{M_2}C_{L-M_1-M_2}^{M_3}}} \, .
\label{lamm1m2m3}
\end{equation}
Hence the entanglement entropy $S_L(n,M_1,M_2,M_3)$ for a block, consisting of $n$ contiguous lattice sites, is given by 
\begin{align}
	S_L(n,M_1,M_2,M_3)=-\sum_{k_1=0}^{\min(M_1,n/2)}\sum_{k_2=0}^{\min(M_2,n/2)}\sum_{k_3=0}^{\min(M_3,n-k_1-k_2)}\Lambda(L,k_1,k_2,k_3,M_1,M_2,M_3)
	\log_{2}\Lambda(L,k_1,k_2,k_3,M_1,M_2,M_3) \,. 
	\label{snkm1m2m3}
\end{align}
Here $\Lambda(L,k_1,k_2,k_3,M_1,M_2,M_3)=[\lambda(L,k_1,k_2,k_3,M_1,M_2,M_3)]^2$ are the eigenvalues of the reduced density matrix $\rho_L(n,M_1,M_2,M_3)$.

An exact Schmidt decomposition may be similarly performed for the degenerate ground states $|\, L,M_1,M_2,M_3\rangle_4$ and  $| \, L,M_1,M_3,M_6\rangle_4$ in Eq.~(\ref{lm1m2m3g}) and Eq.~(\ref{lm1m3m6g}) (cf.  Sec.~D of the SM).
Accordingly,  the entanglement entropy for a block consisting of $n$ contiguous lattice sites follows.
For convenience, we denote $| \, L,M_1,M_3,M_6\rangle_4$ as $| \, L,M_1^*,M_2^*,M_3^*\rangle_4$, with $M_1^*=M_1$, $M_2^*=M_3$ and $M_3^*=M_6$, respectively. 
The filling factors are respectively defined  as $f_1=M_1/L, f_3=M_3/L$, and $f_6=M_6/L$, or $f_1^*=M_1^*/L$, $f_2^*=M_2^*/L$ and $f_3^*=M_3^*/L$. 
Following exactly the same reasoning as that for the staggered $\rm{SU}(3)$ ferromagnetic spin-1 biquadratic model~\cite{golden}, the  set of all fillings for atypical generalized highest weight states becomes dense in the thermodynamic limit. For completeness, we repeat this discussion in Appendix~C. Consequently, it is justified to limit to the atypical degenerate ground states generated from the action of the lowering operators on atypical generalized highest weight states.

\section{Entanglement entropy}

The presence of an exact Schimdt decomposition for an atypical degenerate ground state greatly simplifies the evaluation of the entanglement entropy. This in turn makes it possible to confirm our previous prediction for the scaling behavior of the entanglement entropy for both the orthonormal basis states generated from the highest weight state~\cite{FMGM,finitesize} and a linear combination of them~\cite{john}. In addition, we also need to extend our discussion of this scaling behavior to other atypical (periodic) degenerate ground states.

As already argued in Ref.~\cite{finitesize}, a sequence of degenerate ground states generated from the repeated action of the lowering operator(s) on the highest weight state are scale-invariant, which act as  orthonormal basis  states in a subspace within the ground state subspace that sense the presence of type-B GMs. For the model under investigation, these degenerate ground states {\it only} accommodate atypical degenerate ground states with the period being up to two.  Particularly, the entanglement entropy $S_{\!\!f}(L,n)$ for a given filling $f$ is given by
\begin{equation}
S_{\!\!f}(L,n)=\frac{N_B}{2} \log_2\frac{n(L-n)}{L} +S_{\!\!f0}.
\label{slnf}
\end{equation}
Note that an additive non-universal constant $S_{\!\!f0}$ has been introduced for this  particular type of scale-invariant state, and the subscript
$f$ refers to a set of fillings $f_1, f_2$, and $f_3$, or $f_1^*$, $f_2^*$ and $f_3^*$.
Here the fillings $f_\alpha$ and $f_\alpha^*$ ($\alpha=1,2,3$)  are assumed to be either non-zero or the maximum. 
In addition,  we note that the scaling relation (\ref{slnf}) is also valid for all atypical (periodic) degenerate ground states, with the period $p$ any integer that divides $L$, given that the argument in Ref.~\cite{finitesize} still applies.

\begin{figure}[t]
	\centering
\includegraphics[width=0.35\textwidth]{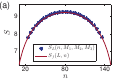}
\includegraphics[width=0.35\textwidth]{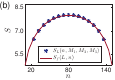}
\includegraphics[width=0.35\textwidth]{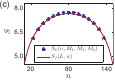}
\caption{ (a) The entanglement entropy $S_L(n,M_1,M_2,M_3)$ vs $n$ against $S_{\!\!f}(L,n)$ vs $n$ for the degenerate ground
state $|L,M_1,M_2,M_3\rangle_2$ in Eq.~(\ref{lm1m2m3}).  The fillings $f_1$, $f_2$ and $f_3$ are chosen to be $f_1=1/5$, $f_2=1/5$ and $f_3=1/4$.
(b) The entanglement entropy $S_L(n,M_1,M_2,M_3)$ vs $n$ against $S_{\!\!f}(L,n)$ vs $n$ for the degenerate ground state $|L,M_1,M_2,M_3\rangle_4$ in Eq.~(\ref{lm1m2m3g}).  The fillings $f_1$, $f_2$ and $f_3$ are chosen to be $f_1=1/16$, $f_2=1/16$ and $f_3=1/8$.
(c) The entanglement entropy $S_L(n,M_1,M_3,M_6)$ vs $n$ against $S_{\!\!f}(L,n)$ vs $n$ for the degenerate ground state $|L,M_1,M_3,M_6\rangle_4$ in Eq.~(\ref{lm1m3m6g}). The fillings $f_1^*$, $f_2^*$ and $f_3^*$ are chosen to be  $f_1^*=1/16$, $f_2^*=1/16$ and $f_3^*=1/16$.
For each case system size $L=160$ with  $n$ ranging from 24 to 136. 
A best fitting procedure yields (a) $S_{\!\!f0}=1.2560$, (b) $S_{\!\!f0}= -0.0034$ and (c) $S_{\!\!f0}=-0.3094$, with the relative errors less than $1.5\%$.
}
\label{comparesu4}
\end{figure}

Plots of the entanglement entropy vs $n$ for the states given in equations (\ref{lm1m2m3}), (\ref{lm1m2m3g}) and (\ref{lm1m3m6g}) are shown in Fig.~\ref{comparesu4} for system size $L=160$, when $n$ ranges from 24 to 136, with the period $p$ being 2, 4 and 4, respectively. 
For each case the results are compared directly against the universal finite-size scaling function $S_{\!\!f}(L,n)$ vs $n$.  The numerical data fall on the curve $S_{\!\!f}(L,n)$, with the relative errors less than $1.5\%$.
Here, we regard $S_{\!\!f}(L,n)$ as a function of $n$ for fixed $L$ and $f$. As a consequence, our numerical check confirms that  the scaling relation (\ref{slnf}) works for all atypical (periodic) degenerate ground states. 

It is worthwhile to mention that this scaling relation (\ref{slnf}) is a special case of a general scaling relation of the entanglement entropy for a linear combination of generalized coherent states generated by the elements of the symmetry group on the highest weight state, as shown in Ref.~\cite{john}. Note that such a linear combination may be expressed in terms of the orthonormal basis states. Mathematically, we have
\begin{equation}
S_{\!\!f}(L,n)=\frac{d_f}{2} \log_2\frac{n(L-n)}{L} +S_{\!\!f0},
\label{slnf1}
\end{equation}
where $d_f$ denotes the fractal dimension for a support  to form a linear combination of  generalized coherent states that sense the presence of type-B GMs.  In Appendix~D, we have numerically checked that this general scaling relation works for a linear combination of generalized coherent states, with support being a Cantor set. This confirmed our prediction in Ref.~\cite{john}.
Here we remark that our numerical check also confirms the theoretical prediction for odd $L$, though the data presented here is restricted to even $L$.

We now turn to the scaling behavior  of the entanglement entropy $S_{\!\!f}(n)$ in the thermodynamic limit.
Indeed, Eq.~(\ref{slnf}) becomes a logarithmic scaling function $S_{\!\!f}(n)$ with the block size $n$ when $L \rightarrow \infty$, if one is restricted to the afore-mentioned orthonormal basis  states spanned an irreducible representation space within the ground state subspace. That is, we have 
\begin{equation}
S_{\!\!f}(n)=\frac{N_B}{2}\log_2n+S_{\!\!f0}.
\end{equation}
This is consistent with a generic but heuristic argument presented in Refs.~\cite{FMGM,finitesize}. Combined with the prediction made by Castro-Alvaredo and Doyon~\cite{doyon}  for the spin-$1/2$ Heisenberg ferromagnetic model and a general argument made in Ref.~\cite{john},
\begin{equation}
S_{\!\!f}(n)=\frac{d_f}{2}\log_2n+S_{\!\!f0},
\end{equation}
we conclude that the fractal dimension $d_f$ is identical to the number of type-B GMs for a sequence of orthonormal basis states in an irreducible representation space of the symmetry group, which are constructed as degenerate ground states from the repeated action of the lowering operators on the highest weight state. Namely $d_f=N_B$ with $N_B=3$ for $L$ even and $N_B=2$ for $L$ odd (cf. Appendix~D for a detailed discussion about the identification of the fractal dimension of a support and the number of type-B GMs). An alternative confirmation for this logarithmic scaling behavior is given in Sec.~E of the SM.

\section{Conclusion}

An extensive investigation has been carried out for the staggered ${\rm SU}(4)$ ferromagnetic spin-orbital model.  The highly degenerate ground states arise from the SSB pattern ${\rm SU(4)} \rightarrow {\rm U(1)} \times {\rm U(1)} \times {\rm U(1)}$, with three type-B GMs,  for even system size and 
from the SSB pattern ${\rm SO(4)} \sim {\rm SU(2)} \times {\rm SU(2)}  \rightarrow {\rm U(1)} \times {\rm U(1)}$, with two type-B GMs,  for odd system size. 
As demonstrated, these degenerate ground states are scale-invariant,  reflecting an abstract fractal underlying the ground state subspace, with the fractal dimension $d_f$ being identified with the number of type-B GMs $N_B$. Namely, $d_f = N_B =3$ for  even system size  and  $d_f = N_B =2$ for odd system size, if PBCs are adopted.

The entanglement entropy of a sequence of orthonormal basis states for the irreducible representation space of the symmetry group has been shown to exhibit a universal finite system-size scaling behavior -- a finite system-size counterpart of the statement that the entanglement entropy  follows a logarithmical relation with the block size in the thermodynamic limit. As a consequence, the model serves as a counter-example to the widely discussed speculation that scale invariance implies conformal invariance~\cite{polyakov}.

In addition,  the ground state degeneracies have been identified to be the two Fibonacci-Lucas sequences, with their
asymptotic forms being a logarithmic spiral when the system size is large enough. 
This in turn implies that the residual entropy is non-zero, and given exactly by $S_{\!r} = -2 \ln R $, where $R=(\! \sqrt{6}-\!\sqrt{2})/2$.

We conclude by recalling that this non-zero residual entropy for the staggered ${\rm SU}(4)$ ferromagnetic spin-orbital model is a consequence of the number of ground states being exponentially large with increasing system size. 
This is a feature in common with the staggered ${\rm SU}(3)$ ferromagnetic spin-1 biquadratic model~\cite{golden}, and as we have observed in Appendix~B, in common with the general staggered ${\rm SU}(2s+1)$ spin-$s$ ferromagnetic model~\cite{barber}.
In contrast, the number of ground states of the ${\rm SO}(2s+1)$ ferromagnetic model is  polynomial in the system size~\cite{son}.
The underlying reason for this fundamental difference is that the ground state space of the ${\rm SO}(2s+1)$ ferromagnetic model is spanned by polynomially many degenerate ground states, originated from the repeated action of the generators of the symmetry group ${\rm SO}(2s+1)$ on the highest weight state, while the ground state space of the general staggered ${\rm SU}(2s+1)$ spin-$s$ ferromagnetic model consists of subspace constructed by actions of lowering operators on the highest weight state, together with a large amount of generalized highest weight states which can be understood by emergent subsystem invertible symmetry tailored to ground states~\cite{dimertrimer, emergentflatband}. 
As further confirmed in the present study, the associated entanglement entropy exhibits universal finite system-size scaling behavior, irrespective of the exponential or polynomial degeneracy of the ferromagnetic ground states, as long as we {\it only} focus on atypical (periodic) degenerate ground states. However, a notable difference between the exponential and polynomial degeneracies lies in the fact that, for the former, atypical degenerate ground states emerge, with the period $p$ ranging from 1 to infinity in the thermodynamic limit, in addition to typical (non-periodic) degenerate ground states. This is a characteristic feature for this type of novel quantum state of matter.

\section{Acknowledgements}

We thank John Fjaerestad for enlightening discussions.
Q.-Q. Shi is supported in part by the National Natural Science Foundation of China (Grant No. 12505003).

\setcounter{equation}{0}
\renewcommand{\theequation}{A\arabic{equation}}

\section*{Appendix A} \label{AppA}

In this Appendix we outline the unitary equivalence between the staggered ${\rm SU}(4)$ ferromagnetic spin-orbital model and the staggered ${\rm SU}(4)$ spin-$3/2$ ferromagnetic chain. 
The former model is described by Hamiltonian~(\ref{hamist}) 
\begin{equation}
	\mathscr{H}=\sum_{j=1}^{L-1}\left(S_j \cdot S_{j+1}-\frac{1}{4}\right) \left(  T_j \cdot T_{j+1}-\frac{1}{4}\right), \label{hamsu3su3sm}
\end{equation}
with OBCs chosen for simplicity.

This model constitutes one realization of the same representation of the Temperley-Lieb algebra~\cite{tla,baxterbook,martin}, a special case of the construction made in Ref.~\cite{barber} for the ${\rm SU(2s+1)}$  spin-$s$ chain.
More precisely, it is unitarily equivalent to the ${\rm SU(4)}$ spin-$3/2$ ferromagnetic chain, with Hamiltonian
\begin{equation}
	\mathscr{H}=\sum_{j=1}^{L-1} \left[\frac{31}{24} \textbf{S}_j \cdot \textbf{S}_{j+1}-\frac{5}{18}(\textbf{S}_j \cdot \textbf{S}_{j+1})^2-\frac{2}{9}(\textbf{S}_j \cdot \textbf{S}_{j+1})^3\right].
	\label{hamsuNsm}
\end{equation}
Here $\textbf{S}_j=(S^x_{j},S^y_{j},S^z_{j})$ represents the spin-$3/2$ operators at site $j$.
Specifically, the unitary transformation connecting the Hamiltonian (\ref{hamsu3su3sm}) to Hamiltonian (\ref{hamsuNsm}) takes the form 
$V =\otimes_k V_{2k-1}\otimes V_{2k}$ ($k=1,\ldots,L/2$ for even $L$), where $V_{2k-1}=V_o$ and  $V_{2k}=V_e $, with $V_o$ and $V_e$ the local unitary transformations
\begin{equation}
	V_o = \begin{pmatrix}
		1 & 0 & 0 & 0\\
		0 & 1 & 0 & 0\\
		0 & 0 & 0 &1\\
		0 & 0 & 1 & 0\\
	\end{pmatrix}
\quad 
\mathrm{and}
\quad
	V_e = \begin{pmatrix}
		0 & 1 & 0 & 0\\
		1 & 0 & 0 & 0\\
		0 & 0 & 1 &0\\
		0 & 0 & 0 & 1\\
	\end{pmatrix} \,.
\end{equation}

The unitary equivalence between the spin-orbital model (\ref{hamsu3su3sm}) and the spin chain (\ref{hamsuNsm}) makes it possible to take advantage of the machinery of the quantum Yang-Baxter equation~\cite{baxterbook,sutherlandb,mccoy}, developed for quantum integrable models, to investigate the staggered ${\rm SU}(4)$ ferromagnetic spin-orbital model (\ref{hamsu3su3sm}), although our approach here is {solely} based on SSB  with type-B GMs.

Here we remark that it is of interest to understand SSB with type-B GMs in the context of the Bethe Ansatz, which remains to be explored. 
In addition, other representations of the Temperley-Lieb operators
are possible, with the antiferromagnetic $16$-state Potts model being an example. 
Note that the latter is not unitarily equivalent to the ${\rm SU(4)}$  spin-$3/2$ ferromagnetic chain and the staggered $\rm SU(4)$ ferromagnetic spin-orbital model, if one restricts to the same size $L$. However, a unitary equivalence between the antiferromagnetic $16$-state Potts model and the staggered $\rm SU(4)$ ferromagnetic spin-orbital model has been established, if system size is doubled for the latter, as shown in Ref.~\cite{frac}.

\setcounter{equation}{0}
\renewcommand{\theequation}{B\arabic{equation}}
\renewcommand{\thetable}{B\arabic{table}}
\section*{Appendix B} \label{AppB}

In this Appendix we discuss the appearance of Fibonacci-Lucas sequences in the staggered ${\rm SU}(4)$ ferromagnetic spin-orbital model. 
We begin by recalling our earlier discussion about the Fibonacci-Lucas sequences in the context of the staggered $\rm{SU}(3)$ ferromagnetic spin-1 biquadratic model~\cite{golden}. As is well known, the two Fibonacci-Lucas sequences~\cite{Fibonacci} follow from a quadratic characteristic equation of the form $X^2-PX+Q=0$, where $P$ and $Q$ are co-prime integers.
The roots $u$ and $v$ of the characteristic equation are such that $u+v=P$ and $uv=Q$.
The two Fibonacci-Lucas sequences $U_m(P,Q)$ and $V_m(P,Q)$, with $m$ a non-negative integer, are then 
\begin{equation}
	U_m(P,Q)=\frac{u^m-v^m}{u-v}, \quad 
	V_m(P,Q)=u^m+v^m.
	\label{uvpqsm}
\end{equation}
They satisfy the three-term recursive relations
\begin{align}
U_{m+2}(P,Q)&=P U_{m+1}(P,Q)-Q U_m(P,Q),\nonumber\\
V_{m+2}(P,Q)&=P V_{m+1}(P,Q)-Q V_m(P,Q),\nonumber
\end{align}
with $U_0(P,Q)=0$, $U_1(P,Q)=1$, $V_0(P,Q)=2$ and $V_1(P,Q)=P$.
A key feature is that they exhibit self-similarities, 
given that one may build rectangles with the ratio $r=\sqrt{u}/\sqrt{v}$ between the length and the width, as an extension of the golden rectangles.

As argued in Refs.~\cite{FMGM,golden}, for a quantum many-body system undergoing SSB with type-B GMs, highly degenerate ground states admit an exact Schmidt decomposition, thus exhibiting the self-similarities, if all the orthonormal basis states spanning an irreducible representation space of the symmetry group are treated as a whole.
In other words, a fractal structure underlies the ground state subspace, which has been reflected in the fact that the ground state degeneracies ${\rm dim }(\Omega_L^{\rm OBC})$ and ${\rm dim }(\Omega_L^{\rm PBC})$  exhibit such self-similarities. 
Indeed, both ${\rm dim }(\Omega_L^{\rm OBC})$ and ${\rm dim }(\Omega_L^{\rm PBC})$ are integers. Then they may be expressed as an integer in terms of an irrational number, namely the Binet formula~\cite{binet}. Assume that the ground state degeneracies ${\rm dim }(\Omega_L^{\rm OBC})$ and ${\rm dim }(\Omega_L^{\rm PBC})$ satisfy a three-term recursive
relation, with the coefficients determined from their values for small $L$, as shown in Table I. 

\begin{table}[ht]
	\centering  
    \setlength{\tabcolsep}{8pt}
	\begin{tabular}{|c|c|c|}
	\hline
	\cline{1-3}
		$L$ & ${\rm dim }(\Omega_L^{\rm OBC})$ &${\rm dim} (\Omega_L^{\rm PBC})$\\
	\hline
        2& 15 & 15\\
	3&  56 & 52\\
	  4&  209 & 194\\
	  5&  780 &724\\
	  6&  2911 &2702\\
	  7&  10864 &10084\\
        8&  40545 &37634\\
	\hline
       \end{tabular}
		\caption{Finite-size ground state degeneracies ${\rm dim }(\Omega_L^{\rm OBC})$  and  ${\rm dim }(\Omega_L^{\rm PBC})$ for the staggered ${\rm SU}(4)$ ferromagnetic spin-orbital model under OBCs and PBCs. }
\end{table}

In this way the ground state degeneracies ${\rm dim }(\Omega_L^{\rm OBC})$ 
and ${\rm dim }(\Omega_L^{\rm PBC})$ constitute the Fibonacci-Lucas sequences $U_m(P,Q)$ and $V_m(P,Q)$, 
respectively, with $P=4$ and $Q=1$,
which in turn may be recognized as those for the ground state degeneracies
${\rm dim }(\Omega_L^{\rm OBC})$  for $L\geq 2$ and ${\rm dim }(\Omega_L^{\rm PBC})$ for $L\geq 3$, 
if $m$ is identified to be $L+1$ for OBCs and $L$ for PBCs.
Hence we have
\begin{align}
	{\rm dim }(\Omega_L^{\rm OBC})&=(R^{-2L-2}-R^{2L+2})/(R^{-2}-R^{2}), \quad L\geq2, \cr
	{\rm dim }(\Omega_L^{\rm PBC})&=R^{-2L}+R^{2L}, \quad  L\geq3.
	\label{gdRsm}
\end{align}
Here $R=(\!\sqrt{6}-\!\sqrt{2})/2$ is an extension of the (inverse) golden ratio. 

The observation that ${\rm dim }(\Omega_L^{\rm PBC})$ is identified as $V_L(4,1)$ is only valid for $L\geq3$. Physically, this is due to the fact that the Hamiltonian under PBCs is identical to that under OBCs if $L=2$. Actually, $V_2(4,1)=14$  in contrast to ${\rm dim }(\Omega_2^{\rm PBC})=15$. A simple explanation for this discrepancy lies in the fact that $V_2(4,1)$ double-counts the removal of the ${\rm SU}(4)$ singlet state under PBCs, in contrast to $U_m(4,1)$ with $m=L+1=3$, which counts the removal of the ${\rm SU}(4)$ singlet state only {\it once}, because there are two distinct ways of wrapping the singlet state around the ring under PBCs. Hence ${\rm dim }(\Omega_2^{\rm PBC})$ does not fit into the Fibonacci-Lucas sequence $V_L(4,1)$ in this particular case.
We thus arrive at the results given in equation~(\ref{degens}).

The non-zero residual entropy $S_{\!r} = -2 \ln R $ follows from the ground state degeneracies being exponential with $L$, similar to the staggered $\rm{SU}(3)$ ferromagnetic spin-1 biquadratic model~\cite{klumper} (see also Refs.~\cite{saleur,katsura}).

We finish by remarking that the ground state degeneracies ${\rm dim }(\Omega_L^{\rm OBC})$ and ${\rm dim }(\Omega_L^{\rm PBC})$ in the general staggered ${\rm SU}(2s+1)$ spin-$s$ ferromagnetic  model~\cite{barber} are also observed to be given by the two Fibonacci-Lucas sequences. That is, they take the same form as Eq.~(\ref{gdRsm}), with now $R=(\!\sqrt{2s+3}-\sqrt{2s-1})/2$. 
Equivalently, the two Fibonacci-Lucas sequences are characterized in terms of Eq.~(\ref{uvpqsm}), with $P=2s+1$ and $Q=1$. Additionally, we point out an intriguing relation between $U_m(P,Q)$ and $V_m(P,Q)$, namely $V_m(P,Q) = P U_m(P,Q) - 2 U_{m-1}(P,Q)$, if $Q=1$.

\section*{Appendix C} \label{AppC}

In this Appendix we provide some additional remarks, which mirror the corresponding discussion for the staggered $\rm{SU}(3)$ ferromagnetic spin-1 biquadratic model~\cite{golden}. As stated in the main text, there is a hierarchical structure emerging behind the exponentially many degenerate ground states, with the most fundamental being the atypical generalized highest weight states. 
Within the framework of the spin-orbital model considered here, one may construct an atypical generalized highest weight state in terms of local states $|\!\uparrow_{s}\uparrow_{t}\rangle$ and $|\!\downarrow_{s}\uparrow_{t}\rangle$, which we recall are the eigenvectors with eigenvalues $\{1/2,1/2\}$ and $\{-1/2,1/2\}$ of $\{S_j^z,T_j^z\}$. 
The filling $f_{d}=N_{d}/L$ is defined for any generalized highest weight state, where $N_{d}$ counts the number of local states $|\!\downarrow_{s}\uparrow_{t}\rangle$ in a specific generalized highest weight state. 
Following the periodicity of an atypical generalized highest weight state, its filling $f_{d}$ becomes $f_{d}=N_d^p/p$, where $N_d^p$ is the number of local states $|\!\downarrow_{s}\uparrow_{t}\rangle$ in one (emergent) unit cell consisting of $p$ adjacent lattice sites, which is distinguished from the filling defined in the main text.

The full set of  fillings for atypical generalized highest weight states constitutes a dense subset within the space of all possible fillings. Consequently, it suffices to only focus on this set in the thermodynamic limit, including all possible values of the period $p$. 
	
We consider two atypical generalized highest weight states with the periods $p_1$ and $p_2$, with respective fillings $f_{d,1}=N_{d,1}^p/p_1$ and $f_{d,2}=N_{d,2}^p/p_2$. Generally, we assume that $f_{d,1} < f_{d,2}$. It is straightforwardly to verify that the filling $f_d=N_d^p/p$, with $p=2p_1 p_2$ and $N_d^p=N_{d,1}^p p_2 +N_{d,2}^p p_1$, satisfies the inequality 
$f_{d,1} < f_d < f_{d,2}$, setting $f_d = 1/2(f_{d,1}+f_{d,2})$.
It is thus implied that for any two atypical generalized highest weight states corresponding to given fillings $f_{d,1}$ and $f_{d,2}$, no matter how close the two fillings are, there is always one atypical generalized highest weight state with a filling that lies between the two fillings. It follows that the set of all the fillings of atypical generalized highest weight states is dense. 
	
Since all atypical degenerate ground states can be generated from the action of the lowering operators on all atypical generalized highest weight states, we can further conclude that the set for all the fillings of atypical degenerate ground states is dense over the entire space of all physically allowed fillings in the thermodynamic limit.

\renewcommand{\theequation}{C\arabic{equation}}
\renewcommand{\thefigure}{C\arabic{figure}}
\setcounter{equation}{0}
\setcounter{figure}{0}
\section*{Appendix D} \label{AppD}

In this Appendix we follow the related calculations for the staggered $\rm{SU}(3)$ ferromagnetic spin-1 biquadratic model~\cite{golden}. Specifically, as given in Appendix E therein.

In Ref.~\cite{john}, an alternative approach is developed to identify the fractal dimension with the number of type-B GMs for the orthonormal basis states in an irreducible representation space of the symmetry group. 
Generically, we need to adapt the conventional definition of generalized coherent states~\cite{gilmore} to be suitable for our discussion about SSB in the staggered $\rm SU(4)$ ferromagnetic spin-orbital model. Indeed, they are defined to be a subset of degenerate factorized ground states, namely exponentially many generalized highest weight states. Hence they are over-complete. 
A linear combination of  such an over-complete set of non-orthonormal basis states on any fractal $C$ as a subset in the coset space can be formed, with a Cantor set as a typical example. Interestingly,
the entanglement entropy $S(L,n)$ for a linear combination of generalized coherent states on a fractal $C$  scales as
\begin{equation}
S(L,n)=\frac{d_f}{2}\log_2{\frac{n(L-n)}{n}}+S_{\!0}.\label{dfsln} 
\end{equation}
Here $S(L,n)=S(L,L-n)$, with $S_{\!0}$ a model-dependent constant, and $d_f$ is the fractal dimension of the fractal $C$, which acts as a support to form a linear combination. 
As argued~\cite{john},  any support may be approximated in terms of a fractal decomposable into a set of the Cantor sets.

In general, a Cantor set may be created step by step.  
Starting from the interval  $[0,1]$, by dividing the interval $[0,1]$ into $1/r$ parts, and removing $1/r -N$ subintervals, we have $C[N,r;1]$ at the first iteration. 
By definition, $1/r$ is always a positive integer and $1/r > N$. 
Repeating the same procedure, and removing the subintervals in the same way as done for $C[N,r;1]$, we have $C[N,r;2]$ at the second iteration. 
Repeating the same procedure $k$ times, we have $C[N,r;k]$ at the $k$-th iteration, thus a (generalized) Cantor set is generated, denoted as $C[N,r;\{k\}]$. The number of subintervals in a Cantor set $C[N,r;\{k\}]$ is $N^k$ at the $k$-th iteration, with the fractal dimension $d_f$ for the Cantor set $C[N,r;\{k\}]$ being $d_f= -\ln N/\ln r$.

In this work, we should make a distinction between odd and even $L$ under PBCs, arising from the even-odd parity effect. 
Although we only focus on PBCs, the discussion below for even $L$ is valid  under OBCs, as the same staggered $\rm{SU(4)}$ group is preserved.

\subsubsection{Entanglement entropy scaling under PBCs: odd $L$'s}

For odd $L$'s, the ferromagnetic spin-orbital model (\ref{hamist}) under PBCs possesses the uniform symmetry group  ${\rm SO}(4) \sim \rm{SU(2)}\times \rm{SU(2)}$, generated by two copies of the ${\rm SU(2)}$ generators, denoted as $S^{x},S^{y}$ and $S^{z}$, and $T^{x},T^{y}$ and $T^{z}$. Their explicit forms are
$S^x=\sum_jS^x_j$, $S^y=\sum_jS^x_j$ and $S^z=\sum_j S^z_j$, and $T^x=\sum_jT^x_j$, $T^y=\sum_jT^y_j$ and $T^z=\sum_j T^z_j$, with raising and lowering operators
 $S_\pm=(S^x\pm iS^y)/\sqrt{2}$ and  $T_\pm=(T^x\pm iT^y)/\sqrt{2}$.

The coset space is diffeomorphic to $S^2\times S^2$, which accommodates spin coherent states $|\psi(\theta_s,\theta_t,\phi_s,\phi_t)\rangle$  expressed in terms of the spherical coordinates $\theta_s, \theta_t \in [0,\pi]$ and $\phi_s, \phi_t \in [0,2\pi]$ on each of the two spheres $S^2$. Mathematically, we have

\begin{equation*}
|\psi(\theta_s,\theta_t,\phi_s,\phi_t)\rangle=|v(\theta_s,\theta_t,\phi_s,\phi_t)\rangle_1 \cdots |v(\theta_s,\theta_t,\phi_s,\phi_t)\rangle_j \cdots |v(\theta_s,\theta_t,\phi_s,\phi_t)\rangle_L,
\end{equation*}
where
\begin{equation*}
|v(\theta_s,\theta_t,\phi_s,\phi_t)\rangle_j=\exp(i\phi_s S^z_{j})\exp(i\theta_s S^y_{j})\;\exp(i\phi_t T^z_{j})\exp(i\theta_t T^y_{j})\;|{\rm{hws}}\rangle_j.
\end{equation*}
Here $|{\rm{hws}}\rangle_j$ represents the local component of the highest weight state at a lattice site $j$, where $|\rm{hws}\rangle \equiv \ket{\otimes_{j=1}^{L} \{\uparrow_{s}\uparrow_{t}\}_{j}}$, with $\vert \uparrow_{sj} \rangle$ and $\vert \uparrow_{tj} \rangle$ representing the eigenvectors  of $S^z_j$ and $T^z_j$, with the eigenvalue $1/2$. Indeed, a linear combination of a set of the overcomplete basis states $|\psi(\theta_s,\theta_t,\phi_s,\phi_t)\rangle$ may be formed on a fractal $C$,  with the Cantor sets $C[N_s,r_s;\{k_s\}]$ and  $C[N_t,r_t;\{k_t\}]$ as examples. 
As a convention, $C[N_s,r_s;\{k_s\}]$ and  $C[N_t,r_t;\{k_t\}]$ may be understood as the image under the mapping $\phi_s,\phi_t: [0,1] \rightarrow S^1$, defined as $\phi_{s/t} (\xi) = 2\pi \xi$, or alternatively, under the mapping $\theta_s,\theta_t: [0,1] \rightarrow S^1$, defined as $\theta_{s/t} (\xi) = \pi/2 \xi$, where $\xi \in [0,1]$. 
Considering the symmetric structure, the two mappings defined above are sufficient. 
For brevity, we do not make any distinction between the image of a fractal under the mapping $\phi_{s/t}$ or $\theta_{s/t}$ and a fractal itself here and afterward.

For a SSB pattern from the  $\rm{SU(2)}\times \rm{SU(2)}$ symmetry group to the residual symmetry group $\rm{U(1)}\times \rm{U(1)}$, one may construct highly degenerate ground states $|L,M_s,M_t\rangle$ from the repeated action of the lowering operators $S_-$  and $T_-$ on the highest weight state $|{\rm hws}\rangle$:
$|L,M_s,M_t\rangle=1/Z(L,M_s,M_t) S_-^{M_s} T_-^{M_t}|{\rm hws}\rangle$ ($M_s,M_t=0$, $1$, $2$, \ldots, $L$), where $Z(L,M_s,M_t)$ is introduced to ensure a normalized $|L,M_s,M_t\rangle$, given $Z(L,M_s,M_t)=M_s!M_t!\sqrt{C_{L}^{M_s}C_{L}^{M_t}}$. 
Actually, it constitutes a set of orthonormal basis states in the coset space $CP^{1}\times CP^{1}$, which is diffeomorphic to $S^2\times S^2$.
It turns out that the orthonormal basis states $|L,M_s,M_t\rangle$ span an irreducible representation of the symmetry group ${\rm SU(2)} \times {\rm SU(2)}$, with the dimension being $(L+1)^2$.
Indeed, one may factorize the orthonormal basis states $|L,M_s,M_t\rangle$ as $|L,M_s,M_t\rangle = |L,M_s\rangle|L,M_t\rangle$.

An exact Schmidt decomposition may be performed in the orthonormal basis states $|L,M_s,M_t\rangle$:
\begin{equation}
 |L,M_s,M_t\rangle=\sum_{k_s,k_t=0}^{n}\lambda(L,n,k_s,k_t,M_s,M_t) 
|n,k_s,k_t\rangle|L-n,M_s-k_s,M_t-k_t\rangle,	\label{lmschmiditso}
\end{equation}
where the Schmidt coefficients $\lambda(L,n,k_s,k_t,M_s,M_t)$ are 
\begin{equation*}
\lambda(L,n,k_s,k_t,M_s,M_t)=\sqrt{\frac{C_{n}^{k_s}C_{n}^{k_t}C_{L-n}^{M_s-k_s}C_{L-n}^{M_t-k_t}} {C_{L}^{M_s}C_{L}^{M_t}}}.
\end{equation*}

We turn to the connection between the fractal dimension $d_f$ and the number of type-B GMs $N_B$ for the orthonormal basis states  $|L,M_s,M_t\rangle$ ($M_s=0$, \ldots, $L$ and $M_t=0$, \ldots, $L$), which span the ground state subspace.  
Now $|\psi(\theta_s,\phi_s; \theta_t,\phi_t)\rangle$ may be expanded into a linear combination
in terms of $|L,M_s,M_t\rangle$: 
\begin{equation*}
|\psi(\theta_s,\phi_s; \theta_t,\phi_t)\rangle=\sum_{M_s,M_t=0}^{L}a_{LM_s,M_t}(\theta_s,\phi_s; \theta_t,\phi_t)|L,M_s,M_t\rangle, 
\end{equation*}
where $a_{LM_sM_t}$ are complex numbers, which are formally equal to $a_{LM_sM_t}(\theta_s,\phi_s; \theta_t,\phi_t)= \langle L,M_s,M_t |\psi(\theta_s,\phi_s; \theta_t,\phi_t)\rangle$. It is readily seen that $a_{LM_sM_t}(\theta_s,\phi_s; \theta_t,\phi_t)$ is proportional to $\exp(i\phi_s(L/2-M_s))\exp(i\phi_t(L/2-M_t))$, i.e, 
\begin{equation*}
a_{LM_sM_t}(\theta_s,\phi_s; \theta_t,\phi_t)=b_{LM_s}(\theta_s)b_{LM_t}(\theta_t)\exp(i\phi_s(L/2-M_s))\exp(i\phi_t(L/2-M_t)).
\end{equation*}
The explicit expression for $b_{LM_{s/t}}(\theta_{s/t})$ is 
\begin{equation*}
b_{LM_{s/t}}(\theta_{s/t})=(-1)^{M_{s/t}}\sqrt{C_L^{M_{s/t}}} \cos(\frac{\theta_{s/t}}{2})^{L-M_{s/t}}\sin(\frac{\theta_{s/t}}{2})^{M_{s/t}}.
\end{equation*}
Given $1/(2\pi)\int_{0}^{2\pi} d\phi_{s/t} \exp(i\phi_{s/t} (L/2-M_{s/t})) = \delta_{L/2\;M_{s/t}}$,  $|L,M_s,M_t\rangle$ are expressed in terms of $|\psi_s(\theta_s,\phi_s; \theta_t,\phi_t)\rangle$ on  $S^1 \times S^1$ with fixed $\theta_s$ and $\theta_t$,
\begin{equation}
|L,M_s,M_t\rangle  =\frac{1}{b_{LM_{s}}(\theta_{s})b_{LM_{t}}(\theta_{t})} \int_{0}^{2\pi} d\phi_s  \int_{0}^{2\pi} d\phi_t 
\exp \left(-i\phi_s(L/2-M_s)\right)\exp \left(-i\phi_t(L/2-M_t)\right) |\psi(\theta_s,\phi_s; \theta_t,\phi_t)\rangle.  \nonumber
\end{equation}
 This representation is a linear combination of the overcomplete basis states $|\psi(\theta_s,\phi_s; \theta_t,\phi_t)\rangle$, where the coefficients only involve a phase factor. 
The fractal dimension $d_f$ is thus equal to 2 for the orthonormal basis states $|L,M_s,M_t\rangle$,  identical to the number of type-B GMs for odd $L$'s.

We plot the entanglement entropy $S(L,n)$ versus $n$ for the ${\rm SO}(4)$ ferromagnetic spin-orbital model under PBCs for odd system size. Here a degenerate ground state is chosen to be $|L,M_s,M_t\rangle$, where $L=201$, and $M_s$ and $M_t$ take different values, as indicated. The data are shown to obey the scaling relation (\ref{slnf}), with the prefactor half the number of GMs $N_B=2$, with a relative error less than $2\%$.

\begin{figure}[h!]
	\centering
	\includegraphics[width=0.5\textwidth]{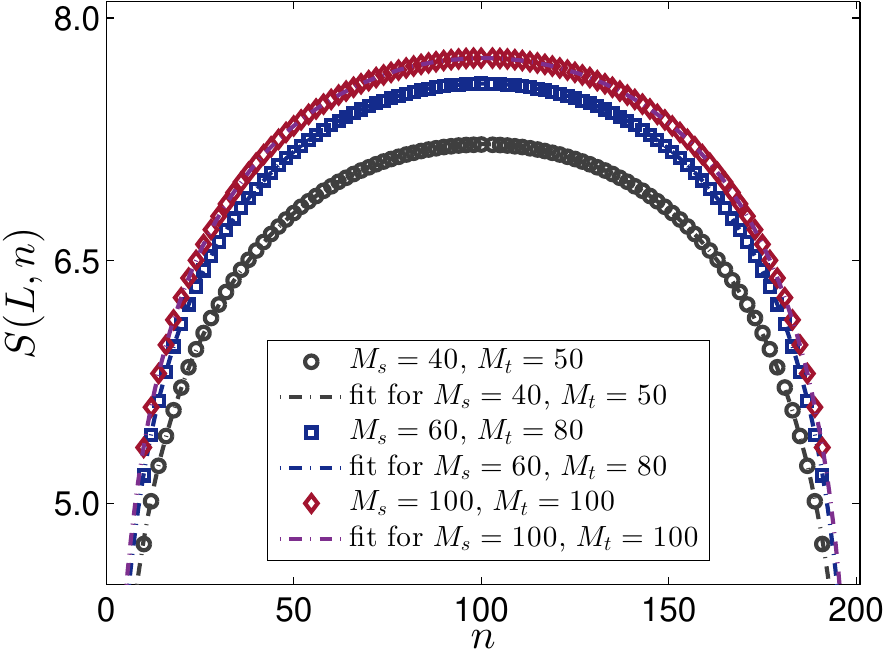}
	\caption{The entanglement entropy $S(L,n)$ versus $n$ for the ${\rm SO}(4)$ ferromagnetic spin-orbital model under PBCs when $L$ is odd. Here a degenerate ground state is chosen to be $|L,M_s,M_t\rangle$, where $L=201$, and $M_s$ and $M_t$ take different values, as indicated. The data are shown to obey the scaling relation (\ref{slnf}), with the prefactor half the number of GMs $N_B=2$.} 
	\label{C1}
\end{figure}

\begin{figure}[h!]
	\centering
	\includegraphics[width=0.5\textwidth]{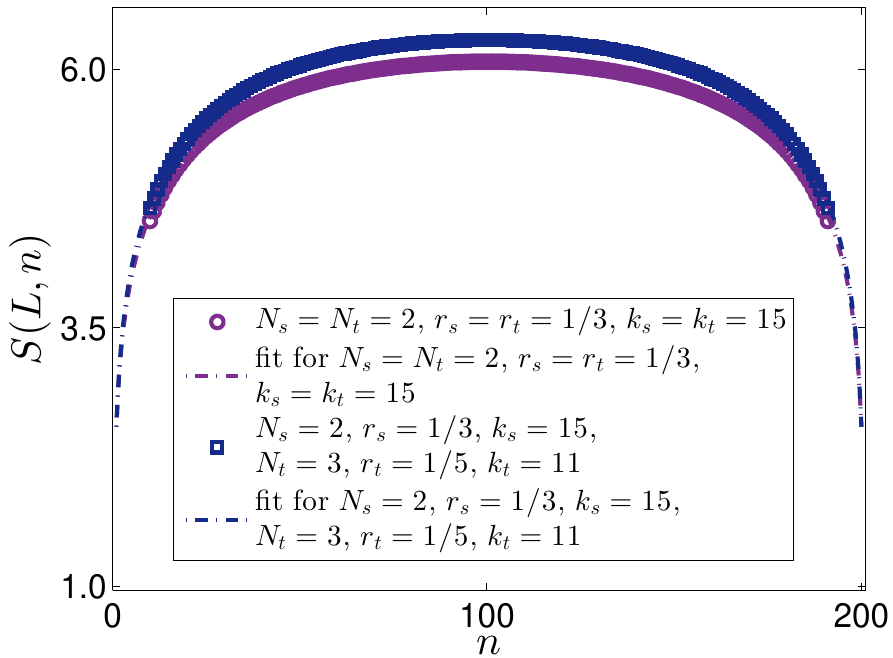}
	\caption{The entanglement entropy $S(L,n)$ versus $n$ for the ${\rm SO}(4)$ ferromagnetic spin-orbital model under PBCs when $L$ is odd.  Here a degenerate ground state $|\Phi_C(\theta_s,\theta_t)\rangle$ is chosen as a linear combination on the Cantor sets $C[N_s,r_s;\{k_s\}]$ and $C[N_t,r_t;\{k_t\}]$, as defined in Eq.(\ref{lcsu4}) with equal coefficients,  where $L=201$, and $\theta_s=\pi/2$ and $\theta_t=\pi/2$. Note that $N_s$, $r_s$, $k_s$, $N_t$, $r_t$ and $k_t$ take different values, as indicated.  The data are shown to obey the scaling relation (\ref{dfsln}), with the prefactor half the sum of the fractal dimensions $d_{f,s}$ and $d_{f,t}$.}
	\label{C2}
\end{figure}

Then we consider a linear combination on a fractal decomposable into a set of the Cantor sets on the circles by setting $\theta_s$ and $\theta_t$ to be certain values in the interval $(0,\pi)$, whereas $\phi_1$ and $\phi_2$ vary from 0 to $2\pi$.
Specifically, a linear combination on a fractal consisting of two Cantor sets $C[N_s;r_s;\{k_s\}]$ and  $C[N_t;r_t;\{k_t\}]$ located on
the two circles with fixed  $\theta_s$ and $\theta_t$ takes the form
\begin{align}
|\Phi_C(\theta_s,\theta_{t})\rangle=&\frac{1}{Z_{C}} \sum_{\substack{\phi_{s,\beta} \in C[N_s;r_s;\{k_s\}],\\\;\phi_{t,\gamma} \in C[N_t;r_t;\{k_t\}]}} c(\phi_{s,\beta},\;\phi_{t,\gamma}) |\psi(\theta_s,\phi_{s,\beta};\theta_{t},\phi_{t,\gamma})\rangle, \label{lcsu4}
\end{align}
where $c(\phi_{s,\beta},\;\phi_{t,\gamma})$ ($\beta =1,2,\ldots,N_s^{k_s}$, $\gamma =1,2,\ldots,N_t^{k_t}$) are complex numbers, and ${Z_C}$ is a normalization factor that ensures $|\Phi_C(\theta_s,\theta_{t})\rangle$ is normalized.  
Restrictions are imposed on the coefficients $c(\phi_{s,\beta},\;\phi_{t,\gamma})$ in the linear combinations in the previous study~\cite{john}.  For brevity, we have assumed that equal coefficients for any  $\phi_{s,\beta}$ and $\phi_{t,\gamma}$, i.e., $c(\phi_{s,\beta},\;\phi_{t,\gamma})=1$, and the sum over $\phi_{s,\beta}$ and $\phi_{t,\gamma}$ is carried out for all the subintervals for chosen $k_s$ and $k_t$. 
As shown in Fig.~\ref{C2},  the entanglement entropy $S(L,n)$  versus $n$  for a degenerate ground state  $|\Phi_C(\theta_s,\theta_{t})\rangle$, as a linear combination with equal coefficients, on the Cantor sets $C[N_s,r_s;\{k_s\}]$ and $C[N_t,r_t;\{k_t\}]$ with $\theta_s=\pi/2$ and $\theta_t=\pi/2$ in the ferromagnetic spin-orbital model when $L=201$, with  $N_s=N_t=2$, $r_s=r_t=1/3$ at the iteration $k_s=k_t=15$, and $N_s=2$,  $r_s=1/3$ at the iteration $k_s=15$, $N_t=3$, $r_t=1/5$ at the iteration $k_t=11$. 
In this example the prefactor in Eq.~(\ref{dfsln}) is observed to be half the sum of the fractal dimensions $d_{f,s}$ and $d_{f,t}$ of the Cantor sets $C[N_s,r_s;\{k_s\}]$ and $C[N_t,r_t;\{k_t\}]$, where $d_{f,{s}}=-\ln N_{s}/\ln r_{s}$ and $d_{f,{t}}=-\ln N_{t}/\ln r_{t}$, with relative errors less than 3$\%$.

\subsubsection{Entanglement entropy scaling under PBCs:  even $L$'s}

As already noted in the main text, when $L$ is even, the symmetry group is the staggered ${\rm SU}(4)$ group. Hence three ${\rm SU}(2)$ subgroups may be introduced, each of which is associated with one of the three type-B GMs as a result of SSB.
One may define the three generators for each ${\rm SU}(2)$ subgroups as $\Sigma_{\alpha}^x$, $\Sigma_{\alpha}^y$ and $\Sigma_{\alpha}^z$ ($\alpha=1,2,3$), with $\Sigma_{\alpha,j}^x=(E_{\alpha,j}+F_{\alpha,j})/2$, $\Sigma_{\alpha,j}^y=-i(E_{\alpha,j}-F_{\alpha,j})/2$ and  $\Sigma_{\alpha,j}^z=H_{\alpha,j}/2$. 
The generators $\Sigma_{\alpha}^x$, $\Sigma_{\alpha}^y$ and $\Sigma_{\alpha}^z$ satisfy  $[\Sigma_{\alpha}^{x},\Sigma_{\alpha}^{y}]=i\Sigma_{\alpha}^{z}$, $[\Sigma_{\alpha}^{y},\Sigma_{\alpha}^{z}]=i\Sigma_{\alpha}^{x}$ and $[\Sigma_{\alpha}^{z},\Sigma_{\alpha}^{x}]=i\Sigma_{\alpha}^{y}$ ($\alpha=1,2,3$).
Note that the set of the overcomplete basis states consists of factorized (unentangled) ground states  $|\psi(\theta_1,\phi_1;\theta_{2},\phi_{2};\theta_{3},\phi_{3})\rangle$, which are parameterized in terms of  $\theta_\alpha \in [0,\pi]$ and $\phi_\alpha \in [0,2\pi]$ ($\alpha=1,2,3$). 
Explicitly, we have
\begin{equation*}
|\psi(\theta_1,\phi_1;\theta_{2},\phi_{2};\theta_{3},\phi_{3})\rangle=|v(\theta_1,\phi_1;\theta_{2},\phi_{2};\theta_{3},\phi_{3})\rangle_1 \cdots |v(\theta_1,\phi_1;\theta_{2},\phi_{2};\theta_{3},\phi_{3})\rangle_{2l-1} |v(\theta_1,\phi_1;\theta_{2},\phi_{2};\theta_{3},\phi_{3})\rangle_{2l}\cdots |v(\theta_1,\phi_1;\theta_{2},\phi_{2};;\theta_{3},\phi_{3})\rangle_L,
\end{equation*}
with
\begin{align*}
|v(\theta_1,\phi_1;\theta_{2},\phi_{2};\theta_{3},\phi_{3})\rangle_{2l-1/2l}=&\exp(i\phi_{3} \Sigma_{3,2l-1/2l}^{z})\exp(i\theta_{3} \Sigma_{3,2l-1/2l}^{y})\exp(i\phi_{2} \Sigma_{2,2l-1/2l}^{z}) ~ \times\nonumber \\
	&\exp(i\theta_{2} \Sigma_{2,2l-1/2l}^{y})\exp(i\phi_1 \Sigma_{1,2l-1/2l}^{z})\exp(i\theta_1 \Sigma_{1,2l-1/2l}^{y})\;|\uparrow_s\uparrow_t\rangle_{2l-1/2l}.
\end{align*}
Here  $l=1,2,\ldots,L/2$. 
Indeed, $|\psi(\theta_1,\phi_1;\theta_2,\phi_2;\theta_{3},\phi_{3})\rangle$ may be regarded as a variant of an extension of the spin coherent states~\cite{gilmore}. 
The parameter space is a $3$-dimensional (complex) manifold, denoted as $CP^{3}_{\pm}$. 
Specifically, at the  $(2l-1)$-th lattice site, we have
	\begin{equation*}
	 |v(\theta_1,\phi_1;\theta_2,\phi_2;\theta_3,\phi_3)\rangle_{2l-1}= 
	 \begin{pmatrix} 
	 	\cos\frac{\theta_1}{2}\cos\frac{\theta_2}{2}\cos\frac{\theta_3}{2}\exp(i\frac{\phi_1 + \phi_2+ \phi_3}{2})\\
	 -\sin\frac{\theta_1}{2}\exp(-i\frac{\phi_1}{2}) \\ 
		-\cos\frac{\theta_1}{2}\sin\frac{\theta_2}{2}\exp(i\frac{\phi_1 - \phi_2}{2}) \\
	-\cos\frac{\theta_1}{2}\cos\frac{\theta_2}{2}\sin\frac{\theta_3}{2}\exp(i\frac{\phi_1 +\phi_2 - \phi_3}{2})  
	\end{pmatrix},
	\end{equation*}
and at the $2l$-th lattice site, we have
	\begin{equation*}
|v(\theta_1,\phi_1;\theta_2,\phi_2;\theta_3,\phi_3)\rangle_{2l}= \begin{pmatrix} 
	\cos\frac{\theta_3}{2}\exp(i\frac{\phi_3}{2})\\
	0\\
	0\\
	\sin\frac{\theta_3}{2}\exp(-i\frac{\phi_3}{2}) 
	 \end{pmatrix}.
	\end{equation*}

The orthonormal basis states $|L,M_1,M_2,M_3\rangle_2$ in Eq.~(\ref{lm1m2m3}), with period 2, where $M_1=0, \ldots, L/2$, $M_2=0, \ldots, L/2$, $M_3=0, \ldots, L$, span an irreducible representation space of the symmetry group in the ground state subspace. They are generated from the repeated action of the lowering operators $F_1$, $F_2$ and $F_3$ on the highest weight state $| \, {\rm hws}\rangle= \vert\otimes_{k=1}^{L}  \{\uparrow_s\uparrow_t\}_{\;k}\rangle$.  Hence $|\psi(\theta_1,\phi_1;\theta_2,\phi_2;\theta_3,\phi_3)\rangle$ may be expanded into a linear combination in terms of $|L,M_1,M_2,M_3\rangle_2$, namely  \begin{equation*}
|\psi(\theta_1,\phi_1;\theta_2,\phi_2; \theta_3,\phi_3)\rangle=\sum_{M_1=0}^{L/2}\sum_{M_2=0}^{L/2}\sum_{M_3=0}^{L}a_{LM_1M_2M_3}(\theta_1,\phi_1;\theta_2,\phi_2;\theta_3,\phi_3) |L,M_1,M_2,M_3\rangle_2, 
\end{equation*}
where $a_{LM_1M_2M_3}$ are complex numbers, which are formally defined as follows 
\begin{equation*}
a_{LM_1M_2M_3}(\theta_1,\phi_1;\theta_2,\phi_2;\theta_3,\phi_3)= \langle L,M_1,M_2,M_3 |\psi(\theta_1,\phi_1;\theta_2,\phi_2;\theta_3,\phi_3)\rangle.
\end{equation*}
A detailed calculation yields 
\begin{equation*}
a_{LM_1M_2M_3}(\theta_1,\phi_1;\theta_2,\phi_2;\theta_3,\phi_3)=b_{LM_1M_2M_3}(\theta_1,\theta_2,\theta_3)\exp(i\phi_1/2(L/2-2M_1))\exp(i\phi_1/2(L/2-M_1-2M_2))\exp(i\phi_2/2(L-M_1-M_2-2M_3)).
\end{equation*}
Here $b_{LM_1M_2M_3}(\theta_1,\theta_2,\theta_3)$ is given by 
\begin{align*}
b_{LM_1M_2M_3}(\theta_1,\theta_2,\theta_3)= & \, 
(-1)^{M_1+M_2+M_3}(\sin\frac{\theta_1}{2})^{M_1}(\cos\frac{\theta_1}{2})^{\frac{L}{2}-M_1}
(\sin\frac{\theta_2}{2})^{M_2}(\cos\frac{\theta_2}{2})^{\frac{L}{2}-M_1-M_2}	
(\sin\frac{\theta_3}{2})^{M_3}(\cos\frac{\theta_3}{2})^{L-M_1-M_2-M_3}\nonumber \\
& \times \sqrt{C_{L/2}^{M_1}C_{L/2-M_1}^{M_2}C_{L-M_1-M_2}^{M_3}}.
\end{align*}
	
Taking into account 
\begin{equation*}
1/(2\pi)\int_{0}^{2\pi} d\phi_1 \exp(i\phi_1(L/4-M_1)) = \delta_{L/4\;M_1},
\end{equation*}
\begin{equation*}
1/(2\pi)\int_{0}^{2\pi} d\phi_2 \exp(i\phi_2(L/4-M_1/2-M_2)) = \delta_{L/4\;M_1/2+M_2}, 
\end{equation*}
and 
\begin{equation*}
1/(2\pi)\int_{0}^{2\pi} d\phi_3 \exp(i\phi_3(L/2-M_1/2-M_2/2-M_3)) = \delta_{L\;M_1+M_2+2M_3}, 
\end{equation*}
one may rewrite the state $|L,M_1,M_2,M_3\rangle_2$ as follows
\begin{align*}
	|L,M_1,M_2,M_3\rangle_2=& \frac {1} {b_{LM_1M_2M_3}(\theta_1,\theta_2,\theta_3)}\int_{0}^{2\pi} d\phi_1 \exp(-i\phi_1(\frac{L}{4}-M_1)) \int_{0}^{2\pi} d\phi_2 \exp(-i\phi_2(\frac{L}{4}-\frac{M_1}{2}-M_2)) \\
	& \times \int_{0}^{2\pi} d\phi_3 \exp(-i\phi_2(\frac{L}{2}-\frac{M_1}{2}-\frac{M_2}{2}-M_3))
	|\psi(\theta_1,\phi_1;\theta_2,\phi_2;\theta_3,\phi_3)\rangle,
	\label{lmexp}
\end{align*}
where $\theta_1$, $\theta_2$ and $\theta_3$ in $|\psi(\theta_1,\phi_1;\theta_2,\phi_2;\theta_3,\phi_3)\rangle$ are fixed ($0<\theta_1<\pi$, $0<\theta_2<\pi$ and $0<\theta_3<\pi$), whereas $\phi_1$, $\phi_2$ and $\phi_3$ are varied. Indeed, this is a linear combination of the overcomplete basis states $|\psi(\theta_1,\phi_1;\theta_2,\phi_2;\theta_3,\phi_3)\rangle$, with only a phase factor involved in the coefficients.
Consequently, the fractal dimension $d_f$ for the orthonormal basis states $|L,M_1,M_2,M_3\rangle_2$ is equal to 3, identical to the number of type-B GMs $N_B$.

We now consider a linear combination on a decomposable fractal.
Mathematically, such  a linear combination on a fractal consisting of three Cantor sets $C[N_1;r_1;\{k_1\}]$,   $C[N_2;r_2;\{k_2\}]$ and  $C[N_3;r_3;\{k_3\}]$ located on the three circles with fixed  $\theta_1$, $\theta_2$  and $\theta_3$ takes the form
\begin{equation}
|\Phi_C(\theta_1,\theta_{2},\theta_{3})\rangle=
\frac{1}{Z_{C}} \sum_{\substack{\phi_{1,\beta} \in C[N_1;r_1;\{k_1\}],\;\\\phi_{2,\gamma} \in C[N_2;r_2;\{k_2\}]
\;\\\phi_{3,\gamma} \in C[N_3;r_3;\{k_3\}]}} 
c(\phi_{1,\beta},\;\phi_{2,\gamma}\;\phi_{3,\eta})  |\psi(\theta_1,\phi_{1,\beta};\theta_{2},\phi_{2,\gamma};\theta_{3},\phi_{3,\gamma})\rangle. 
\label{lcsu4}
\end{equation}
Here $\phi_1$, $\phi_2$  and $\phi_3$ vary from 0 to $2\pi$. Note that $c(\phi_{1,\beta},\;\phi_{2,\gamma},\;\phi_{3,\eta})$ ($\beta =1,2,\ldots,N_1^{k_1}$, $\gamma =1,2,\ldots,N_2^{k_2}$, $\eta =1,2,\ldots,N_3^{k_3}$) are complex numbers, and ${Z_C}$ is a normalization factor to ensure that $|\Phi_C(\theta_1,\theta_{2},\theta_{3})\rangle$ has been normalized. For brevity, we focus on the case with equal coefficients, namely, $c(\phi_{1,\beta},\;\phi_{2,\gamma},\;\phi_{3,\eta})=1$.

\begin{figure}[ht!]
	\centering
	\includegraphics[width=0.55\textwidth]{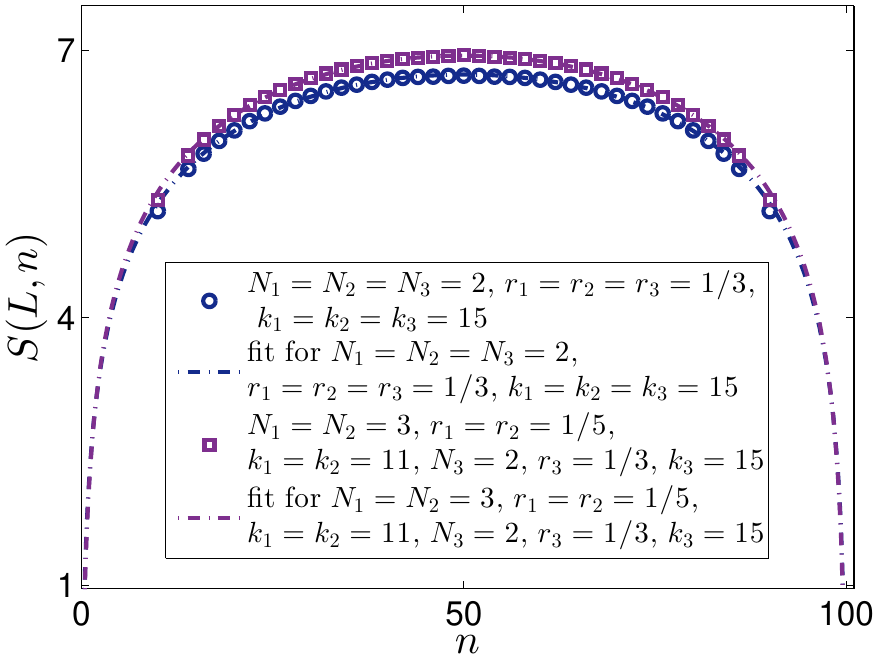}
	\caption{The entanglement entropy $S(L,n)$ versus $n$ for the staggered ${\rm SU}(4)$ ferromagnetic spin-orbital model when $L$ is even. Here a degenerate ground state, $|\Phi_C(\theta_1,\theta_{2},\theta_{3})\rangle$, as a linear
combination with equal coefficients, on a fractal decomposable into three Cantor sets $C[N_1,r_1;\{k_1\}]$, $C[N_2,r_2;\{k_2\}]$ and $C[N_3,r_3;\{k_3\}]$. Here we have taken $L=100$ and $\theta_1=\theta_2=\theta_3=\pi/2$. Meanwhile, we consider two cases: in the first case, we have $N_1=N_2=N_3=2$, $r_1=r_2=r_3=1/3$ and $k_1=k_2=k_3=15$, with the data labeled by circles; in the second case, we have $N_1=N_2=3$, $r_1=r_2=1/5$ and $k_1=k_2=11$, whereas $N_3=2$, $r_3=1/3$ and $k_3=15$,  with the data labeled by squares.}
	\label{C3}
\end{figure}

In Fig.~\ref{C3}, we plot the entanglement entropy $S(L,n)$  versus $n$  for a degenerate ground state  $|\Phi_C(\theta_1,\theta_2,\theta_3)\rangle$, as a linear combination with equal coefficients, on a decomposable fractal that consists of three Cantor sets, located on three circles, with $\theta_1=\theta_2=\theta_3=\pi/2$ when $L=100$. Here the Cantor sets are $C[N_1,r_1;\{k_1\}]$, $C[N_2,r_2;\{k_2\}]$ and $C[N_3,r_3;\{k_3\}]$, with $d_{f1}= -\ln N_1/\ln r_1$, $d_{f2} = -\ln N_2/\ln r_2$ and $d_{f3} = -\ln N_3/\ln r_3$. Here we tackle two cases: (i) $N_1=N_2=N_3=2$, $r_1=r_2=r_3=1/3$ and $k_1=k_2=k_3=15$; (ii) $N_1=N_2=3$, $r_1=r_2=1/5$ and $k_1=k_2=11$, whereas $N_3=2$, $r_3=1/3$ and $k_3=15$. In both cases, we are led to conclude that the prefactor in Eq.~(\ref{dfsln}) is equal to half the sum of the fractal dimensions $d_{f1}$, $d_{f2}$ and $d_{f3}$, with relative errors less than 4$\%$.

\end{document}


\title{Supplementary Material for ``Type-B Goldstone modes and a logarithmic spiral in the staggered $\rm SU(4)$ ferromagnetic spin-orbital model"}

\author{Qian-Qian Shi}
\affiliation{Centre for Modern Physics, Chongqing University, Chongqing 400044, The People's Republic of China}

\author{Huan-Qiang Zhou}
\affiliation{Centre for Modern Physics, Chongqing University, Chongqing 400044, The People's Republic of China}

\author{Ian P. McCulloch}
\affiliation{Department of Physics, National Tsing Hua University, Hsinchu 30013, Taiwan}
\affiliation{School of Mathematics and Physics, The University of Queensland, St. Lucia, QLD 4072, Australia}
\affiliation{Centre for Modern Physics, Chongqing University, Chongqing 400044, The People's Republic of China}

\author{Murray T. Batchelor}
\affiliation{Mathematical Sciences Institute, The Australian National University, Canberra ACT 2601, Australia}
\affiliation{Centre for Modern Physics, Chongqing University, Chongqing 400044, The People's Republic of China}
\maketitle

This Supplementary Material sets out further details relevant to the main article.  Unless indicated otherwise 
equation numbers and references herein refer to the main article.

\subsection{Fifteen generators for the staggered  ${\rm SU}(4)$ symmetry group}

The staggered  ${\rm SU}(4)$ symmetry group is generated by the three Cartan generators $H_1$, $H_2$ and $H_3$, the six raising operators $E_1$, $E_2$, $E_3$, $E_4$, $E_5$ and $E_6$, and
the six lowering operators $F_1$, $F_2$, $F_3$, $E_4$, $F_5$ and $F_6$, which may be expressed in terms of
the spin-$1/2$ operators $S_{\!j}$ and the orbital pseudo spin-$1/2$ operators
$T_{\!j}$  for the staggered ${\rm SU}(4)$ ferromagnetic spin-orbital model (A1), and in terms of
the spin-$3/2$ operators for the staggered ${\rm SU}(4)$ spin-$3/2$ ferromagnetic model (A2). 
Here, we restrict our discussions to the staggered ${\rm SU}(4)$ ferromagnetic spin-orbital model (A1).

It is the presence of such a staggered ${\rm SU}(4)$ symmetry group that accounts for the peculiarities of the staggered ${\rm SU}(4)$ ferromagnetic spin-orbital model (A1).  Here, we stress that it is necessary to distinguish the local components of the generators between odd and even sites, due to the staggered nature. 
For completeness, we present their explicit expressions as follows.

Define $S_{\pm,j}=S^x_j\pm iS^y_j$ and $T_{\pm,j}=T^x_j\pm iT^y_j$.
At an odd site, the local components of the generators for the staggered  ${\rm SU}(4)$ symmetry group take the form
\begin{eqnarray*}
	&H_{1,j}=(1/2+S_j^z)(1/2+T_j^z)-(1/2+S_j^z)(1/2-T_j^z),\\
	&H_{2,j}=(1/2+S_j^z)(1/2+T_j^z)-(1/2-S_j^z)(1/2+T_j^z),\\
	&H_{3,j}=(1/2+S_j^z)(1/2+T_j^z)-(1/2-S_j^z)(1/2-T_j^z),\\
	&E_{1,j}=(1/2+S_j^z) T_{+,j}, \quad E_{2,j}=S_{+,j}(1/2+T_j^z),\\
	&E_{3,j}=S_{+,j} T_{+,j},\quad E_{4,j}=S_{+,j} T_{-,j},\\
	&E_{5,j}=(1/2-S_j^z)T_{+,j},\quad E_{6,j}=S_{+,j} (1/2-T_j^z),\\
	&F_{1,j}=(1/2+S_j^z)T_{-,j},\quad F_{2,j}=S_{-,j}(1/2+T_j^z),\\
	&F_{3,j}=S_{-,j} T_{-,j},\quad F_{4,j}=S_{-,j}T_{+,j},\\
	&F_{5,j}=(1/2-S_j^z) T_{-,j},\quad F_{6,j}=S_{-,j} (1/2-T_j^z)\,.
\end{eqnarray*}
At an even site,  the local components of the generators for the staggered ${\rm SU}(4)$ symmetry group take the form
\begin{eqnarray*}
	&H_{1,j}=(1/2-S_j^z)(1/2+T_j^z)-(1/2-S_j^z)(1/2-T_j^z),\\
	&H_{2,j}=(1/2+S_j^z)(1/2-T_j^z)-(1/2-S_j^z)(1/2-T_j^z),\\
	&H_{3,j}=(1/2+S_j^z)(1/2+T_j^z)-(1/2-S_j^z)(1/2-T_j^z),\\
	&E_{1,j}=(1/2-S_j^z) T_{+,j},\quad E_{2,j}=S_{+,j} (1/2-T_j^z),\\
	&E_{3,j}=-S_{+,j} T_{+,j},\quad E_{4,j}=-S_{+,j}  T_{-,j},\\
	&E_{5,j}=(1/2+S_j^z) T_{+,j},\quad E_{6,j}=S_{+,j} (1/2+T_j^z),\\
	&F_{1,j}=(1/2-S_j^z) T_{-,j},\quad F_{2,j}=S_{-,j} (1/2-T_j^z),\\
	&F_{3,j}=-S_{-,j} T_{-,j},\quad F_{4,j}=-S_{-,j} T_{+,j},\\
	&F_{5,j}=(1/2+S_j^z) T_{-,j},\quad F_{6,j}=S_{-,j}(1/2+T_j^z)\,.
\end{eqnarray*}

The fifteen generators of the staggered  ${\rm SU}(4)$ symmetry group satisfy $[H_\alpha,E_\alpha]=2E_\alpha$, $[H_\alpha,F_\alpha]=-2F_\alpha$ and $[E_\alpha,F_\alpha]=H_\alpha$ for $\alpha=1$, $2$ and $3$, in addition to other commutation relations:
\begin{eqnarray} 
	&[H_1,E_2]=E_2, \quad [H_1,E_3]=E_3,  \quad[H_1,E_4]=-E_4,   \nonumber\\  &[H_1,E_5]=0,  \quad[H_1,E_6]=-E_6,  \quad [H_1,F_2]=-F_2,\nonumber\\
	&[H_1,F_3]=-F_3, \quad [H_1,F_4]=F_4, \quad  [H_1,F_5]=0, \nonumber\\
	&[H_1,F_6]=F_6, \quad [H_2,E_1]=E_1,\quad  [H_2,E_3]=E_3,\nonumber\\  &[H_2,E_4]=E_4, \quad [H_2,E_5]=-E_5,   \quad [H_2,E_6]=0,\nonumber\\ &[H_2,F_1]=-F_1, \quad [H_2,F_3]=-F_3,\quad [H_2,F_4]=-F_4, \nonumber\\ &[H_2,F_5]=F_5, \quad [H_2,F_6]=0,  \quad   [H_3,E_1]=E_1,\nonumber\\ &[H_3,E_2]=E_2, \quad [H_3,E_4]=0, \quad  [H_3,E_5]=E_5, \nonumber\\ &[H_3,E_6]=E_6, \quad[H_3,F_1]=-F_1, \quad  [H_3,F_2]=-F_2,\nonumber\\
	&[H_3,F_4]=0, \quad [H_3,F_5]=-F_5,\quad  [H_3,F_6]=-F_5,\nonumber\\ &[E_1,E_2]=0, \quad [E_1,E_3]=0,  \quad   [E_1,E_4]=E_2,\nonumber\\  &[E_1,E_5]=0, \quad [E_1,E_6]=-E_3, \quad [E_1,F_2]=-F_4,\nonumber\\
	&[E_1,F_3]=F_6, \quad [E_1,F_4]=0,  \quad[E_1,F_5]=0,\nonumber\\
	&[E_1,F_6]=0, \quad [E_2,E_3]=0, \quad [E_2,E_4]=0,  \nonumber\\
	&[E_2,E_5]=E_3, \quad [E_2,E_6]=0,\quad [E_2,F_1]=-E_4, \nonumber\\
	&[E_2,F_3]=-F_5, \quad [E_2,F_4]=E_1,  \quad [E_2,F_5]=0, \nonumber\\
	&[E_2,F_6]=0, \quad [E_3,E_4]=0, \quad [E_3,E_5]=0, \nonumber\\
    	&[E_3,E_6]=0, \quad [E_3,F_1]=E_6, \quad[E_3,F_3]=-E_5, \nonumber\\
        	&[E_3,F_4]=0, \quad [E_3,F_5]=E_2,  \quad  [E_3,F_6]=-E_1,\nonumber\\
            &[E_4,E_5]=-E_6, \quad [E_4,E_6]=0, \quad[E_4,F_1]=0, \nonumber\\
            &[E_4,F_2]=F_1, \quad [E_4,F_3]=0, \quad[E_4,F_4]=H_2-H_1,  \nonumber\\
            &[E_4,F_5]=0, \quad [E_4,F_6]=F_5,  \quad [E_5,E_6]=0,\nonumber
\end{eqnarray}
    \begin{eqnarray}
         &[E_5,F_1]=0, \quad [E_5,F_2]=0, \quad  [E_5,F_3]=F_2, \nonumber\\
	&[E_5,F_4]=0, \quad [E_5,F_5]=H_3-H_2, \quad[E_5,F_6]=-F_4,  \nonumber\\  &[E_6,F_1]=0, \quad [E_6,F_2]=0, \quad [E_6,F_3]=-F_1, \nonumber\\ &[E_6,F_4]=E_5, \quad [E_6,F_5]=-E_4, \quad [E_6,F_6]=H_3-H_1, \nonumber\\
	&[F_1,F_2]=0, \quad [F_1,F_3]=0,  \quad [F_1,F_4]=-F_2,  \nonumber\\
	&[F_1,F_5]=0, \quad [F_1,F_6]=F_3,  \quad [F_2,F_3]=0, \nonumber\\
	&[F_2,F_4]=0, \quad [F_2,F_5]=-F_3, \quad [F_2,F_6]=0, \nonumber\\
	&[F_3,F_4]=0, \quad [F_3,F_5]=0, \quad[F_3,F_6]=0,\nonumber\\
	&[F_4,F_5]=F_6, \quad [F_4,F_6]=0, \quad [F_5,F_6]=0. \nonumber
\end{eqnarray}

\subsection{Ground state subspaces $\Omega_L^{\rm OBC}$ and $\Omega_L^{\rm PBC}$}

When the system size $L$ is even, then the Hamiltonian possesses the staggered symmetry group ${\rm SU}(4)$ under both PBCs and OBCs.
Additionally, the time-reversal symmetry $K_s$ in the spin sector and time-reversal symmetry $K_t$ in the orbital sector, together with an additional $Z_2$ symmetry operation $\sigma$ -- the one-site translation operation $T_1$ under PBCs, and the bond-centred inversion operation $I_b$ under OBCs, are preserved. 
When $L$ is odd, the symmetry group under OBCs is the same as that when $L$ is even, while the Hamiltonian under PBCs does not possess the staggered symmetry group ${\rm SU}(4)$. 
Thus, it leads to a parity effect between even $L$ and odd $L$.
Actually, the symmetry group reduces to ${\rm SO(4)} \approx {\rm SU(2)}\times{\rm SU(2)}$, generated by $S^x$, $S^y$ and $S^z$ and  $T^x$, $T^y$ and $T^z$.
For simplicity, we restrict ourselves here to even $L$.

All of the linearly independent degenerate ground states in the staggered ${\rm SU}(4)$ ferromagnetic spin-orbital model may be constructed,
similar to those for the staggered ${\rm SU}(3)$ ferromagnetic spin-1 biquadratic model, under both PBCs and OBCs~[18].

We restrict ourselves to $L= 4$ for illustration.
There are 108 distinct factorized ground states under OBCs and 84 distinct factorized ground states under PBCs.
As it turns out, one may choose $| \! \uparrow_s\uparrow_t\uparrow_s\uparrow_t\uparrow_s\uparrow_t\uparrow_s\uparrow_t\rangle$ as the  highest weight states $| \, {\rm hws}\rangle$, together with two generalized highest weight states $| \, {\rm hws}\rangle_g^1$ and $| \, {\rm hws}\rangle_g^2$ for PBCs and three generalized highest weight states $| \, {\rm hws}\rangle_g^1$, $|\, {\rm hws}\rangle_g^2$  and $| \, {\rm hws}\rangle_g^3$ for OBCs.
Here,
$| \, {\rm hws}\rangle_g^1=| \! \downarrow_s\uparrow_t\uparrow_s\uparrow_t\uparrow_s\uparrow_t\uparrow_s\uparrow_t\rangle$, $| \, {\rm hws}\rangle_g^2=| \!\downarrow_s\uparrow_t\downarrow_s\uparrow_t\uparrow_s\uparrow_t\uparrow_s\uparrow_t\rangle$ and
$|\, {\rm hws}\rangle_g^3=| \! \downarrow_s\downarrow_t\downarrow_s\uparrow_t\uparrow_s\uparrow_t\uparrow_s\uparrow_t\rangle$.

For $L=4$, there are 84 linearly independent ground states generated from the repeated action of $F_1$, $F_2$ and $F_3$, combining with $K_s$ and $K_t$, together with  $\sigma$ under OBCs and under PBCs, on the highest weight state $| \, {\rm hws}\rangle$, with the eigenvalue of $S^z$ and $T^z$ being 2 and 2, respectively.
Accordingly, the subspace $V_0$ is spanned by a set of 84 linearly independent ground states.
Then, one may perform the repeated action of $F_1$, $F_2$ and $F_3$, combining with $K_s$ and $K_t$, and $\sigma$ under OBCs and under PBCs, respectively, on the $1$-th generalized highest weight state $| \, {\rm hws}\rangle_g^1$, with the eigenvalue of $S^z$ and $T^z$ being 1 and 2, respectively, yielding a sequence of the degenerate ground states, with a total number of 174 linearly independent ground states.
Thus an extra $90$ linearly independent ground states are generated,  spanning the subspace $V_1$.
Then the repeated action of $F_1$, $F_2$ and $F_3$, combining with $K_s$ and $K_t$, and $\sigma$ under OBCs and under PBCs, on the $2$-th generalized highest weight state $| \, {\rm hws}\rangle_g^2$, with the eigenvalue of $S^z$ and $T^z$ being 0 and 2, respectively, generates a sequence of the degenerate ground states, with total number 194 linearly independent ground states.
Hence,  an extra $20$ linearly independent ground states are generated, spanning the subspace $V_2$.
Therefore, the ground state subspace for $L=4$ under PBCs is exhausted.
That is, $\Omega_L^{\rm PBC}=V_0\oplus V_1\oplus V_2$, with ${\rm dim }(\Omega_L^{\rm PBC})=194$ for $L=4$.

For OBCs, one may perform the repeated action of $F_1$, $F_2$ and $F_3$, combining with $K_s$ and $K_t$, and $I_b$, on the generalized highest weight state $| \, {\rm hws}\rangle_g^3$, with the eigenvalue of $S^z$ and $T^z$ being 0 and 1, respectively, which generates a sequence of degenerate ground states, with the number of linearly independent ground states being 209.
There are now an extra $35$ linearly independent ground states, spanning the subspace $V_3$.
Then, the ground state subspace for $L=4$ under OBCs is exhausted.
That is, $\Omega_L^{\rm OBC}=V_0\oplus V_1\oplus V_2\oplus V_3$, with ${\rm dim }(\Omega_L^{\rm OBC})=209$ for $L=4$.

Our construction offers a way to explain why the ground state degeneracies depend on the boundary conditions adopted, i.e., ${\rm dim }(\Omega_L^{\rm OBC})$ and ${\rm dim }(\Omega_L^{\rm PBC})$ are different.

The explicit expressions for the above mentioned linearly independent ground states are available upon request.

\subsection{A derivation of $Z_2(L,M_1,M_2,M_3)$,  $Z_4(L,M_1,M_2,M_3)$ and $Z_4(L,M_1,M_3,M_6)$ for the highly degenerate ground states in Eq.~(4),  Eq.~(6) and Eq.~(8)}\label{normsu31z}

We note that evaluating the norm for a degenerate ground state generated from the repeated action of the lowering operators $F_1$, $F_2$ and $F_3$, combining with $K_s$ and $K_t$, and the additional $Z_2$ symmetry operator $\sigma$, on a generic generalized highest weight state is highly non-trivial, though possible in principle, if one increases the period up to the system size $L$ itself.
Here we take a derivation of the norms for the degenerate ground states in Eq.~(4), Eq.~(6), and Eq.~(8), with the period being two, four and four, respectively, as examples.

(a) For the degenerate ground states $| \, L,M_1,M_2,M_3\rangle_2$ given in Eq.~(4), we introduce a unit cell consisting of the two nearest-neighbor sites, due to the staggered nature of the symmetry group $\rm{SU(4)}$.
Therefore,  eight possible configurations in the unit cell are related: $|\! \uparrow_s\uparrow_t\uparrow_s\uparrow_t\rangle$, $| \! \downarrow_s\uparrow_t\uparrow_s\uparrow_t\rangle$, $| \!\uparrow_s\downarrow_t\uparrow_s\uparrow_t\rangle$, $| \! \downarrow_s\downarrow_t\uparrow_s\uparrow_t\rangle$, $| \! \uparrow_s\uparrow_t\downarrow_s\downarrow_t\rangle$, $| \! \downarrow_s\uparrow_t\downarrow_s\downarrow_t\rangle$, $| \! \uparrow_s\downarrow_t\downarrow_s\downarrow_t\rangle$ and $| \!\downarrow_s\downarrow_t\downarrow_s\downarrow_t\rangle$. 
We point out that $L$ is required to be a multiple of two.

One may rewrite $| \, L,M_1,M_2,M_3\rangle_2$ as
\begin{widetext}
	\begin{align}
		| \, L,M_1,M_2,M_3\rangle_2=&\frac{1}{Z_2(L,M_1,M_2,M_3)}{\sum}'_{N_{\downarrow_s\uparrow_t\uparrow_s\uparrow_t},N_{\uparrow_s\downarrow_t\uparrow_s\uparrow_t}, N_{\downarrow_s\downarrow_t\uparrow_s\uparrow_t},N_{\downarrow_s\uparrow_t\downarrow_s\downarrow_t},\atop N_{\uparrow_s\downarrow_t\downarrow_s\downarrow_t}, N_{\uparrow_s\uparrow_t\downarrow_s\downarrow_t}, N_{\downarrow_s\downarrow_t\downarrow_s\downarrow_t},N_{\uparrow_s\uparrow_t\uparrow_s\uparrow_t}}\nonumber \\
		&(-1)^{N_{\downarrow_s\uparrow_t\downarrow_s\downarrow_t}+N_{\uparrow_s\downarrow_t\downarrow_s\downarrow_t}+N_{\uparrow_s\uparrow_t\downarrow_s\downarrow_t}+N_{\downarrow_s\downarrow_t\downarrow_s\downarrow_t}}\sum_P|\underbrace{\downarrow_s\uparrow_t\uparrow_s\uparrow_t...\downarrow_s\uparrow_t\uparrow_s\uparrow_t}_{N_{\downarrow_s\uparrow_t\uparrow_s\uparrow_t}}
		|\underbrace{\uparrow_s\downarrow_t\uparrow_s\uparrow_t...\uparrow_s\downarrow_t\uparrow_s\uparrow_t}_{N_{\uparrow_s\downarrow_t\uparrow_s\uparrow_t}}
		|\underbrace{\downarrow_s\downarrow_t\uparrow_s\uparrow_t...\downarrow_s\downarrow_t\uparrow_s\uparrow_t}_{N_{\downarrow_s\downarrow_t\uparrow_s\uparrow_t}}\!\nonumber \\
		&|\!\underbrace{\downarrow_s\uparrow_t\downarrow_s\downarrow_t\!...\!\downarrow_s\uparrow_t\downarrow_s\downarrow_t}_{N_{\downarrow_s\uparrow_t\downarrow_s\downarrow_t}}\! |\!\underbrace{\uparrow_s\downarrow_t\downarrow_s\downarrow_t\!...\!\uparrow_s\downarrow_t\downarrow_s\downarrow_t}_{N_{\uparrow_s\downarrow_t\downarrow_s\downarrow_t}}\!
		|\!\underbrace{\uparrow_s\uparrow_t\downarrow_s\downarrow_t\!...\!\uparrow_s\uparrow_t\downarrow_s\downarrow_t}_{N_{\uparrow_s\uparrow_t\downarrow_s\downarrow_t}}\!
		|\!\underbrace{\downarrow_s\downarrow_t\downarrow_s\downarrow_t\!...\!\downarrow_s\downarrow_t\downarrow_s\downarrow_t}_{N_{\downarrow_s\downarrow_t\downarrow_s\downarrow_t}}\!
		|\!\underbrace{\uparrow_s\uparrow_t\uparrow_s\uparrow_t\!...\!\uparrow_s\uparrow_t\uparrow_s\uparrow_t}_{N_{\uparrow_s\uparrow_t\uparrow_s\uparrow_t}}\rangle,
		\label{lm1m2m3sm}
	\end{align}
where the sum $\sum_{P}$ is taken over all the permutations $P$ for a given partition $\{N_{\downarrow_s\uparrow_t\uparrow_s\uparrow_t},N_{\uparrow_s\downarrow_t\uparrow_s\uparrow_t}, N_{\downarrow_s\downarrow_t\uparrow_s\uparrow_t},N_{\downarrow_s\uparrow_t\downarrow_s\downarrow_t}, N_{\uparrow_s\downarrow_t\downarrow_s\downarrow_t}, N_{\uparrow_s\uparrow_t\downarrow_s\downarrow_t}, N_{\downarrow_s\downarrow_t\downarrow_s\downarrow_t},N_{\uparrow_s\uparrow_t\uparrow_s\uparrow_t}\}$, the sum ${\sum}'$ is taken over all the possible values of $N_{\downarrow_s\uparrow_t\uparrow_s\uparrow_t}$, $N_{\uparrow_s\downarrow_t\uparrow_s\uparrow_t}$, $N_{\downarrow_s\downarrow_t\uparrow_s\uparrow_t}$, $N_{\downarrow_s\uparrow_t\downarrow_s\downarrow_t}$,  $N_{\uparrow_s\downarrow_t\downarrow_s\downarrow_t}$,  $N_{\uparrow_s\uparrow_t\downarrow_s\downarrow_t}$,  $N_{\downarrow_s\downarrow_t\downarrow_s\downarrow_t}$, and $N_{\uparrow_s\uparrow_t\uparrow_s\uparrow_t}$, subject to the constraints: $N_{\downarrow_s\uparrow_t\uparrow_s\uparrow_t}+N_{\downarrow_s\uparrow_t\downarrow_s\downarrow_t}=M_1$,
$N_{\uparrow_s\downarrow_t\uparrow_s\uparrow_t}+N_{\uparrow_s\downarrow_t\downarrow_s\downarrow_t}=M_2$,  $N_{\downarrow_s\uparrow_t\downarrow_s\downarrow_t}+N_{\uparrow_s\downarrow_t\downarrow_s\downarrow_t}+
N_{\uparrow_s\uparrow_t\downarrow_s\downarrow_t}+N_{\downarrow_s\downarrow_t\uparrow_s\uparrow_t}+
N_{\downarrow_s\downarrow_t\downarrow_s\downarrow_t}=M_3$, and $N_{\downarrow_s\uparrow_t\uparrow_s\uparrow_t},N_{\downarrow_s\uparrow_t\uparrow_s\uparrow_t}+ N_{\downarrow_s\downarrow_t\uparrow_s\uparrow_t}+N_{\downarrow_s\uparrow_t\downarrow_s\downarrow_t}+ N_{\uparrow_s\downarrow_t\downarrow_s\downarrow_t}+ N_{\uparrow_s\uparrow_t\downarrow_s\downarrow_t}+ N_{\downarrow_s\downarrow_t\downarrow_s\downarrow_t}+N_{\uparrow_s\uparrow_t\uparrow_s\uparrow_t}=L/4$, and $N_{\downarrow_s\uparrow_t\uparrow_s\uparrow_t}$, $N_{\uparrow_s\downarrow_t\uparrow_s\uparrow_t}$, $N_{\downarrow_s\downarrow_t\uparrow_s\uparrow_t}$, $N_{\downarrow_s\uparrow_t\downarrow_s\downarrow_t}$,  $N_{\uparrow_s\downarrow_t\downarrow_s\downarrow_t}$,  $N_{\uparrow_s\uparrow_t\downarrow_s\downarrow_t}$,  $N_{\downarrow_s\downarrow_t\downarrow_s\downarrow_t}$. Here, $N_{\uparrow_s\uparrow_t\uparrow_s\uparrow_t}$ denote the numbers of the unit cells in the configurations $| \!\downarrow_s\uparrow_t\uparrow_s\uparrow_t\rangle$, $| \! \uparrow_s\downarrow_t\uparrow_s\uparrow_t\rangle$, $| \! \downarrow_s\downarrow_t\uparrow_s\uparrow_t\rangle$, $| \!\downarrow_s\uparrow_t\downarrow_s\downarrow_t\rangle$,  $| \! \uparrow_s\downarrow_t\downarrow_s\downarrow_t\rangle$,  $| \!\uparrow_s\uparrow_t\downarrow_s\downarrow_t\rangle$,  $| \! \downarrow_s\downarrow_t\downarrow_s\downarrow_t\rangle$, and $| \!\uparrow_s\uparrow_t\uparrow_s\uparrow_t\rangle$, respectively.
In addition,  $Z_2(L,M_1,M_2,M_3)$ is introduced to ensure a normalized $| \, L,M_1,M_2,M_3\rangle_2$, which takes the form
	\begin{equation}
		Z_2(L,M_1,M_2,M_3)=M_1!M_2!M_3!\sqrt{K_2(L,M_1,M_2,M_3)} \, 
	\end{equation}
with
	\begin{align}
		K_2(L,M_1,M_2,M_3)=&{\sum}'_{N_{\downarrow_s\uparrow_t\uparrow_s\uparrow_t},N_{\uparrow_s\downarrow_t\uparrow_s\uparrow_t}, N_{\downarrow_s\downarrow_t\uparrow_s\uparrow_t},N_{\downarrow_s\uparrow_t\downarrow_s\downarrow_t},\atop N_{\uparrow_s\downarrow_t\downarrow_s\downarrow_t}, N_{\uparrow_s\uparrow_t\downarrow_s\downarrow_t}, N_{\downarrow_s\downarrow_t\downarrow_s\downarrow_t},N_{\uparrow_s\uparrow_t\uparrow_s\uparrow_t}}\nonumber\\ &C_{L/4}^{N_{\downarrow_s\uparrow_t\uparrow_s\uparrow_t}}C_{L/4-N_{\downarrow_s\uparrow_t\uparrow_s\uparrow_t}}^{N_{\uparrow_s\downarrow_t\uparrow_s\uparrow_t}}
		C_{L/4-N_{\downarrow_s\uparrow_t\uparrow_s\uparrow_t}-N_{\uparrow_s\downarrow_t\uparrow_s\uparrow_t}}^{N_{\downarrow_s\downarrow_t\uparrow_s\uparrow_t}}	 C_{L/4-N_{\downarrow_s\uparrow_t\uparrow_s\uparrow_t}-N_{\uparrow_s\downarrow_t\uparrow_s\uparrow_t}-N_{\downarrow_s\downarrow_t\uparrow_s\uparrow_t}}^	 {N_{\downarrow_s\uparrow_t\downarrow_s\downarrow_t}}
		C_{L/4-N_{\downarrow_s\uparrow_t\uparrow_s\uparrow_t}-N_{\uparrow_s\downarrow_t\uparrow_s\uparrow_t}-N_{\downarrow_s\downarrow_t\uparrow_s\uparrow_t}-N_{\downarrow_s\uparrow_t\downarrow_s\downarrow_t}}^{N_{\uparrow_s\downarrow_t\downarrow_s\downarrow_t}}\nonumber \\
		&C_{L/4-N_{\downarrow_s\uparrow_t\uparrow_s\uparrow_t}-N_{\uparrow_s\downarrow_t\uparrow_s\uparrow_t}-N_{\downarrow_s\downarrow_t\uparrow_s\uparrow_t}-N_{\downarrow_s\uparrow_t\downarrow_s\downarrow_t}-N_{\uparrow_s\downarrow_t\downarrow_s\downarrow_t}}^{N_{\uparrow_s\uparrow_t\downarrow_s\downarrow_t}}
		C_{L/4-N_{\downarrow_s\uparrow_t\uparrow_s\uparrow_t}-N_{\uparrow_s\downarrow_t\uparrow_s\uparrow_t}-N_{\downarrow_s\downarrow_t\uparrow_s\uparrow_t}-N_{\downarrow_s\uparrow_t\downarrow_s\downarrow_t}-N_{\uparrow_s\downarrow_t\downarrow_s\downarrow_t}-N_{\uparrow_s\uparrow_t\downarrow_s\downarrow_t}}^{N_{\downarrow_s\downarrow_t\downarrow_s\downarrow_t}}\nonumber\\ &C_{L/4-N_{\downarrow_s\uparrow_t\uparrow_s\uparrow_t}-N_{\uparrow_s\downarrow_t\uparrow_s\uparrow_t}-N_{\downarrow_s\downarrow_t\uparrow_s\uparrow_t}-N_{\downarrow_s\uparrow_t\downarrow_s\downarrow_t}-N_{\uparrow_s\downarrow_t\downarrow_s\downarrow_t}-N_{\uparrow_s\uparrow_t\downarrow_s\downarrow_t}-N_{\downarrow_s\downarrow_t\downarrow_s\downarrow_t}}^{N_{\uparrow_s\uparrow_t\uparrow_s\uparrow_t}}.
	\end{align}
It may be simplified as
	\begin{equation}
		Z_2(L,M_1,M_2,M_3)=M_1!M_2!M_3!\sqrt{C_{L/2}^{M_1}C_{L/2-M_1}^{M_2}C_{L-M_1-M_2}^{M_3}}.
		\label{zlm1m2m3sm}
	\end{equation}
We have
$H_1 | \, L,M_1,M_2,M_3\rangle_2=(L/2-2M_1-M_2-M_3) | \, L,M_1,M_2,M_3\rangle_2$, $H_2 | \, L,M_1,M_2,M_3\rangle_2=(L/2-M_1-2M_2-M_3) | \, L,M_1,M_2,M_3\rangle_2$, and  $H_3 | \, L,M_1,M_2,M_3\rangle_2=(L-M_1-M_2-2M_3) | \, L,M_1,M_2,M_3\rangle_2$, as long as $M_1+M_2\leq L/2$ and $M_1+M_2+M_3\leq L$.

(b) For the degenerate ground states $| \, L,M_1,M_2,M_3\rangle_4$ given in Eq.~(6), we introduce a unit cell consisting of the four nearest-neighbor sites.
Accordingly, sixteen possible configurations are included: $| \! \downarrow_s\uparrow_t\uparrow_s\uparrow_t\uparrow_s\uparrow_t\uparrow_s\uparrow_t\rangle$, $| \! \downarrow_s\uparrow_t\downarrow_s\downarrow_t\uparrow_s\uparrow_t\uparrow_s\uparrow_t\rangle$, $| \!\downarrow_s\uparrow_t\uparrow_s\uparrow_t\downarrow_s\uparrow_t\uparrow_s\uparrow_t\rangle$, $| \!\downarrow_s\uparrow_t\uparrow_s\uparrow_t\uparrow_s\downarrow_t\uparrow_s\uparrow_t\rangle$, $| \!\downarrow_s\uparrow_t\uparrow_s\uparrow_t\downarrow_s\downarrow_t\uparrow_s\uparrow_t\rangle$,
$| \! \downarrow_s\uparrow_t\uparrow_s\uparrow_t\uparrow_s\uparrow_t\downarrow_s\downarrow_t\rangle$,
$| \! \downarrow_s\uparrow_t\downarrow_s\downarrow_t\downarrow_s\uparrow_t\uparrow_s\uparrow_t\rangle$,
$| \! \downarrow_s\uparrow_t\downarrow_s\downarrow_t\uparrow_s\downarrow_t\uparrow_s\uparrow_t\rangle$,
$| \! \downarrow_s\uparrow_t\downarrow_s\downarrow_t\downarrow_s\downarrow_t\uparrow_s\uparrow_t\rangle$, $| \!\downarrow_s\uparrow_t\downarrow_s\downarrow_t\uparrow_s\uparrow_t\downarrow_s\downarrow_t\rangle$, $| \!\downarrow_s\uparrow_t\uparrow_s\uparrow_t\downarrow_s\uparrow_t\downarrow_s\downarrow_t\rangle$, $| \!\downarrow_s\uparrow_t\uparrow_s\uparrow_t\uparrow_s\downarrow_t\downarrow_s\downarrow_t\rangle$, $| \!\downarrow_s\uparrow_t\uparrow_s\uparrow_t\downarrow_s\downarrow_t\downarrow_s\downarrow_t\rangle$, $| \!\downarrow_s\uparrow_t\downarrow_s\downarrow_t\downarrow_s\uparrow_t\downarrow_s\downarrow_t\rangle$, $| \!\downarrow_s\uparrow_t\downarrow_s\downarrow_t\uparrow_s\downarrow_t\downarrow_s\downarrow_t\rangle$, and $| \!\downarrow_s\uparrow_t\downarrow_s\downarrow_t\downarrow_s\downarrow_t\downarrow_s\downarrow_t\rangle$ in the unit cell.
Here $L$ is required to be a multiple of four.
	
$| \, L,M_1,M_2,M_3\rangle_4$ may be rewritten as follows
\begin{align}
| \, L,M_1,M_2,M_3\rangle_4=&\frac{1}{Z_4(L,M_1,M_2,M_3)}{\sum}'_{N_{\downarrow_s\uparrow_t\uparrow_s\uparrow_t\uparrow_s\uparrow_t\uparrow_s\uparrow_t},
N_{\downarrow_s\uparrow_t\downarrow_s\downarrow_t\uparrow_s\uparrow_t\uparrow_s\uparrow_t},N_{\downarrow_s\uparrow_t\uparrow_s\uparrow_t\downarrow_s\uparrow_t\uparrow_s\uparrow_t},N_{\downarrow_s\uparrow_t\uparrow_s\uparrow_t\uparrow_s\downarrow_t\uparrow_s\uparrow_t},N_{\downarrow_s\uparrow_t\uparrow_s\uparrow_t\downarrow_s\downarrow_t\uparrow_s\uparrow_t},N_{\downarrow_s\uparrow_t\uparrow_s\uparrow_t\uparrow_s\uparrow_t\downarrow_s\downarrow_t},N_{\downarrow_s\uparrow_t\downarrow_s\downarrow_t\downarrow_s\uparrow_t\uparrow_s\uparrow_t},N_{\downarrow_s\uparrow_t\downarrow_s\downarrow_t\uparrow_s\downarrow_t\uparrow_s\uparrow_t},\atop N_{\downarrow_s\uparrow_t\downarrow_s\downarrow_t\downarrow_s\downarrow_t\uparrow_s\uparrow_t},
N_{\downarrow_s\uparrow_t\downarrow_s\downarrow_t\uparrow_s\uparrow_t\downarrow_s\downarrow_t},
N_{\downarrow_s\uparrow_t\uparrow_s\uparrow_t\downarrow_s\uparrow_t\downarrow_s\downarrow_t},
N_{\downarrow_s\uparrow_t\uparrow_s\uparrow_t\uparrow_s\downarrow_t\downarrow_s\downarrow_t},
N_{\downarrow_s\uparrow_t\uparrow_s\uparrow_t\downarrow_s\downarrow_t\downarrow_s\downarrow_t},
N_{\downarrow_s\uparrow_t\downarrow_s\downarrow_t\downarrow_s\uparrow_t\downarrow_s\downarrow_t},
N_{\downarrow_s\uparrow_t\downarrow_s\downarrow_t\uparrow_s\downarrow_t\downarrow_s\downarrow_t},
N_{\downarrow_s\uparrow_t\downarrow_s\downarrow_t\downarrow_s\downarrow_t\downarrow_s\downarrow_t}
}\nonumber \\
&(-1)^{\epsilon}\sum_{P}
|\underbrace{\downarrow_s\uparrow_t\uparrow_s\uparrow_t\uparrow_s\uparrow_t\uparrow_s\uparrow_t}_{N_{\downarrow_s\uparrow_t\uparrow_s\uparrow_t\uparrow_s\uparrow_t\uparrow_s\uparrow_t}} |\underbrace{\downarrow_s\uparrow_t\downarrow_s\downarrow_t\uparrow_s\uparrow_t\uparrow_s\uparrow_t}_{N_{\downarrow_s\uparrow_t\downarrow_s\downarrow_t\uparrow_s\uparrow_t\uparrow_s\uparrow_t}} |\underbrace{\downarrow_s\uparrow_t\uparrow_s\uparrow_t\downarrow_s\uparrow_t\uparrow_s\uparrow_t}_{N_{\downarrow_s\uparrow_t\uparrow_s\uparrow_t\downarrow_s\uparrow_t\uparrow_s\uparrow_t}} |\underbrace{\downarrow_s\uparrow_t\uparrow_s\uparrow_t\uparrow_s\downarrow_t\uparrow_s\uparrow_t}_{N_{\downarrow_s\uparrow_t\uparrow_s\uparrow_t\uparrow_s\downarrow_t\uparrow_s\uparrow_t}} |\underbrace{\downarrow_s\uparrow_t\uparrow_s\uparrow_t\downarrow_s\downarrow_t\uparrow_s\uparrow_t}_{N_{\downarrow_s\uparrow_t\uparrow_s\uparrow_t\downarrow_s\downarrow_t\uparrow_s\uparrow_t}}
		\nonumber \\
&|\underbrace{\downarrow_s\uparrow_t\uparrow_s\uparrow_t\uparrow_s\uparrow_t\downarrow_s\downarrow_t}_{N_{\downarrow_s\uparrow_t\uparrow_s\uparrow_t\uparrow_s\uparrow_t\downarrow_s\downarrow_t}}|\underbrace{\downarrow_s\uparrow_t\downarrow_s\downarrow_t\downarrow_s\uparrow_t\uparrow_s\uparrow_t}_{N_{\downarrow_s\uparrow_t\downarrow_s\downarrow_t\downarrow_s\uparrow_t\uparrow_s\uparrow_t}}
|\underbrace{\downarrow_s\uparrow_t\downarrow_s\downarrow_t\uparrow_s\downarrow_t\uparrow_s\uparrow_t}_{N_{\downarrow_s\uparrow_t\downarrow_s\downarrow_t\uparrow_s\downarrow_t\uparrow_s\uparrow_t}}
|\underbrace{\downarrow_s\uparrow_t\downarrow_s\downarrow_t\downarrow_s\downarrow_t\uparrow_s\uparrow_t}_{N_{\downarrow_s\uparrow_t\downarrow_s\downarrow_t\downarrow_s\downarrow_t\uparrow_s\uparrow_t}} |\underbrace{\downarrow_s\uparrow_t\downarrow_s\downarrow_t\uparrow_s\uparrow_t\downarrow_s\downarrow_t}_{N_{\downarrow_s\uparrow_t\downarrow_s\downarrow_t\uparrow_s\uparrow_t\downarrow_s\downarrow_t}} |\underbrace{\downarrow_s\uparrow_t\uparrow_s\uparrow_t\downarrow_s\uparrow_t\downarrow_s\downarrow_t}_{N_{\downarrow_s\uparrow_t\uparrow_s\uparrow_t\downarrow_s\uparrow_t\downarrow_s\downarrow_t}} \nonumber \\ &|\underbrace{\downarrow_s\uparrow_t\uparrow_s\uparrow_t\uparrow_s\downarrow_t\downarrow_s\downarrow_t}_{N_{\downarrow_s\uparrow_t\uparrow_s\uparrow_t\uparrow_s\downarrow_t\downarrow_s\downarrow_t}} |\underbrace{\downarrow_s\uparrow_t\uparrow_s\uparrow_t\downarrow_s\downarrow_t\downarrow_s\downarrow_t}_{N_{\downarrow_s\uparrow_t\uparrow_s\uparrow_t\downarrow_s\downarrow_t\downarrow_s\downarrow_t}} |\underbrace{\downarrow_s\uparrow_t\downarrow_s\downarrow_t\downarrow_s\uparrow_t\downarrow_s\downarrow_t}_{N_{\downarrow_s\uparrow_t\downarrow_s\downarrow_t\downarrow_s\uparrow_t\downarrow_s\downarrow_t}} |\underbrace{\downarrow_s\uparrow_t\downarrow_s\downarrow_t\uparrow_s\downarrow_t\downarrow_s\downarrow_t}_{N_{\downarrow_s\uparrow_t\downarrow_s\downarrow_t\uparrow_s\downarrow_t\downarrow_s\downarrow_t}}
|\underbrace{\downarrow_s\uparrow_t\downarrow_s\downarrow_t\downarrow_s\downarrow_t\downarrow_s\downarrow_t}_{N_{\downarrow_s\uparrow_t\downarrow_s\downarrow_t\downarrow_s\downarrow_t\downarrow_s\downarrow_t}}
\rangle,
\label{lm1m2m3gsm}
\end{align}
with the sum $\sum_{P}$ taken over all the permutations $P$ for a given partition $\{N_{\downarrow_s\uparrow_t\uparrow_s\uparrow_t\uparrow_s\uparrow_t\uparrow_s\uparrow_t}$,  
$N_{\downarrow_s\uparrow_t\downarrow_s\downarrow_t\uparrow_s\uparrow_t\uparrow_s\uparrow_t}$,  
$N_{\downarrow_s\uparrow_t\uparrow_s\uparrow_t\downarrow_s\uparrow_t\uparrow_s\uparrow_t}$, $N_{\downarrow_s\uparrow_t\uparrow_s\uparrow_t\uparrow_s\downarrow_t\uparrow_s\uparrow_t}$, $N_{\downarrow_s\uparrow_t\uparrow_s\uparrow_t\downarrow_s\downarrow_t\uparrow_s\uparrow_t}$, $N_{\downarrow_s\uparrow_t\uparrow_s\uparrow_t\uparrow_s\uparrow_t\downarrow_s\downarrow_t}$, $N_{\downarrow_s\uparrow_t\downarrow_s\downarrow_t\downarrow_s\uparrow_t\uparrow_s\uparrow_t}$, $N_{\downarrow_s\uparrow_t\downarrow_s\downarrow_t\uparrow_s\downarrow_t\uparrow_s\uparrow_t}$, $N_{\downarrow_s\uparrow_t\downarrow_s\downarrow_t\downarrow_s\downarrow_t\uparrow_s\uparrow_t}$, $N_{\downarrow_s\uparrow_t\downarrow_s\downarrow_t\uparrow_s\uparrow_t\downarrow_s\downarrow_t}$, 	$N_{\downarrow_s\uparrow_t\uparrow_s\uparrow_t\downarrow_s\uparrow_t\downarrow_s\downarrow_t}$, $N_{\downarrow_s\uparrow_t\uparrow_s\uparrow_t\uparrow_s\downarrow_t\downarrow_s\downarrow_t}$, $N_{\downarrow_s\uparrow_t\uparrow_s\uparrow_t\downarrow_s\downarrow_t\downarrow_s\downarrow_t}$, $N_{\downarrow_s\uparrow_t\downarrow_s\downarrow_t\downarrow_s\uparrow_t\downarrow_s\downarrow_t}$, $N_{\downarrow_s\uparrow_t\downarrow_s\downarrow_t\uparrow_s\downarrow_t\downarrow_s\downarrow_t}$, $N_{\downarrow_s\uparrow_t\downarrow_s\downarrow_t\downarrow_s\downarrow_t\downarrow_s\downarrow_t}\}$, and the sum ${\sum}'$ taken over all the possible values of $N_{\downarrow_s\uparrow_t\uparrow_s\uparrow_t\uparrow_s\uparrow_t\uparrow_s\uparrow_t}$, 	$N_{\downarrow_s\uparrow_t\downarrow_s\downarrow_t\uparrow_s\uparrow_t\uparrow_s\uparrow_t}$, 	$N_{\downarrow_s\uparrow_t\uparrow_s\uparrow_t\downarrow_s\uparrow_t\uparrow_s\uparrow_t}$, 	$N_{\downarrow_s\uparrow_t\uparrow_s\uparrow_t\uparrow_s\downarrow_t\uparrow_s\uparrow_t}$, $N_{\downarrow_s\uparrow_t\uparrow_s\uparrow_t\downarrow_s\downarrow_t\uparrow_s\uparrow_t}$, 	$N_{\downarrow_s\uparrow_t\uparrow_s\uparrow_t\uparrow_s\uparrow_t\downarrow_s\downarrow_t}$, 	$N_{\downarrow_s\uparrow_t\downarrow_s\downarrow_t\downarrow_s\uparrow_t\uparrow_s\uparrow_t}$, 	$N_{\downarrow_s\uparrow_t\downarrow_s\downarrow_t\uparrow_s\downarrow_t\uparrow_s\uparrow_t}$, $N_{\downarrow_s\uparrow_t\downarrow_s\downarrow_t\downarrow_s\downarrow_t\uparrow_s\uparrow_t}$, $N_{\downarrow_s\uparrow_t\downarrow_s\downarrow_t\uparrow_s\uparrow_t\downarrow_s\downarrow_t}$, 	$N_{\downarrow_s\uparrow_t\uparrow_s\uparrow_t\downarrow_s\uparrow_t\downarrow_s\downarrow_t}$, 	$N_{\downarrow_s\uparrow_t\uparrow_s\uparrow_t\uparrow_s\downarrow_t\downarrow_s\downarrow_t}$, $N_{\downarrow_s\uparrow_t\uparrow_s\uparrow_t\downarrow_s\downarrow_t\downarrow_s\downarrow_t}$, $N_{\downarrow_s\uparrow_t\downarrow_s\downarrow_t\downarrow_s\uparrow_t\downarrow_s\downarrow_t}$, $N_{\downarrow_s\uparrow_t\downarrow_s\downarrow_t\uparrow_s\downarrow_t\downarrow_s\downarrow_t}$, 	and $N_{\downarrow_s\uparrow_t\downarrow_s\downarrow_t\downarrow_s\downarrow_t\downarrow_s\downarrow_t}$, subject to the constraints: $N_{\downarrow_s\uparrow_t\uparrow_s\uparrow_t\downarrow_s\uparrow_t\uparrow_s\uparrow_t}+N_{\downarrow_s\uparrow_t\downarrow_s\downarrow_t\downarrow_s\uparrow_t\uparrow_s\uparrow_t}+ N_{\downarrow_s\uparrow_t\uparrow_s\uparrow_t\downarrow_s\uparrow_t\downarrow_s\downarrow_t}+N_{\downarrow_s\uparrow_t\downarrow_s\downarrow_t\downarrow_s\uparrow_t\downarrow_s\downarrow_t}=M_1$,  $N_{\downarrow_s\uparrow_t\uparrow_s\uparrow_t\uparrow_s\downarrow_t\uparrow_s\uparrow_t}+N_{\downarrow_s\uparrow_t\downarrow_s\downarrow_t\uparrow_s\downarrow_t\uparrow_s\uparrow_t}+ N_{\downarrow_s\uparrow_t\uparrow_s\uparrow_t\uparrow_s\downarrow_t\downarrow_s\downarrow_t}+N_{\downarrow_s\uparrow_t\downarrow_s\downarrow_t\uparrow_s\downarrow_t\downarrow_s\uparrow_t}=M_2$,   $N_{\downarrow_s\uparrow_t\downarrow_s\downarrow_t\uparrow_s\uparrow_t\uparrow_s\uparrow_t}+N_{\downarrow_s\uparrow_t\uparrow_s\uparrow_t\downarrow_s\downarrow_t\uparrow_s\uparrow_t}+ N_{\downarrow_s\uparrow_t\uparrow_s\uparrow_t\uparrow_s\uparrow_t\downarrow_s\downarrow_t} +N_{\downarrow_s\uparrow_t\downarrow_s\downarrow_t\downarrow_s\uparrow_t\uparrow_s\uparrow_t}+N_{\downarrow_s\uparrow_t\downarrow_s\downarrow_t\uparrow_s\downarrow_t\uparrow_s\uparrow_t}+ +2N_{\downarrow_s\uparrow_t\uparrow_s\uparrow_t\downarrow_s\downarrow_t\downarrow_s\downarrow_t}+2N_{\downarrow_s\uparrow_t\downarrow_s\downarrow_t\downarrow_s\downarrow_t\uparrow_s\uparrow_t} +2N_{\downarrow_s\uparrow_t\downarrow_s\downarrow_t\uparrow_s\downarrow_t\downarrow_s\downarrow_t}+2N_{\downarrow_s\uparrow_t\downarrow_s\downarrow_t\downarrow_s\uparrow_t\downarrow_s\downarrow_t} +3N_{\downarrow_s\uparrow_t\downarrow_s\downarrow_t\downarrow_s\downarrow_t\downarrow_s\downarrow_t}=M_3$,   and $N_{\downarrow_s\uparrow_t\uparrow_s\uparrow_t\uparrow_s\uparrow_t\uparrow_s\uparrow_t}+N_{\downarrow_s\uparrow_t\downarrow_s\downarrow_t\uparrow_s\uparrow_t\uparrow_s\uparrow_t}+ N_{\downarrow_s\uparrow_t\uparrow_s\uparrow_t\downarrow_s\uparrow_t\uparrow_s\uparrow_t}+N_{\downarrow_s\uparrow_t\uparrow_s\uparrow_t\uparrow_s\downarrow_t\uparrow_s\uparrow_t}+N_{\downarrow_s\uparrow_t\uparrow_s\uparrow_t\downarrow_s\downarrow_t\uparrow_s\uparrow_t}+N_{\downarrow_s\uparrow_t\uparrow_s\uparrow_t\uparrow_s\uparrow_t\downarrow_s\downarrow_t}+N_{\downarrow_s\uparrow_t\downarrow_s\downarrow_t\downarrow_s\uparrow_t\uparrow_s\uparrow_t}+N_{\downarrow_s\uparrow_t\downarrow_s\downarrow_t\uparrow_s\downarrow_t\uparrow_s\uparrow_t}+N_{\downarrow_s\uparrow_t\downarrow_s\downarrow_t\downarrow_s\downarrow_t\uparrow_s\uparrow_t}+N_{\downarrow_s\uparrow_t\downarrow_s\downarrow_t\uparrow_s\uparrow_t\downarrow_s\downarrow_t}+N_{\downarrow_s\uparrow_t\uparrow_s\uparrow_t\downarrow_s\uparrow_t\downarrow_s\downarrow_t}+N_{\downarrow_s\uparrow_t\uparrow_s\uparrow_t\uparrow_s\downarrow_t\downarrow_s\downarrow_t}+N_{\downarrow_s\uparrow_t\uparrow_s\uparrow_t\downarrow_s\downarrow_t\downarrow_s\downarrow_t}+N_{\downarrow_s\uparrow_t\downarrow_s\downarrow_t\downarrow_s\uparrow_t\downarrow_s\downarrow_t}+N_{\downarrow_s\uparrow_t\downarrow_s\downarrow_t\uparrow_s\downarrow_t\downarrow_s\downarrow_t}+N_{\downarrow_s\uparrow_t\downarrow_s\downarrow_t\downarrow_s\downarrow_t\downarrow_s\downarrow_t}=L/4$.
Here, $\epsilon=N_{\downarrow_s\uparrow_t\downarrow_s\downarrow_t\uparrow_s\uparrow_t\uparrow_s\uparrow_t}+ N_{\downarrow_s\uparrow_t\downarrow_s\downarrow_t\downarrow_s\uparrow_t\uparrow_s\uparrow_t}+ N_{\downarrow_s\uparrow_t\downarrow_s\downarrow_t\uparrow_s\downarrow_t\uparrow_s\uparrow_t}+ N_{\downarrow_s\uparrow_t\downarrow_s\downarrow_t\downarrow_s\downarrow_t\uparrow_s\uparrow_t}+ 	N_{\downarrow_s\uparrow_t\downarrow_s\downarrow_t\uparrow_s\uparrow_t\downarrow_s\downarrow_t}+ N_{\downarrow_s\uparrow_t\uparrow_s\uparrow_t\downarrow_s\uparrow_t\downarrow_s\downarrow_t}+ N_{\downarrow_s\uparrow_t\uparrow_s\uparrow_t\uparrow_s\downarrow_t\downarrow_s\downarrow_t}+ N_{\downarrow_s\uparrow_t\uparrow_s\uparrow_t\downarrow_s\downarrow_t\downarrow_s\downarrow_t}$, where  $N_{\downarrow_s\uparrow_t\uparrow_s\uparrow_t\uparrow_s\uparrow_t\uparrow_s\uparrow_t}$, $N_{\downarrow_s\uparrow_t\downarrow_s\downarrow_t\uparrow_s\uparrow_t\uparrow_s\uparrow_t}$, $N_{\downarrow_s\uparrow_t\uparrow_s\uparrow_t\downarrow_s\uparrow_t\uparrow_s\uparrow_t}$, $N_{\downarrow_s\uparrow_t\uparrow_s\uparrow_t\uparrow_s\downarrow_t\uparrow_s\uparrow_t}$, $N_{\downarrow_s\uparrow_t\uparrow_s\uparrow_t\downarrow_s\downarrow_t\uparrow_s\uparrow_t}$, $N_{\downarrow_s\uparrow_t\uparrow_s\uparrow_t\uparrow_s\uparrow_t\downarrow_s\downarrow_t}$, $N_{\downarrow_s\uparrow_t\downarrow_s\downarrow_t\downarrow_s\uparrow_t\uparrow_s\uparrow_t}$, $N_{\downarrow_s\uparrow_t\downarrow_s\downarrow_t\uparrow_s\downarrow_t\uparrow_s\uparrow_t}$, $N_{\downarrow_s\uparrow_t\downarrow_s\downarrow_t\downarrow_s\downarrow_t\uparrow_s\uparrow_t}$, $N_{\downarrow_s\uparrow_t\downarrow_s\downarrow_t\uparrow_s\uparrow_t\downarrow_s\downarrow_t}$, $N_{\downarrow_s\uparrow_t\uparrow_s\uparrow_t\downarrow_s\uparrow_t\downarrow_s\downarrow_t}$, $N_{\downarrow_s\uparrow_t\uparrow_s\uparrow_t\uparrow_s\downarrow_t\downarrow_s\downarrow_t}$, $N_{\downarrow_s\uparrow_t\uparrow_s\uparrow_t\downarrow_s\downarrow_t\downarrow_s\downarrow_t}$, $N_{\downarrow_s\uparrow_t\downarrow_s\downarrow_t\downarrow_s\uparrow_t\downarrow_s\downarrow_t}$, $N_{\downarrow_s\uparrow_t\downarrow_s\downarrow_t\uparrow_s\downarrow_t\downarrow_s\downarrow_t}$, and $N_{\downarrow_s\uparrow_t\downarrow_s\downarrow_t\downarrow_s\downarrow_t\downarrow_s\downarrow_t}$ denote the numbers of unit cells in the configurations $| \! \downarrow_s\uparrow_t\uparrow_s\uparrow_t\uparrow_s\uparrow_t\uparrow_s\uparrow_t\rangle$, $| \!\downarrow_s\uparrow_t\downarrow_s\downarrow_t\uparrow_s\uparrow_t\uparrow_s\uparrow_t\rangle$, $| \!\downarrow_s\uparrow_t\uparrow_s\uparrow_t\downarrow_s\uparrow_t\uparrow_s\uparrow_t\rangle$, $| \!\downarrow_s\uparrow_t\uparrow_s\uparrow_t\uparrow_s\downarrow_t\uparrow_s\uparrow_t\rangle$, $| \!\downarrow_s\uparrow_t\uparrow_s\uparrow_t\downarrow_s\downarrow_t\uparrow_s\uparrow_t\rangle$, $| \! \downarrow_s\uparrow_t\uparrow_s\uparrow_t\uparrow_s\uparrow_t\downarrow_s\downarrow_t\rangle$,  $|\! \downarrow_s\uparrow_t\downarrow_s\downarrow_t\downarrow_s\uparrow_t\uparrow_s\uparrow_t\rangle$, $| \! \downarrow_s\uparrow_t\downarrow_s\downarrow_t\uparrow_s\downarrow_t\uparrow_s\uparrow_t\rangle$, $| \! \downarrow_s\uparrow_t\downarrow_s\downarrow_t\downarrow_s\downarrow_t\uparrow_s\uparrow_t\rangle$, $| \!\downarrow_s\uparrow_t\downarrow_s\downarrow_t\uparrow_s\uparrow_t\downarrow_s\downarrow_t\rangle$, $| \!\downarrow_s\uparrow_t\uparrow_s\uparrow_t\downarrow_s\uparrow_t\downarrow_s\downarrow_t\rangle$, $| \!\downarrow_s\uparrow_t\uparrow_s\uparrow_t\uparrow_s\downarrow_t\downarrow_s\downarrow_t\rangle$, $| \!\downarrow_s\uparrow_t\uparrow_s\uparrow_t\downarrow_s\downarrow_t\downarrow_s\downarrow_t\rangle$, $| \!\downarrow_s\uparrow_t\downarrow_s\downarrow_t\downarrow_s\uparrow_t\downarrow_s\downarrow_t\rangle$, $| \!\downarrow_s\uparrow_t\downarrow_s\downarrow_t\uparrow_s\downarrow_t\downarrow_s\downarrow_t\rangle$, and $| \!\downarrow_s\uparrow_t\downarrow_s\downarrow_t\downarrow_s\downarrow_t\downarrow_s\downarrow_t\rangle$, respectively.
In addition, $Z_4(L,M_1,M_2,M_3)$ is introduced to ensure a normalized $| \, L,M_1,M_2,M_3\rangle_4$. It takes the form
	\begin{equation}
		Z_4(L,M_1,M_2,M_3)=M_1!M_2!M_3!\sqrt{K_4(L,M_1,M_2,M_3)},
	\end{equation}
	with
	\begin{align}
		&K_4(L,M_1,M_2,M_3)\nonumber \\
		=&{\sum}'_{N_{\downarrow_s\uparrow_t\uparrow_s\uparrow_t\uparrow_s\uparrow_t\uparrow_s\uparrow_t},
			N_{\downarrow_s\uparrow_t\downarrow_s\downarrow_t\uparrow_s\uparrow_t\uparrow_s\uparrow_t},
			N_{\downarrow_s\uparrow_t\uparrow_s\uparrow_t\downarrow_s\uparrow_t\uparrow_s\uparrow_t},
			N_{\downarrow_s\uparrow_t\uparrow_s\uparrow_t\uparrow_s\downarrow_t\uparrow_s\uparrow_t},
			N_{\downarrow_s\uparrow_t\uparrow_s\uparrow_t\downarrow_s\downarrow_t\uparrow_s\uparrow_t},
			N_{\downarrow_s\uparrow_t\uparrow_s\uparrow_t\uparrow_s\uparrow_t\downarrow_s\downarrow_t},
			N_{\downarrow_s\uparrow_t\downarrow_s\downarrow_t\downarrow_s\uparrow_t\uparrow_s\uparrow_t},
			N_{\downarrow_s\uparrow_t\downarrow_s\downarrow_t\uparrow_s\downarrow_t\uparrow_s\uparrow_t},
			\atop
			N_{\downarrow_s\uparrow_t\downarrow_s\downarrow_t\downarrow_s\downarrow_t\uparrow_s\uparrow_t},
			N_{\downarrow_s\uparrow_t\downarrow_s\downarrow_t\uparrow_s\uparrow_t\downarrow_s\downarrow_t},
			N_{\downarrow_s\uparrow_t\uparrow_s\uparrow_t\downarrow_s\uparrow_t\downarrow_s\downarrow_t},
			N_{\downarrow_s\uparrow_t\uparrow_s\uparrow_t\uparrow_s\downarrow_t\downarrow_s\downarrow_t},
			N_{\downarrow_s\uparrow_t\uparrow_s\uparrow_t\downarrow_s\downarrow_t\downarrow_s\downarrow_t},
			N_{\downarrow_s\uparrow_t\downarrow_s\downarrow_t\downarrow_s\uparrow_t\downarrow_s\downarrow_t},
			N_{\downarrow_s\uparrow_t\downarrow_s\downarrow_t\uparrow_s\downarrow_t\downarrow_s\downarrow_t},
			N_{\downarrow_s\uparrow_t\downarrow_s\downarrow_t\downarrow_s\downarrow_t\downarrow_s\downarrow_t}
		}\nonumber\\
		&C_{L/4}^{N_{\downarrow_s\uparrow_t\uparrow_s\uparrow_t\uparrow_s\uparrow_t\uparrow_s\uparrow_t}}
		C_{L/4-N_{\downarrow_s\uparrow_t\uparrow_s\uparrow_t\uparrow_s\uparrow_t\uparrow_s\uparrow_t}}^{N_{\downarrow_s\uparrow_t\downarrow_s\downarrow_t\uparrow_s\uparrow_t\uparrow_s\uparrow_t}}
		C_{L/4-N_{\downarrow_s\uparrow_t\uparrow_s\uparrow_t\uparrow_s\uparrow_t\uparrow_s\uparrow_t}-N_{\downarrow_s\uparrow_t\downarrow_s\downarrow_t\uparrow_s\uparrow_t\uparrow_s\uparrow_t}}
		^{N_{\downarrow_s\uparrow_t\uparrow_s\uparrow_t\downarrow_s\uparrow_t\uparrow_s\uparrow_t}}C_{L/4-N_{\downarrow_s\uparrow_t\uparrow_s\uparrow_t\uparrow_s\uparrow_t\uparrow_s\uparrow_t}-N_{\downarrow_s\uparrow_t\downarrow_s\downarrow_t\uparrow_s\uparrow_t\uparrow_s\uparrow_t}
			-N_{\downarrow_s\uparrow_t\uparrow_s\uparrow_t\downarrow_s\uparrow_t\uparrow_s\uparrow_t}}^{N_{\downarrow_s\uparrow_t\uparrow_s\uparrow_t\uparrow_s\downarrow_t\uparrow_s\uparrow_t}}
		\nonumber \\
		&C_{L/4-N_{\downarrow_s\uparrow_t\uparrow_s\uparrow_t\uparrow_s\uparrow_t\uparrow_s\uparrow_t}-N_{\downarrow_s\uparrow_t\downarrow_s\downarrow_t\uparrow_s\uparrow_t\uparrow_s\uparrow_t}
			-N_{\downarrow_s\uparrow_t\uparrow_s\uparrow_t\downarrow_s\uparrow_t\uparrow_s\uparrow_t}-N_{\downarrow_s\uparrow_t\uparrow_s\uparrow_t\uparrow_s\downarrow_t\uparrow_s\uparrow_t}}
		^{N_{\downarrow_s\uparrow_t\uparrow_s\uparrow_t\downarrow_s\downarrow_t\uparrow_s\uparrow_t}}
		C_{L/4-N_{\downarrow_s\uparrow_t\uparrow_s\uparrow_t\uparrow_s\uparrow_t\uparrow_s\uparrow_t}-N_{\downarrow_s\uparrow_t\downarrow_s\downarrow_t\uparrow_s\uparrow_t\uparrow_s\uparrow_t}
			-N_{\downarrow_s\uparrow_t\uparrow_s\uparrow_t\downarrow_s\uparrow_t\uparrow_s\uparrow_t}-N_{\downarrow_s\uparrow_t\uparrow_s\uparrow_t\uparrow_s\downarrow_t\uparrow_s\uparrow_t}-N_{\downarrow_s\uparrow_t\uparrow_s\uparrow_t\downarrow_s\downarrow_t\uparrow_s\uparrow_t}}
		^{N_{\downarrow_s\uparrow_t\uparrow_s\uparrow_t\uparrow_s\uparrow_t\downarrow_s\downarrow_t}}
		\nonumber \\
		&C_{L/4-N_{\downarrow_s\uparrow_t\uparrow_s\uparrow_t\uparrow_s\uparrow_t\uparrow_s\uparrow_t}-N_{\downarrow_s\uparrow_t\downarrow_s\downarrow_t\uparrow_s\uparrow_t\uparrow_s\uparrow_t}
			-N_{\downarrow_s\uparrow_t\uparrow_s\uparrow_t\downarrow_s\uparrow_t\uparrow_s\uparrow_t}-N_{\downarrow_s\uparrow_t\uparrow_s\uparrow_t\uparrow_s\downarrow_t\uparrow_s\uparrow_t}-N_{\downarrow_s\uparrow_t\uparrow_s\uparrow_t\downarrow_s\downarrow_t\uparrow_s\uparrow_t}
			-N_{\downarrow_s\uparrow_t\uparrow_s\uparrow_t\uparrow_s\uparrow_t\downarrow_s\downarrow_t}}
		^{N_{\downarrow_s\uparrow_t\downarrow_s\downarrow_t\downarrow_s\uparrow_t\uparrow_s\uparrow_t}}
		\nonumber \\
		&C_{L/4-N_{\downarrow_s\uparrow_t\uparrow_s\uparrow_t\uparrow_s\uparrow_t\uparrow_s\uparrow_t}-N_{\downarrow_s\uparrow_t\downarrow_s\downarrow_t\uparrow_s\uparrow_t\uparrow_s\uparrow_t}
			-N_{\downarrow_s\uparrow_t\uparrow_s\uparrow_t\downarrow_s\uparrow_t\uparrow_s\uparrow_t}-N_{\downarrow_s\uparrow_t\uparrow_s\uparrow_t\uparrow_s\downarrow_t\uparrow_s\uparrow_t}-N_{\downarrow_s\uparrow_t\uparrow_s\uparrow_t\downarrow_s\downarrow_t\uparrow_s\uparrow_t}
			-N_{\downarrow_s\uparrow_t\uparrow_s\uparrow_t\uparrow_s\uparrow_t\downarrow_s\downarrow_t}-N_{\downarrow_s\uparrow_t\downarrow_s\downarrow_t\downarrow_s\uparrow_t\uparrow_s\uparrow_t}}
		^{N_{\downarrow_s\uparrow_t\downarrow_s\downarrow_t\uparrow_s\downarrow_t\uparrow_s\uparrow_t}}
		\nonumber \\
		&C_{L/4-N_{\downarrow_s\uparrow_t\uparrow_s\uparrow_t\uparrow_s\uparrow_t\uparrow_s\uparrow_t}-N_{\downarrow_s\uparrow_t\downarrow_s\downarrow_t\uparrow_s\uparrow_t\uparrow_s\uparrow_t}
			-N_{\downarrow_s\uparrow_t\uparrow_s\uparrow_t\downarrow_s\uparrow_t\uparrow_s\uparrow_t}-N_{\downarrow_s\uparrow_t\uparrow_s\uparrow_t\uparrow_s\downarrow_t\uparrow_s\uparrow_t}-N_{\downarrow_s\uparrow_t\uparrow_s\uparrow_t\downarrow_s\downarrow_t\uparrow_s\uparrow_t}
			-N_{\downarrow_s\uparrow_t\uparrow_s\uparrow_t\uparrow_s\uparrow_t\downarrow_s\downarrow_t}-N_{\downarrow_s\uparrow_t\downarrow_s\downarrow_t\downarrow_s\uparrow_t\uparrow_s\uparrow_t}
			-N_{\downarrow_s\uparrow_t\downarrow_s\downarrow_t\uparrow_s\downarrow_t\uparrow_s\uparrow_t}}^{N_{\downarrow_s\uparrow_t\downarrow_s\downarrow_t\downarrow_s\downarrow_t\uparrow_s\uparrow_t}}
		\nonumber \\
		&C_{L/4-N_{\downarrow_s\uparrow_t\uparrow_s\uparrow_t\uparrow_s\uparrow_t\uparrow_s\uparrow_t}-N_{\downarrow_s\uparrow_t\downarrow_s\downarrow_t\uparrow_s\uparrow_t\uparrow_s\uparrow_t}
			-N_{\downarrow_s\uparrow_t\uparrow_s\uparrow_t\downarrow_s\uparrow_t\uparrow_s\uparrow_t}-N_{\downarrow_s\uparrow_t\uparrow_s\uparrow_t\uparrow_s\downarrow_t\uparrow_s\uparrow_t}-N_{\downarrow_s\uparrow_t\uparrow_s\uparrow_t\downarrow_s\downarrow_t\uparrow_s\uparrow_t}
			-N_{\downarrow_s\uparrow_t\uparrow_s\uparrow_t\uparrow_s\uparrow_t\downarrow_s\downarrow_t}-N_{\downarrow_s\uparrow_t\downarrow_s\downarrow_t\downarrow_s\uparrow_t\uparrow_s\uparrow_t}
			-N_{\downarrow_s\uparrow_t\downarrow_s\downarrow_t\uparrow_s\downarrow_t\uparrow_s\uparrow_t}-N_{\downarrow_s\uparrow_t\downarrow_s\downarrow_t\downarrow_s\downarrow_t\uparrow_s\uparrow_t}}
		^{N_{\downarrow_s\uparrow_t\downarrow_s\downarrow_t\uparrow_s\uparrow_t\downarrow_s\downarrow_t}}
		\nonumber \\
		&C_{L/4-N_{\downarrow_s\uparrow_t\uparrow_s\uparrow_t\uparrow_s\uparrow_t\uparrow_s\uparrow_t}-N_{\downarrow_s\uparrow_t\downarrow_s\downarrow_t\uparrow_s\uparrow_t\uparrow_s\uparrow_t}
			-N_{\downarrow_s\uparrow_t\uparrow_s\uparrow_t\downarrow_s\uparrow_t\uparrow_s\uparrow_t}-N_{\downarrow_s\uparrow_t\uparrow_s\uparrow_t\uparrow_s\downarrow_t\uparrow_s\uparrow_t}-
			N_{\downarrow_s\uparrow_t\uparrow_s\uparrow_t\downarrow_s\downarrow_t\uparrow_s\uparrow_t}-N_{\downarrow_s\uparrow_t\uparrow_s\uparrow_t\uparrow_s\uparrow_t\downarrow_s\downarrow_t}
			\atop
			-N_{\downarrow_s\uparrow_t\downarrow_s\downarrow_t\downarrow_s\uparrow_t\uparrow_s\uparrow_t}
			-N_{\downarrow_s\uparrow_t\downarrow_s\downarrow_t\uparrow_s\downarrow_t\uparrow_s\uparrow_t}-N_{\downarrow_s\uparrow_t\downarrow_s\downarrow_t\downarrow_s\downarrow_t\uparrow_s\uparrow_t}
			-N_{\downarrow_s\uparrow_t\downarrow_s\downarrow_t\uparrow_s\uparrow_t\downarrow_s\downarrow_t}}^{N_{\downarrow_s\uparrow_t\uparrow_s\uparrow_t\downarrow_s\uparrow_t\downarrow_s\downarrow_t}}
		\nonumber \\
		&C_{L/4-N_{\downarrow_s\uparrow_t\uparrow_s\uparrow_t\uparrow_s\uparrow_t\uparrow_s\uparrow_t}-N_{\downarrow_s\uparrow_t\downarrow_s\downarrow_t\uparrow_s\uparrow_t\uparrow_s\uparrow_t}
			-N_{\downarrow_s\uparrow_t\uparrow_s\uparrow_t\downarrow_s\uparrow_t\uparrow_s\uparrow_t}-N_{\downarrow_s\uparrow_t\uparrow_s\uparrow_t\uparrow_s\downarrow_t\uparrow_s\uparrow_t}-
			N_{\downarrow_s\uparrow_t\uparrow_s\uparrow_t\downarrow_s\downarrow_t\uparrow_s\uparrow_t}
			-N_{\downarrow_s\uparrow_t\uparrow_s\uparrow_t\uparrow_s\uparrow_t\downarrow_s\downarrow_t}-N_{\downarrow_s\uparrow_t\downarrow_s\downarrow_t\downarrow_s\uparrow_t\uparrow_s\uparrow_t}
			\atop
			-N_{\downarrow_s\uparrow_t\downarrow_s\downarrow_t\uparrow_s\downarrow_t\uparrow_s\uparrow_t}-N_{\downarrow_s\uparrow_t\downarrow_s\downarrow_t\downarrow_s\downarrow_t\uparrow_s\uparrow_t}
			-N_{\downarrow_s\uparrow_t\downarrow_s\downarrow_t\uparrow_s\uparrow_t\downarrow_s\downarrow_t}-N_{\downarrow_s\uparrow_t\uparrow_s\uparrow_t\downarrow_s\uparrow_t\downarrow_s\downarrow_t}}
		^{N_{\downarrow_s\uparrow_t\uparrow_s\uparrow_t\uparrow_s\downarrow_t\downarrow_s\downarrow_t}}
		\nonumber \\
		&C_{L/4-N_{\downarrow_s\uparrow_t\uparrow_s\uparrow_t\uparrow_s\uparrow_t\uparrow_s\uparrow_t}-N_{\downarrow_s\uparrow_t\downarrow_s\downarrow_t\uparrow_s\uparrow_t\uparrow_s\uparrow_t}
			-N_{\downarrow_s\uparrow_t\uparrow_s\uparrow_t\downarrow_s\uparrow_t\uparrow_s\uparrow_t}-N_{\downarrow_s\uparrow_t\uparrow_s\uparrow_t\uparrow_s\downarrow_t\uparrow_s\uparrow_t}-
			N_{\downarrow_s\uparrow_t\uparrow_s\uparrow_t\downarrow_s\downarrow_t\uparrow_s\uparrow_t}
			-N_{\downarrow_s\uparrow_t\uparrow_s\uparrow_t\uparrow_s\uparrow_t\downarrow_s\downarrow_t}-N_{\downarrow_s\uparrow_t\downarrow_s\downarrow_t\downarrow_s\uparrow_t\uparrow_s\uparrow_t}
			\atop
			-N_{\downarrow_s\uparrow_t\downarrow_s\downarrow_t\uparrow_s\downarrow_t\uparrow_s\uparrow_t}-N_{\downarrow_s\uparrow_t\downarrow_s\downarrow_t\downarrow_s\downarrow_t\uparrow_s\uparrow_t}
			-N_{\downarrow_s\uparrow_t\downarrow_s\downarrow_t\uparrow_s\uparrow_t\downarrow_s\downarrow_t}-N_{\downarrow_s\uparrow_t\uparrow_s\uparrow_t\downarrow_s\uparrow_t\downarrow_s\downarrow_t}-
			N_{\downarrow_s\uparrow_t\uparrow_s\uparrow_t\uparrow_s\downarrow_t\downarrow_s\downarrow_t}}
		^{N_{\downarrow_s\uparrow_t\uparrow_s\uparrow_t\downarrow_s\downarrow_t\downarrow_s\downarrow_t}}
		\nonumber \\
		&C_{L/4-N_{\downarrow_s\uparrow_t\uparrow_s\uparrow_t\uparrow_s\uparrow_t\uparrow_s\uparrow_t}-N_{\downarrow_s\uparrow_t\downarrow_s\downarrow_t\uparrow_s\uparrow_t\uparrow_s\uparrow_t}
			-N_{\downarrow_s\uparrow_t\uparrow_s\uparrow_t\downarrow_s\uparrow_t\uparrow_s\uparrow_t}-N_{\downarrow_s\uparrow_t\uparrow_s\uparrow_t\uparrow_s\downarrow_t\uparrow_s\uparrow_t}-N_{\downarrow_s\uparrow_t\uparrow_s\uparrow_t\downarrow_s\downarrow_t\uparrow_s\uparrow_t}
			-N_{\downarrow_s\uparrow_t\uparrow_s\uparrow_t\uparrow_s\uparrow_t\downarrow_s\downarrow_t}-N_{\downarrow_s\uparrow_t\downarrow_s\downarrow_t\downarrow_s\uparrow_t\uparrow_s\uparrow_t}
			-N_{\downarrow_s\uparrow_t\downarrow_s\downarrow_t\uparrow_s\downarrow_t\uparrow_s\uparrow_t}
			\atop
			-N_{\downarrow_s\uparrow_t\downarrow_s\downarrow_t\downarrow_s\downarrow_t\uparrow_s\uparrow_t}
			-N_{\downarrow_s\uparrow_t\downarrow_s\downarrow_t\uparrow_s\uparrow_t\downarrow_s\downarrow_t}-N_{\downarrow_s\uparrow_t\uparrow_s\uparrow_t\downarrow_s\uparrow_t\downarrow_s\downarrow_t}-
			N_{\downarrow_s\uparrow_t\uparrow_s\uparrow_t\uparrow_s\downarrow_t\downarrow_s\downarrow_t}-N_{\downarrow_s\uparrow_t\uparrow_s\uparrow_t\downarrow_s\downarrow_t\downarrow_s\downarrow_t}}
		^{N_{\downarrow_s\uparrow_t\downarrow_s\downarrow_t\downarrow_s\uparrow_t\downarrow_s\downarrow_t}}
		\nonumber \\
		&C_{L/4-N_{\downarrow_s\uparrow_t\uparrow_s\uparrow_t\uparrow_s\uparrow_t\uparrow_s\uparrow_t}-N_{\downarrow_s\uparrow_t\downarrow_s\downarrow_t\uparrow_s\uparrow_t\uparrow_s\uparrow_t}
			-N_{\downarrow_s\uparrow_t\uparrow_s\uparrow_t\downarrow_s\uparrow_t\uparrow_s\uparrow_t}-N_{\downarrow_s\uparrow_t\uparrow_s\uparrow_t\uparrow_s\downarrow_t\uparrow_s\uparrow_t}-N_{\downarrow_s\uparrow_t\uparrow_s\uparrow_t\downarrow_s\downarrow_t\uparrow_s\uparrow_t}
			-N_{\downarrow_s\uparrow_t\uparrow_s\uparrow_t\uparrow_s\uparrow_t\downarrow_s\downarrow_t}-N_{\downarrow_s\uparrow_t\downarrow_s\downarrow_t\downarrow_s\uparrow_t\uparrow_s\uparrow_t}
			-N_{\downarrow_s\uparrow_t\downarrow_s\downarrow_t\uparrow_s\downarrow_t\uparrow_s\uparrow_t}
			\atop
			-N_{\downarrow_s\uparrow_t\downarrow_s\downarrow_t\downarrow_s\downarrow_t\uparrow_s\uparrow_t}
			-N_{\downarrow_s\uparrow_t\downarrow_s\downarrow_t\uparrow_s\uparrow_t\downarrow_s\downarrow_t}-N_{\downarrow_s\uparrow_t\uparrow_s\uparrow_t\downarrow_s\uparrow_t\downarrow_s\downarrow_t}-
			N_{\downarrow_s\uparrow_t\uparrow_s\uparrow_t\uparrow_s\downarrow_t\downarrow_s\downarrow_t}-N_{\downarrow_s\uparrow_t\uparrow_s\uparrow_t\downarrow_s\downarrow_t\downarrow_s\downarrow_t}-N_{\downarrow_s\uparrow_t\downarrow_s\downarrow_t\downarrow_s\uparrow_t\downarrow_s\downarrow_t}}^{N_{\downarrow_s\uparrow_t\downarrow_s\downarrow_t\uparrow_s\downarrow_t\downarrow_s\downarrow_t}}
		\nonumber \\
		&C_{L/4-N_{\downarrow_s\uparrow_t\uparrow_s\uparrow_t\uparrow_s\uparrow_t\uparrow_s\uparrow_t}-N_{\downarrow_s\uparrow_t\downarrow_s\downarrow_t\uparrow_s\uparrow_t\uparrow_s\uparrow_t}
			-N_{\downarrow_s\uparrow_t\uparrow_s\uparrow_t\downarrow_s\uparrow_t\uparrow_s\uparrow_t}-N_{\downarrow_s\uparrow_t\uparrow_s\uparrow_t\uparrow_s\downarrow_t\uparrow_s\uparrow_t}-N_{\downarrow_s\uparrow_t\uparrow_s\uparrow_t\downarrow_s\downarrow_t\uparrow_s\uparrow_t}
			-N_{\downarrow_s\uparrow_t\uparrow_s\uparrow_t\uparrow_s\uparrow_t\downarrow_s\downarrow_t}-N_{\downarrow_s\uparrow_t\downarrow_s\downarrow_t\downarrow_s\uparrow_t\uparrow_s\uparrow_t}
			-N_{\downarrow_s\uparrow_t\downarrow_s\downarrow_t\uparrow_s\downarrow_t\uparrow_s\uparrow_t}-N_{\downarrow_s\uparrow_t\downarrow_s\downarrow_t\downarrow_s\downarrow_t\uparrow_s\uparrow_t}
			\atop
			-N_{\downarrow_s\uparrow_t\downarrow_s\downarrow_t\uparrow_s\uparrow_t\downarrow_s\downarrow_t}-N_{\downarrow_s\uparrow_t\uparrow_s\uparrow_t\downarrow_s\uparrow_t\downarrow_s\downarrow_t}-
			N_{\downarrow_s\uparrow_t\uparrow_s\uparrow_t\uparrow_s\downarrow_t\downarrow_s\downarrow_t}-N_{\downarrow_s\uparrow_t\uparrow_s\uparrow_t\downarrow_s\downarrow_t\downarrow_s\downarrow_t}-
			N_{\downarrow_s\uparrow_t\downarrow_s\downarrow_t\downarrow_s\uparrow_t\downarrow_s\downarrow_t}-N_{\downarrow_s\uparrow_t\downarrow_s\downarrow_t\uparrow_s\downarrow_t\downarrow_s\downarrow_t}}
		^{N_{\downarrow_s\uparrow_t\downarrow_s\downarrow_t\downarrow_s\downarrow_t\downarrow_s\downarrow_t}}.
	\end{align}
Then, $Z_4(L,M_1,M_2,M_3)$ may be simplified as
	\begin{equation}
		Z_4(L,M_1,M_2,M_3)=M_1!M_2!M_3!\sqrt{C_{L/4}^{M_1}C_{L/4-M_1}^{M_2}C_{3L/4-M_1-M_2}^{M_3}}.
		\label{zlm1m2m3gsm}
	\end{equation}
In addition, we have
$H_1 | \, L,M_1,M_2,M_3\rangle_4=(-2M_1-M_2-M_3)|L,M_1,M_2,M_3\rangle_4$, $H_2 | \, L,M_1,M_2,M_3\rangle_4=(L/4-M_1-2M_2-M_3) | \, L,M_1,M_2,M_3\rangle_4$, and  $H_3 | \, L,M_1,M_2,M_3\rangle_4=(3L/4-M_1-M_2-2M_3) | \, L,M_1,M_2,M_3\rangle_4$, as long as $M_1+M_2\leq L/4$ and $M_1+M_2+M_3\leq 3L/4$.

(c) For the degenerate ground states $| \, L,M_1,M_3,M_6\rangle_4$ given in Eq.~(8),  eighteen possible configurations are introduced: $| \!\downarrow_s\uparrow_t\downarrow_s\uparrow_t\uparrow_s\uparrow_t\uparrow_s\uparrow_t\rangle$, $| \!\downarrow_s\downarrow_t\downarrow_s\uparrow_t\uparrow_s\uparrow_t\uparrow_s\uparrow_t\rangle$, $| \!\downarrow_s\uparrow_t\downarrow_s\uparrow_t\downarrow_s\uparrow_t\uparrow_s\uparrow_t\rangle$, $| \!\downarrow_s\uparrow_t\downarrow_s\uparrow_t\uparrow_s\uparrow_t\uparrow_s\downarrow_t\rangle$, $| \!\downarrow_s\uparrow_t\downarrow_s\uparrow_t\downarrow_s\downarrow_t\uparrow_s\uparrow_t\rangle$, $| \!\downarrow_s\uparrow_t\downarrow_s\uparrow_t\uparrow_s\uparrow_t\downarrow_s\downarrow_t\rangle$, 
$| \! \downarrow_s\downarrow_t\downarrow_s\uparrow_t\downarrow_s\uparrow_t\uparrow_s\uparrow_t\rangle$,
$| \! \downarrow_s\downarrow_t\downarrow_s\uparrow_t\uparrow_s\uparrow_t\uparrow_s\downarrow_t\rangle$, $| \!\downarrow_s\downarrow_t\downarrow_s\uparrow_t\downarrow_s\downarrow_t\uparrow_s\uparrow_t\rangle$, $| \!\downarrow_s\downarrow_t\downarrow_s\uparrow_t\uparrow_s\uparrow_t\downarrow_s\downarrow_t\rangle$, $| \!\downarrow_s\uparrow_t\downarrow_s\uparrow_t\downarrow_s\uparrow_t\uparrow_s\downarrow_t\rangle$, $| \!\downarrow_s\uparrow_t\downarrow_s\uparrow_t\downarrow_s\uparrow_t\downarrow_s\downarrow_t\rangle$, $| \!\downarrow_s\uparrow_t\downarrow_s\uparrow_t\downarrow_s\downarrow_t\uparrow_s\downarrow_t\rangle$, $| \!\downarrow_s\uparrow_t\downarrow_s\uparrow_t\downarrow_s\downarrow_t\downarrow_s\downarrow_t\rangle$, $| \!\downarrow_s\downarrow_t\downarrow_s\uparrow_t\downarrow_s\uparrow_t\uparrow_s\downarrow_t\rangle$, $| \!\downarrow_s\downarrow_t\downarrow_s\uparrow_t\downarrow_s\uparrow_t\downarrow_s\downarrow_t\rangle$, $| \!\downarrow_s\downarrow_t\downarrow_s\uparrow_t\downarrow_s\downarrow_t\uparrow_s\downarrow_t\rangle$ and $| \!\downarrow_s\downarrow_t\downarrow_s\uparrow_t\downarrow_s\downarrow_t\downarrow_s\downarrow_t\rangle$ in a unit cell with four adjacent sites.
Here $L$ is required to be a multiple of four.

One may rewrite $| \, L,M_1,M_3,M_6\rangle_4$ as 
	\begin{align}
		&| \, L,M_1,M_3,M_6\rangle_4 \nonumber \\
		=&\frac{1}{Z_4(L,M_1,M_3,M_6)}{\sum}'_{
			N_{\downarrow_s\uparrow_t\downarrow_s\uparrow_t\uparrow_s\uparrow_t\uparrow_s\uparrow_t}, N_{\downarrow_s\downarrow_t\downarrow_s\uparrow_t\uparrow_s\uparrow_t\uparrow_s\uparrow_t}, N_{\downarrow_s\uparrow_t\downarrow_s\uparrow_t\downarrow_s\uparrow_t\uparrow_s\uparrow_t}, N_{\downarrow_s\uparrow_t\downarrow_s\uparrow_t\uparrow_s\uparrow_t\uparrow_s\downarrow_t}, N_{\downarrow_s\uparrow_t\downarrow_s\uparrow_t\downarrow_s\downarrow_t\uparrow_s\uparrow_t}, N_{\downarrow_s\uparrow_t\downarrow_s\uparrow_t\uparrow_s\uparrow_t\downarrow_s\downarrow_t},
			N_{\downarrow_s\downarrow_t\downarrow_s\uparrow_t\downarrow_s\uparrow_t\uparrow_s\uparrow_t},
			N_{\downarrow_s\downarrow_t\downarrow_s\uparrow_t\uparrow_s\uparrow_t\uparrow_s\downarrow_t}, N_{\downarrow_s\downarrow_t\downarrow_s\uparrow_t\downarrow_s\downarrow_t\uparrow_s\uparrow_t},
			\atop
			N_{\downarrow_s\downarrow_t\downarrow_s\uparrow_t\uparrow_s\uparrow_t\downarrow_s\downarrow_t}, N_{\downarrow_s\uparrow_t\downarrow_s\uparrow_t\downarrow_s\uparrow_t\uparrow_s\downarrow_t}, N_{\downarrow_s\uparrow_t\downarrow_s\uparrow_t\downarrow_s\uparrow_t\downarrow_s\downarrow_t}, N_{\downarrow_s\uparrow_t\downarrow_s\uparrow_t\downarrow_s\downarrow_t\uparrow_s\downarrow_t}, N_{\downarrow_s\uparrow_t\downarrow_s\uparrow_t\downarrow_s\downarrow_t\downarrow_s\downarrow_t}, N_{\downarrow_s\downarrow_t\downarrow_s\uparrow_t\downarrow_s\uparrow_t\uparrow_s\downarrow_t}, N_{\downarrow_s\downarrow_t\downarrow_s\uparrow_t\downarrow_s\uparrow_t\downarrow_s\downarrow_t}, N_{\downarrow_s\downarrow_t\downarrow_s\uparrow_t\downarrow_s\downarrow_t\uparrow_s\downarrow_t}, N_{\downarrow_s\downarrow_t\downarrow_s\uparrow_t\downarrow_s\downarrow_t\downarrow_s\downarrow_t} } \nonumber \\
		&(-1)^{\epsilon}\sum_{P} |\underbrace{\downarrow_s\uparrow_t\downarrow_s\uparrow_t\uparrow_s\uparrow_t\uparrow_s\uparrow_t...\downarrow_s\uparrow_t\downarrow_s\uparrow_t\uparrow_s\uparrow_t\uparrow_s\uparrow_t}_{N_{\downarrow_s\uparrow_t\downarrow_s\uparrow_t\uparrow_s\uparrow_t\uparrow_s\uparrow_t}}
		|\underbrace{\downarrow_s\downarrow_t\downarrow_s\uparrow_t\uparrow_s\uparrow_t\uparrow_s\uparrow_t...\downarrow_s\downarrow_t\downarrow_s\uparrow_t\uparrow_s\uparrow_t\uparrow_s\uparrow_t}_{N_{\downarrow_s\downarrow_t\downarrow_s\uparrow_t\uparrow_s\uparrow_t\uparrow_s\uparrow_t}}
		|\underbrace{\downarrow_s\uparrow_t\downarrow_s\uparrow_t\downarrow_s\uparrow_t\uparrow_s\uparrow_t...\downarrow_s\uparrow_t\downarrow_s\uparrow_t\downarrow_s\uparrow_t\uparrow_s\uparrow_t}_{N_{\downarrow_s\uparrow_t\downarrow_s\uparrow_t\downarrow_s\uparrow_t\uparrow_s\uparrow_t}}
		\nonumber\\
		&	
		|\underbrace{\downarrow_s\uparrow_t\downarrow_s\uparrow_t\uparrow_s\uparrow_t\uparrow_s\downarrow_t...\downarrow_s\uparrow_t\downarrow_s\uparrow_t\uparrow_s\uparrow_t\uparrow_s\downarrow_t}_{N_{\downarrow_s\uparrow_t\downarrow_s\uparrow_t\uparrow_s\uparrow_t\uparrow_s\downarrow_t}}
		|\underbrace{\downarrow_s\uparrow_t\downarrow_s\uparrow_t\downarrow_s\downarrow_t\uparrow_s\uparrow_t...\downarrow_s\uparrow_t\downarrow_s\uparrow_t\downarrow_s\downarrow_t\uparrow_s\uparrow_t}_{N_{\downarrow_s\uparrow_t\downarrow_s\uparrow_t\downarrow_s\downarrow_t\uparrow_s\uparrow_t}}
		|\underbrace{\downarrow_s\uparrow_t\downarrow_s\uparrow_t\uparrow_s\uparrow_t\downarrow_s\downarrow_t...\downarrow_s\uparrow_t\downarrow_s\uparrow_t\uparrow_s\uparrow_t\downarrow_s\downarrow_t}_{N_{\downarrow_s\uparrow_t\downarrow_s\uparrow_t\uparrow_s\uparrow_t\downarrow_s\downarrow_t}}
		\nonumber\\
		&	 |\underbrace{\downarrow_s\downarrow_t\downarrow_s\uparrow_t\downarrow_s\uparrow_t\uparrow_s\uparrow_t...\downarrow_s\downarrow_t\downarrow_s\uparrow_t\downarrow_s\uparrow_t\uparrow_s\uparrow_t}_{N_{\downarrow_s\downarrow_t\downarrow_s\uparrow_t\downarrow_s\uparrow_t\uparrow_s\uparrow_t}}
		|\underbrace{\downarrow_s\downarrow_t\downarrow_s\uparrow_t\uparrow_s\uparrow_t\uparrow_s\downarrow_t...\downarrow_s\downarrow_t\downarrow_s\uparrow_t\uparrow_s\uparrow_t\uparrow_s\downarrow_t}_{N_{\downarrow_s\downarrow_t\downarrow_s\uparrow_t\uparrow_s\uparrow_t\uparrow_s\downarrow_t}}
		|\underbrace{\downarrow_s\downarrow_t\downarrow_s\uparrow_t\downarrow_s\downarrow_t\uparrow_s\uparrow_t...\downarrow_s\downarrow_t\downarrow_s\uparrow_t\downarrow_s\downarrow_t\uparrow_s\uparrow_t}_{N_{\downarrow_s\downarrow_t\downarrow_s\uparrow_t\downarrow_s\downarrow_t\uparrow_s\uparrow_t}}
		\nonumber\\
		&	
		|\underbrace{\downarrow_s\downarrow_t\downarrow_s\uparrow_t\uparrow_s\uparrow_t\downarrow_s\downarrow_t...\downarrow_s\downarrow_t\downarrow_s\uparrow_t\uparrow_s\uparrow_t\downarrow_s\downarrow_t}_{N_{\downarrow_s\downarrow_t\downarrow_s\uparrow_t\uparrow_s\uparrow_t\downarrow_s\downarrow_t}}
		|\underbrace{\downarrow_s\uparrow_t\downarrow_s\uparrow_t\downarrow_s\uparrow_t\uparrow_s\downarrow_t...\downarrow_s\uparrow_t\downarrow_s\uparrow_t\downarrow_s\uparrow_t\uparrow_s\downarrow_t}_{N_{\downarrow_s\uparrow_t\downarrow_s\uparrow_t\downarrow_s\uparrow_t\uparrow_s\downarrow_t}}
		|\underbrace{\downarrow_s\uparrow_t\downarrow_s\uparrow_t\downarrow_s\uparrow_t\downarrow_s\downarrow_t...\downarrow_s\uparrow_t\downarrow_s\uparrow_t\downarrow_s\uparrow_t\downarrow_s\downarrow_t}_{N_{\downarrow_s\uparrow_t\downarrow_s\uparrow_t\downarrow_s\uparrow_t\downarrow_s\downarrow_t}}
		\nonumber\\
		&	
		|\underbrace{\downarrow_s\uparrow_t\downarrow_s\uparrow_t\downarrow_s\downarrow_t\uparrow_s\downarrow_t...\downarrow_s\uparrow_t\downarrow_s\uparrow_t\downarrow_s\downarrow_t\uparrow_s\downarrow_t}_{N_{\downarrow_s\uparrow_t\downarrow_s\uparrow_t\downarrow_s\downarrow_t\uparrow_s\downarrow_t}}
		|\underbrace{\downarrow_s\uparrow_t\downarrow_s\uparrow_t\downarrow_s\downarrow_t\downarrow_s\downarrow_t...\downarrow_s\uparrow_t\downarrow_s\uparrow_t\downarrow_s\downarrow_t\downarrow_s\downarrow_t}_{N_{\downarrow_s\uparrow_t\downarrow_s\uparrow_t\downarrow_s\downarrow_t\downarrow_s\downarrow_t}}
		|\underbrace{\downarrow_s\downarrow_t\downarrow_s\uparrow_t\downarrow_s\uparrow_t\uparrow_s\downarrow_t...\downarrow_s\downarrow_t\downarrow_s\uparrow_t\downarrow_s\uparrow_t\uparrow_s\downarrow_t}_{N_{\downarrow_s\downarrow_t\downarrow_s\uparrow_t\downarrow_s\uparrow_t\uparrow_s\downarrow_t}}
		\nonumber\\
		&	 |\underbrace{\downarrow_s\downarrow_t\downarrow_s\uparrow_t\downarrow_s\uparrow_t\downarrow_s\downarrow_t...\downarrow_s\downarrow_t\downarrow_s\uparrow_t\downarrow_s\uparrow_t\downarrow_s\downarrow_t}_{N_{\downarrow_s\downarrow_t\downarrow_s\uparrow_t\downarrow_s\uparrow_t\downarrow_s\downarrow_t}}
		|\underbrace{\downarrow_s\downarrow_t\downarrow_s\uparrow_t\downarrow_s\downarrow_t\uparrow_s\downarrow_t...\downarrow_s\downarrow_t\downarrow_s\uparrow_t\downarrow_s\downarrow_t\uparrow_s\downarrow_t}_{N_{\downarrow_s\downarrow_t\downarrow_s\uparrow_t\downarrow_s\downarrow_t\uparrow_s\downarrow_t}}
		|\underbrace{\downarrow_s\downarrow_t\downarrow_s\uparrow_t\downarrow_s\downarrow_t\downarrow_s\downarrow_t...\downarrow_s\downarrow_t\downarrow_s\uparrow_t\downarrow_s\downarrow_t\downarrow_s\downarrow_t}_{N_{\downarrow_s\downarrow_t\downarrow_s\uparrow_t\downarrow_s\downarrow_t\downarrow_s\downarrow_t}}
		\rangle,
		\label{lm2m3sm}
	\end{align}
with the sum $\sum_{P}$ taken over all the permutations $P$ for a given partition
$\{N_{\downarrow_s\uparrow_t\downarrow_s\uparrow_t\uparrow_s\uparrow_t\uparrow_s\uparrow_t}$, $N_{\downarrow_s\downarrow_t\downarrow_s\uparrow_t\uparrow_s\uparrow_t\uparrow_s\uparrow_t}$, $N_{\downarrow_s\uparrow_t\downarrow_s\uparrow_t\downarrow_s\uparrow_t\uparrow_s\uparrow_t}$, $N_{\downarrow_s\uparrow_t\downarrow_s\uparrow_t\uparrow_s\uparrow_t\uparrow_s\downarrow_t}$, $N_{\downarrow_s\uparrow_t\downarrow_s\uparrow_t\downarrow_s\downarrow_t\uparrow_s\uparrow_t}$, $N_{\downarrow_s\uparrow_t\downarrow_s\uparrow_t\uparrow_s\uparrow_t\downarrow_s\downarrow_t}$,
	$N_{\downarrow_s\downarrow_t\downarrow_s\uparrow_t\downarrow_s\uparrow_t\uparrow_s\uparrow_t}$,
	$N_{\downarrow_s\downarrow_t\downarrow_s\uparrow_t\uparrow_s\uparrow_t\uparrow_s\downarrow_t}$, $N_{\downarrow_s\downarrow_t\downarrow_s\uparrow_t\downarrow_s\downarrow_t\uparrow_s\uparrow_t}$,
	$N_{\downarrow_s\downarrow_t\downarrow_s\uparrow_t\uparrow_s\uparrow_t\downarrow_s\downarrow_t}$, $N_{\downarrow_s\uparrow_t\downarrow_s\uparrow_t\downarrow_s\uparrow_t\uparrow_s\downarrow_t}$, $N_{\downarrow_s\uparrow_t\downarrow_s\uparrow_t\downarrow_s\uparrow_t\downarrow_s\downarrow_t}$, $N_{\downarrow_s\uparrow_t\downarrow_s\uparrow_t\downarrow_s\downarrow_t\uparrow_s\downarrow_t}$, $N_{\downarrow_s\uparrow_t\downarrow_s\uparrow_t\downarrow_s\downarrow_t\downarrow_s\downarrow_t}$, $N_{\downarrow_s\downarrow_t\downarrow_s\uparrow_t\downarrow_s\uparrow_t\uparrow_s\downarrow_t}$, $N_{\downarrow_s\downarrow_t\downarrow_s\uparrow_t\downarrow_s\uparrow_t\downarrow_s\downarrow_t}$, $N_{\downarrow_s\downarrow_t\downarrow_s\uparrow_t\downarrow_s\downarrow_t\uparrow_s\downarrow_t}$, $N_{\downarrow_s\downarrow_t\downarrow_s\uparrow_t\downarrow_s\downarrow_t\downarrow_s\downarrow_t}\}$, and the sum ${\sum}'$ taken over all the possible values of
	$N_{\downarrow_s\uparrow_t\downarrow_s\uparrow_t\uparrow_s\uparrow_t\uparrow_s\uparrow_t}$, $N_{\downarrow_s\downarrow_t\downarrow_s\uparrow_t\uparrow_s\uparrow_t\uparrow_s\uparrow_t}$, $N_{\downarrow_s\uparrow_t\downarrow_s\uparrow_t\downarrow_s\uparrow_t\uparrow_s\uparrow_t}$, $N_{\downarrow_s\uparrow_t\downarrow_s\uparrow_t\uparrow_s\uparrow_t\uparrow_s\downarrow_t}$, $N_{\downarrow_s\uparrow_t\downarrow_s\uparrow_t\downarrow_s\downarrow_t\uparrow_s\uparrow_t}$, $N_{\downarrow_s\uparrow_t\downarrow_s\uparrow_t\uparrow_s\uparrow_t\downarrow_s\downarrow_t}$,
	$N_{\downarrow_s\downarrow_t\downarrow_s\uparrow_t\downarrow_s\uparrow_t\uparrow_s\uparrow_t}$,
	$N_{\downarrow_s\downarrow_t\downarrow_s\uparrow_t\uparrow_s\uparrow_t\uparrow_s\downarrow_t}$, $N_{\downarrow_s\downarrow_t\downarrow_s\uparrow_t\downarrow_s\downarrow_t\uparrow_s\uparrow_t}$,
	$N_{\downarrow_s\downarrow_t\downarrow_s\uparrow_t\uparrow_s\uparrow_t\downarrow_s\downarrow_t}$, $N_{\downarrow_s\uparrow_t\downarrow_s\uparrow_t\downarrow_s\uparrow_t\uparrow_s\downarrow_t}$, $N_{\downarrow_s\uparrow_t\downarrow_s\uparrow_t\downarrow_s\uparrow_t\downarrow_s\downarrow_t}$, $N_{\downarrow_s\uparrow_t\downarrow_s\uparrow_t\downarrow_s\downarrow_t\uparrow_s\downarrow_t}$, $N_{\downarrow_s\uparrow_t\downarrow_s\uparrow_t\downarrow_s\downarrow_t\downarrow_s\downarrow_t}$, $N_{\downarrow_s\downarrow_t\downarrow_s\uparrow_t\downarrow_s\uparrow_t\uparrow_s\downarrow_t}$, $N_{\downarrow_s\downarrow_t\downarrow_s\uparrow_t\downarrow_s\uparrow_t\downarrow_s\downarrow_t}$, $N_{\downarrow_s\downarrow_t\downarrow_s\uparrow_t\downarrow_s\downarrow_t\uparrow_s\downarrow_t}$, and $N_{\downarrow_s\downarrow_t\downarrow_s\uparrow_t\downarrow_s\downarrow_t\downarrow_s\downarrow_t}$
subject to the constraints: $ N_{\downarrow_s\uparrow_t\downarrow_s\uparrow_t\downarrow_s\uparrow_t\uparrow_s\uparrow_t}+
	N_{\downarrow_s\downarrow_t\downarrow_s\uparrow_t\downarrow_s\uparrow_t\uparrow_s\uparrow_t}+
	N_{\downarrow_s\uparrow_t\downarrow_s\uparrow_t\downarrow_s\uparrow_t\uparrow_s\downarrow_t}+ N_{\downarrow_s\uparrow_t\downarrow_s\uparrow_t\downarrow_s\uparrow_t\downarrow_s\downarrow_t}+ N_{\downarrow_s\downarrow_t\downarrow_s\uparrow_t\downarrow_s\uparrow_t\uparrow_s\downarrow_t}+ N_{\downarrow_s\downarrow_t\downarrow_s\uparrow_t\downarrow_s\uparrow_t\downarrow_s\downarrow_t}=M_1$,
	$N_{\downarrow_s\downarrow_t\downarrow_s\uparrow_t\uparrow_s\uparrow_t\uparrow_s\uparrow_t}+ N_{\downarrow_s\uparrow_t\downarrow_s\uparrow_t\uparrow_s\uparrow_t\uparrow_s\downarrow_t}+
	N_{\downarrow_s\downarrow_t\downarrow_s\uparrow_t\downarrow_s\uparrow_t\uparrow_s\uparrow_t}+
	2N_{\downarrow_s\downarrow_t\downarrow_s\uparrow_t\uparrow_s\uparrow_t\uparrow_s\downarrow_t}+ N_{\downarrow_s\downarrow_t\downarrow_s\uparrow_t\downarrow_s\downarrow_t\uparrow_s\uparrow_t}+
	N_{\downarrow_s\downarrow_t\downarrow_s\uparrow_t\uparrow_s\uparrow_t\downarrow_s\downarrow_t}+ N_{\downarrow_s\uparrow_t\downarrow_s\uparrow_t\downarrow_s\uparrow_t\uparrow_s\downarrow_t}+ N_{\downarrow_s\uparrow_t\downarrow_s\uparrow_t\downarrow_s\downarrow_t\uparrow_s\downarrow_t}+ 2N_{\downarrow_s\downarrow_t\downarrow_s\uparrow_t\downarrow_s\uparrow_t\uparrow_s\downarrow_t}+ N_{\downarrow_s\downarrow_t\downarrow_s\uparrow_t\downarrow_s\uparrow_t\downarrow_s\downarrow_t}+ 2N_{\downarrow_s\downarrow_t\downarrow_s\uparrow_t\downarrow_s\downarrow_t\uparrow_s\downarrow_t}+ N_{\downarrow_s\downarrow_t\downarrow_s\uparrow_t\downarrow_s\downarrow_t\downarrow_s\downarrow_t}=M_2$,
	$ N_{\downarrow_s\uparrow_t\downarrow_s\uparrow_t\downarrow_s\downarrow_t\uparrow_s\uparrow_t}+ N_{\downarrow_s\uparrow_t\downarrow_s\uparrow_t\uparrow_s\uparrow_t\downarrow_s\downarrow_t}+ N_{\downarrow_s\downarrow_t\downarrow_s\uparrow_t\downarrow_s\downarrow_t\uparrow_s\uparrow_t}+
	N_{\downarrow_s\downarrow_t\downarrow_s\uparrow_t\uparrow_s\uparrow_t\downarrow_s\downarrow_t}+ N_{\downarrow_s\uparrow_t\downarrow_s\uparrow_t\downarrow_s\uparrow_t\downarrow_s\downarrow_t}+ N_{\downarrow_s\uparrow_t\downarrow_s\uparrow_t\downarrow_s\downarrow_t\uparrow_s\downarrow_t}+ 2N_{\downarrow_s\uparrow_t\downarrow_s\uparrow_t\downarrow_s\downarrow_t\downarrow_s\downarrow_t}+ N_{\downarrow_s\downarrow_t\downarrow_s\uparrow_t\downarrow_s\uparrow_t\downarrow_s\downarrow_t}+ N_{\downarrow_s\downarrow_t\downarrow_s\uparrow_t\downarrow_s\downarrow_t\uparrow_s\downarrow_t}+ 2N_{\downarrow_s\downarrow_t\downarrow_s\uparrow_t\downarrow_s\downarrow_t\downarrow_s\downarrow_t}=M_3$, and
	$N_{\downarrow_s\uparrow_t\downarrow_s\uparrow_t\uparrow_s\uparrow_t\uparrow_s\uparrow_t}+ N_{\downarrow_s\downarrow_t\downarrow_s\uparrow_t\uparrow_s\uparrow_t\uparrow_s\uparrow_t}+ N_{\downarrow_s\uparrow_t\downarrow_s\uparrow_t\downarrow_s\uparrow_t\uparrow_s\uparrow_t}+ N_{\downarrow_s\uparrow_t\downarrow_s\uparrow_t\uparrow_s\uparrow_t\uparrow_s\downarrow_t}+ N_{\downarrow_s\uparrow_t\downarrow_s\uparrow_t\downarrow_s\downarrow_t\uparrow_s\uparrow_t}+ N_{\downarrow_s\uparrow_t\downarrow_s\uparrow_t\uparrow_s\uparrow_t\downarrow_s\downarrow_t}+
	N_{\downarrow_s\downarrow_t\downarrow_s\uparrow_t\downarrow_s\uparrow_t\uparrow_s\uparrow_t}+
	N_{\downarrow_s\downarrow_t\downarrow_s\uparrow_t\uparrow_s\uparrow_t\uparrow_s\downarrow_t}+ N_{\downarrow_s\downarrow_t\downarrow_s\uparrow_t\downarrow_s\downarrow_t\uparrow_s\uparrow_t}+
	N_{\downarrow_s\downarrow_t\downarrow_s\uparrow_t\uparrow_s\uparrow_t\downarrow_s\downarrow_t}+ N_{\downarrow_s\uparrow_t\downarrow_s\uparrow_t\downarrow_s\uparrow_t\uparrow_s\downarrow_t}+ N_{\downarrow_s\uparrow_t\downarrow_s\uparrow_t\downarrow_s\uparrow_t\downarrow_s\downarrow_t}+ N_{\downarrow_s\uparrow_t\downarrow_s\uparrow_t\downarrow_s\downarrow_t\uparrow_s\downarrow_t}+ N_{\downarrow_s\uparrow_t\downarrow_s\uparrow_t\downarrow_s\downarrow_t\downarrow_s\downarrow_t}+ N_{\downarrow_s\downarrow_t\downarrow_s\uparrow_t\downarrow_s\uparrow_t\uparrow_s\downarrow_t}+ N_{\downarrow_s\downarrow_t\downarrow_s\uparrow_t\downarrow_s\uparrow_t\downarrow_s\downarrow_t}+ N_{\downarrow_s\downarrow_t\downarrow_s\uparrow_t\downarrow_s\downarrow_t\uparrow_s\downarrow_t}+ N_{\downarrow_s\downarrow_t\downarrow_s\uparrow_t\downarrow_s\downarrow_t\downarrow_s\downarrow_t}=L/4$.
Here, $\epsilon= N_{\downarrow_s\uparrow_t\downarrow_s\uparrow_t\uparrow_s\uparrow_t\downarrow_s\downarrow_t}+
   N_{\downarrow_s\downarrow_t\downarrow_s\uparrow_t\uparrow_s\uparrow_t\downarrow_s\downarrow_t}+ N_{\downarrow_s\uparrow_t\downarrow_s\uparrow_t\downarrow_s\uparrow_t\downarrow_s\downarrow_t}+ N_{\downarrow_s\uparrow_t\downarrow_s\uparrow_t\downarrow_s\downarrow_t\downarrow_s\downarrow_t}+ N_{\downarrow_s\downarrow_t\downarrow_s\uparrow_t\downarrow_s\uparrow_t\downarrow_s\downarrow_t}+ N_{\downarrow_s\downarrow_t\downarrow_s\uparrow_t\downarrow_s\downarrow_t\downarrow_s\downarrow_t}$, $N_{\downarrow_s\uparrow_t\downarrow_s\uparrow_t\uparrow_s\uparrow_t\uparrow_s\uparrow_t}$, $N_{\downarrow_s\downarrow_t\downarrow_s\uparrow_t\uparrow_s\uparrow_t\uparrow_s\uparrow_t}$, $N_{\downarrow_s\uparrow_t\downarrow_s\uparrow_t\downarrow_s\uparrow_t\uparrow_s\uparrow_t}$, $N_{\downarrow_s\uparrow_t\downarrow_s\uparrow_t\uparrow_s\uparrow_t\uparrow_s\downarrow_t}$, $N_{\downarrow_s\uparrow_t\downarrow_s\uparrow_t\downarrow_s\downarrow_t\uparrow_s\uparrow_t}$, $N_{\downarrow_s\uparrow_t\downarrow_s\uparrow_t\uparrow_s\uparrow_t\downarrow_s\downarrow_t}$,
   $N_{\downarrow_s\downarrow_t\downarrow_s\uparrow_t\downarrow_s\uparrow_t\uparrow_s\uparrow_t}$,
   $N_{\downarrow_s\downarrow_t\downarrow_s\uparrow_t\uparrow_s\uparrow_t\uparrow_s\downarrow_t}$, $N_{\downarrow_s\downarrow_t\downarrow_s\uparrow_t\downarrow_s\downarrow_t\uparrow_s\uparrow_t}$,
   $N_{\downarrow_s\downarrow_t\downarrow_s\uparrow_t\uparrow_s\uparrow_t\downarrow_s\downarrow_t}$, $N_{\downarrow_s\uparrow_t\downarrow_s\uparrow_t\downarrow_s\uparrow_t\uparrow_s\downarrow_t}$, $N_{\downarrow_s\uparrow_t\downarrow_s\uparrow_t\downarrow_s\uparrow_t\downarrow_s\downarrow_t}$, $N_{\downarrow_s\uparrow_t\downarrow_s\uparrow_t\downarrow_s\downarrow_t\uparrow_s\downarrow_t}$, $N_{\downarrow_s\uparrow_t\downarrow_s\uparrow_t\downarrow_s\downarrow_t\downarrow_s\downarrow_t}$, $N_{\downarrow_s\downarrow_t\downarrow_s\uparrow_t\downarrow_s\uparrow_t\uparrow_s\downarrow_t}$, $N_{\downarrow_s\downarrow_t\downarrow_s\uparrow_t\downarrow_s\uparrow_t\downarrow_s\downarrow_t}$, $N_{\downarrow_s\downarrow_t\downarrow_s\uparrow_t\downarrow_s\downarrow_t\uparrow_s\downarrow_t}$, and $N_{\downarrow_s\downarrow_t\downarrow_s\uparrow_t\downarrow_s\downarrow_t\downarrow_s\downarrow_t}$
denote the numbers of the unit cells in the configurations $| \! 
   \downarrow_s\uparrow_t\downarrow_s\uparrow_t\uparrow_s\uparrow_t\uparrow_s\uparrow_t\rangle$, $|\downarrow_s\downarrow_t\downarrow_s\uparrow_t\uparrow_s\uparrow_t\uparrow_s\uparrow_t\rangle$, $| \!\downarrow_s\uparrow_t\downarrow_s\uparrow_t\downarrow_s\uparrow_t\uparrow_s\uparrow_t\rangle$, $| \!\downarrow_s\uparrow_t\downarrow_s\uparrow_t\uparrow_s\uparrow_t\uparrow_s\downarrow_t\rangle$, $| \!\downarrow_s\uparrow_t\downarrow_s\uparrow_t\downarrow_s\downarrow_t\uparrow_s\uparrow_t\rangle$, $| \!\downarrow_s\uparrow_t\downarrow_s\uparrow_t\uparrow_s\uparrow_t\downarrow_s\downarrow_t\rangle$,
	$| \! \downarrow_s\downarrow_t\downarrow_s\uparrow_t\downarrow_s\uparrow_t\uparrow_s\uparrow_t\rangle$,
	$| \! \downarrow_s\downarrow_t\downarrow_s\uparrow_t\uparrow_s\uparrow_t\uparrow_s\downarrow_t\rangle$, $| \!\downarrow_s\downarrow_t\downarrow_s\uparrow_t\downarrow_s\downarrow_t\uparrow_s\uparrow_t\rangle$, $| \!\downarrow_s\downarrow_t\downarrow_s\uparrow_t\uparrow_s\uparrow_t\downarrow_s\downarrow_t\rangle$, $| \!\downarrow_s\uparrow_t\downarrow_s\uparrow_t\downarrow_s\uparrow_t\uparrow_s\downarrow_t\rangle$, $| \!\downarrow_s\uparrow_t\downarrow_s\uparrow_t\downarrow_s\uparrow_t\downarrow_s\downarrow_t\rangle$, $| \!\downarrow_s\uparrow_t\downarrow_s\uparrow_t\downarrow_s\downarrow_t\uparrow_s\downarrow_t\rangle$, $| \!\downarrow_s\uparrow_t\downarrow_s\uparrow_t\downarrow_s\downarrow_t\downarrow_s\downarrow_t\rangle$, $| \!\downarrow_s\downarrow_t\downarrow_s\uparrow_t\downarrow_s\uparrow_t\uparrow_s\downarrow_t\rangle$, $| \!\downarrow_s\downarrow_t\downarrow_s\uparrow_t\downarrow_s\uparrow_t\downarrow_s\downarrow_t\rangle$, $| \!\downarrow_s\downarrow_t\downarrow_s\uparrow_t\downarrow_s\downarrow_t\uparrow_s\downarrow_t\rangle$ and $| \!\downarrow_s\downarrow_t\downarrow_s\uparrow_t\downarrow_s\downarrow_t\downarrow_s\downarrow_t\rangle$ respectively.  
In addition, $Z_4(L,M_1,M_3,M_6)$ is introduced to ensure a normalized $| \, L,M_1,M_3,M_6\rangle_4$,  taking the form
	\begin{equation}
		Z_4(L,M_1,M_3,M_6)=M_1!M_3!M_6!\sqrt{K_4(L,M_1,M_3,M_6)},
	\end{equation}
with
	\begin{align}
		&K_4(L,M_1,M_3,M_6)\nonumber \\
		=&{\sum}'_{
			N_{\downarrow_s\uparrow_t\downarrow_s\uparrow_t\uparrow_s\uparrow_t\uparrow_s\uparrow_t}, N_{\downarrow_s\downarrow_t\downarrow_s\uparrow_t\uparrow_s\uparrow_t\uparrow_s\uparrow_t}, N_{\downarrow_s\uparrow_t\downarrow_s\uparrow_t\downarrow_s\uparrow_t\uparrow_s\uparrow_t}, N_{\downarrow_s\uparrow_t\downarrow_s\uparrow_t\uparrow_s\uparrow_t\uparrow_s\downarrow_t}, N_{\downarrow_s\uparrow_t\downarrow_s\uparrow_t\downarrow_s\downarrow_t\uparrow_s\uparrow_t}, N_{\downarrow_s\uparrow_t\downarrow_s\uparrow_t\uparrow_s\uparrow_t\downarrow_s\downarrow_t},
			N_{\downarrow_s\downarrow_t\downarrow_s\uparrow_t\downarrow_s\uparrow_t\uparrow_s\uparrow_t},
			N_{\downarrow_s\downarrow_t\downarrow_s\uparrow_t\uparrow_s\uparrow_t\uparrow_s\downarrow_t}, N_{\downarrow_s\downarrow_t\downarrow_s\uparrow_t\downarrow_s\downarrow_t\uparrow_s\uparrow_t},
			\atop
			N_{\downarrow_s\downarrow_t\downarrow_s\uparrow_t\uparrow_s\uparrow_t\downarrow_s\downarrow_t}, N_{\downarrow_s\uparrow_t\downarrow_s\uparrow_t\downarrow_s\uparrow_t\uparrow_s\downarrow_t}, N_{\downarrow_s\uparrow_t\downarrow_s\uparrow_t\downarrow_s\uparrow_t\downarrow_s\downarrow_t}, N_{\downarrow_s\uparrow_t\downarrow_s\uparrow_t\downarrow_s\downarrow_t\uparrow_s\downarrow_t}, N_{\downarrow_s\uparrow_t\downarrow_s\uparrow_t\downarrow_s\downarrow_t\downarrow_s\downarrow_t}, N_{\downarrow_s\downarrow_t\downarrow_s\uparrow_t\downarrow_s\uparrow_t\uparrow_s\downarrow_t}, N_{\downarrow_s\downarrow_t\downarrow_s\uparrow_t\downarrow_s\uparrow_t\downarrow_s\downarrow_t}, N_{\downarrow_s\downarrow_t\downarrow_s\uparrow_t\downarrow_s\downarrow_t\uparrow_s\downarrow_t}, N_{\downarrow_s\downarrow_t\downarrow_s\uparrow_t\downarrow_s\downarrow_t\downarrow_s\downarrow_t} }
		\nonumber\\
		&C_{L/4}^{N_{\downarrow_s\uparrow_t\downarrow_s\uparrow_t\uparrow_s\uparrow_t\uparrow_s\uparrow_t}} C_{L/4-N_{\downarrow_s\uparrow_t\downarrow_s\uparrow_t\uparrow_s\uparrow_t\uparrow_s\uparrow_t}}^
		{N_{\downarrow_s\downarrow_t\downarrow_s\uparrow_t\uparrow_s\uparrow_t\uparrow_s\uparrow_t}} C_{L/4-N_{\downarrow_s\uparrow_t\downarrow_s\uparrow_t\uparrow_s\uparrow_t\uparrow_s\uparrow_t}-
			N_{\downarrow_s\downarrow_t\downarrow_s\uparrow_t\uparrow_s\uparrow_t\uparrow_s\uparrow_t}} ^{N_{\downarrow_s\uparrow_t\downarrow_s\uparrow_t\downarrow_s\uparrow_t\uparrow_s\uparrow_t}}
		C_{L/4-N_{\downarrow_s\uparrow_t\downarrow_s\uparrow_t\uparrow_s\uparrow_t\uparrow_s\uparrow_t}-
			N_{\downarrow_s\downarrow_t\downarrow_s\uparrow_t\uparrow_s\uparrow_t\uparrow_s\uparrow_t}-
			N_{\downarrow_s\uparrow_t\downarrow_s\uparrow_t\downarrow_s\uparrow_t\uparrow_s\uparrow_t}} ^{N_{\downarrow_s\uparrow_t\downarrow_s\uparrow_t\uparrow_s\uparrow_t\uparrow_s\downarrow_t}}
		\nonumber\\
		&C_{L/4-N_{\downarrow_s\uparrow_t\downarrow_s\uparrow_t\uparrow_s\uparrow_t\uparrow_s\uparrow_t}-
			N_{\downarrow_s\downarrow_t\downarrow_s\uparrow_t\uparrow_s\uparrow_t\uparrow_s\uparrow_t}-
			N_{\downarrow_s\uparrow_t\downarrow_s\uparrow_t\downarrow_s\uparrow_t\uparrow_s\uparrow_t}-
			N_{\downarrow_s\uparrow_t\downarrow_s\uparrow_t\uparrow_s\uparrow_t\uparrow_s\downarrow_t}} ^{N_{\downarrow_s\uparrow_t\downarrow_s\uparrow_t\downarrow_s\downarrow_t\uparrow_s\uparrow_t}} C_{L/4-N_{\downarrow_s\uparrow_t\downarrow_s\uparrow_t\uparrow_s\uparrow_t\uparrow_s\uparrow_t}-
			N_{\downarrow_s\downarrow_t\downarrow_s\uparrow_t\uparrow_s\uparrow_t\uparrow_s\uparrow_t}-
			N_{\downarrow_s\uparrow_t\downarrow_s\uparrow_t\downarrow_s\uparrow_t\uparrow_s\uparrow_t}-
			N_{\downarrow_s\uparrow_t\downarrow_s\uparrow_t\uparrow_s\uparrow_t\uparrow_s\downarrow_t}-
			N_{\downarrow_s\uparrow_t\downarrow_s\uparrow_t\downarrow_s\downarrow_t\uparrow_s\uparrow_t}}
		^{N_{\downarrow_s\uparrow_t\downarrow_s\uparrow_t\uparrow_s\uparrow_t\downarrow_s\downarrow_t}}
		\nonumber\\
		&C_{L/4-N_{\downarrow_s\uparrow_t\downarrow_s\uparrow_t\uparrow_s\uparrow_t\uparrow_s\uparrow_t}-
			N_{\downarrow_s\downarrow_t\downarrow_s\uparrow_t\uparrow_s\uparrow_t\uparrow_s\uparrow_t}-
			N_{\downarrow_s\uparrow_t\downarrow_s\uparrow_t\downarrow_s\uparrow_t\uparrow_s\uparrow_t}-
			N_{\downarrow_s\uparrow_t\downarrow_s\uparrow_t\uparrow_s\uparrow_t\uparrow_s\downarrow_t}-
			N_{\downarrow_s\uparrow_t\downarrow_s\uparrow_t\downarrow_s\downarrow_t\uparrow_s\uparrow_t}-
			N_{\downarrow_s\uparrow_t\downarrow_s\uparrow_t\uparrow_s\uparrow_t\downarrow_s\downarrow_t}}
		^{N_{\downarrow_s\downarrow_t\downarrow_s\uparrow_t\downarrow_s\uparrow_t\uparrow_s\uparrow_t}}
		\nonumber\\
		&C_{L/4-N_{\downarrow_s\uparrow_t\downarrow_s\uparrow_t\uparrow_s\uparrow_t\uparrow_s\uparrow_t}-
			N_{\downarrow_s\downarrow_t\downarrow_s\uparrow_t\uparrow_s\uparrow_t\uparrow_s\uparrow_t}-
			N_{\downarrow_s\uparrow_t\downarrow_s\uparrow_t\downarrow_s\uparrow_t\uparrow_s\uparrow_t}-
			N_{\downarrow_s\uparrow_t\downarrow_s\uparrow_t\uparrow_s\uparrow_t\uparrow_s\downarrow_t}-
			N_{\downarrow_s\uparrow_t\downarrow_s\uparrow_t\downarrow_s\downarrow_t\uparrow_s\uparrow_t}-
			N_{\downarrow_s\uparrow_t\downarrow_s\uparrow_t\uparrow_s\uparrow_t\downarrow_s\downarrow_t}-
			N_{\downarrow_s\downarrow_t\downarrow_s\uparrow_t\downarrow_s\uparrow_t\uparrow_s\uparrow_t}}
		^{N_{\downarrow_s\downarrow_t\downarrow_s\uparrow_t\uparrow_s\uparrow_t\uparrow_s\downarrow_t}} \nonumber\\
		&C_{L/4-N_{\downarrow_s\uparrow_t\downarrow_s\uparrow_t\uparrow_s\uparrow_t\uparrow_s\uparrow_t}-
			N_{\downarrow_s\downarrow_t\downarrow_s\uparrow_t\uparrow_s\uparrow_t\uparrow_s\uparrow_t}-
			N_{\downarrow_s\uparrow_t\downarrow_s\uparrow_t\downarrow_s\uparrow_t\uparrow_s\uparrow_t}-
			N_{\downarrow_s\uparrow_t\downarrow_s\uparrow_t\uparrow_s\uparrow_t\uparrow_s\downarrow_t}-
			N_{\downarrow_s\uparrow_t\downarrow_s\uparrow_t\downarrow_s\downarrow_t\uparrow_s\uparrow_t}-
			N_{\downarrow_s\uparrow_t\downarrow_s\uparrow_t\uparrow_s\uparrow_t\downarrow_s\downarrow_t}-
			N_{\downarrow_s\downarrow_t\downarrow_s\uparrow_t\downarrow_s\uparrow_t\uparrow_s\uparrow_t}-
			N_{\downarrow_s\downarrow_t\downarrow_s\uparrow_t\uparrow_s\uparrow_t\uparrow_s\downarrow_t}} ^{N_{\downarrow_s\downarrow_t\downarrow_s\uparrow_t\downarrow_s\downarrow_t\uparrow_s\uparrow_t}}
		\nonumber\\
		&C_{L/4-N_{\downarrow_s\uparrow_t\downarrow_s\uparrow_t\uparrow_s\uparrow_t\uparrow_s\uparrow_t}-
			N_{\downarrow_s\downarrow_t\downarrow_s\uparrow_t\uparrow_s\uparrow_t\uparrow_s\uparrow_t}-
			N_{\downarrow_s\uparrow_t\downarrow_s\uparrow_t\downarrow_s\uparrow_t\uparrow_s\uparrow_t}-
			N_{\downarrow_s\uparrow_t\downarrow_s\uparrow_t\uparrow_s\uparrow_t\uparrow_s\downarrow_t}-
			N_{\downarrow_s\uparrow_t\downarrow_s\uparrow_t\downarrow_s\downarrow_t\uparrow_s\uparrow_t}-
			N_{\downarrow_s\uparrow_t\downarrow_s\uparrow_t\uparrow_s\uparrow_t\downarrow_s\downarrow_t}-
			N_{\downarrow_s\downarrow_t\downarrow_s\uparrow_t\downarrow_s\uparrow_t\uparrow_s\uparrow_t}-
			N_{\downarrow_s\downarrow_t\downarrow_s\uparrow_t\uparrow_s\uparrow_t\uparrow_s\downarrow_t}-
			N_{\downarrow_s\downarrow_t\downarrow_s\uparrow_t\downarrow_s\downarrow_t\uparrow_s\uparrow_t}}
		^{N_{\downarrow_s\downarrow_t\downarrow_s\uparrow_t\uparrow_s\uparrow_t\downarrow_s\downarrow_t}} \nonumber\\
		&C_{L/4-N_{\downarrow_s\uparrow_t\downarrow_s\uparrow_t\uparrow_s\uparrow_t\uparrow_s\uparrow_t}-
			N_{\downarrow_s\downarrow_t\downarrow_s\uparrow_t\uparrow_s\uparrow_t\uparrow_s\uparrow_t}-
			N_{\downarrow_s\uparrow_t\downarrow_s\uparrow_t\downarrow_s\uparrow_t\uparrow_s\uparrow_t}-
			N_{\downarrow_s\uparrow_t\downarrow_s\uparrow_t\uparrow_s\uparrow_t\uparrow_s\downarrow_t}-
			N_{\downarrow_s\uparrow_t\downarrow_s\uparrow_t\downarrow_s\downarrow_t\uparrow_s\uparrow_t}-
			N_{\downarrow_s\uparrow_t\downarrow_s\uparrow_t\uparrow_s\uparrow_t\downarrow_s\downarrow_t}-
			\atop
			N_{\downarrow_s\downarrow_t\downarrow_s\uparrow_t\downarrow_s\uparrow_t\uparrow_s\uparrow_t}-
			N_{\downarrow_s\downarrow_t\downarrow_s\uparrow_t\uparrow_s\uparrow_t\uparrow_s\downarrow_t}-
			N_{\downarrow_s\downarrow_t\downarrow_s\uparrow_t\downarrow_s\downarrow_t\uparrow_s\uparrow_t}-
			N_{\downarrow_s\downarrow_t\downarrow_s\uparrow_t\uparrow_s\uparrow_t\downarrow_s\downarrow_t}} ^{N_{\downarrow_s\uparrow_t\downarrow_s\uparrow_t\downarrow_s\uparrow_t\uparrow_s\downarrow_t}} \nonumber\\
		&C_{L/4-N_{\downarrow_s\uparrow_t\downarrow_s\uparrow_t\uparrow_s\uparrow_t\uparrow_s\uparrow_t}-
			N_{\downarrow_s\downarrow_t\downarrow_s\uparrow_t\uparrow_s\uparrow_t\uparrow_s\uparrow_t}-
			N_{\downarrow_s\uparrow_t\downarrow_s\uparrow_t\downarrow_s\uparrow_t\uparrow_s\uparrow_t}-
			N_{\downarrow_s\uparrow_t\downarrow_s\uparrow_t\uparrow_s\uparrow_t\uparrow_s\downarrow_t}-
			N_{\downarrow_s\uparrow_t\downarrow_s\uparrow_t\downarrow_s\downarrow_t\uparrow_s\uparrow_t}-
			N_{\downarrow_s\uparrow_t\downarrow_s\uparrow_t\uparrow_s\uparrow_t\downarrow_s\downarrow_t}-
			\atop
			N_{\downarrow_s\downarrow_t\downarrow_s\uparrow_t\downarrow_s\uparrow_t\uparrow_s\uparrow_t}-
			N_{\downarrow_s\downarrow_t\downarrow_s\uparrow_t\uparrow_s\uparrow_t\uparrow_s\downarrow_t}-
			N_{\downarrow_s\downarrow_t\downarrow_s\uparrow_t\downarrow_s\downarrow_t\uparrow_s\uparrow_t}-
			N_{\downarrow_s\downarrow_t\downarrow_s\uparrow_t\uparrow_s\uparrow_t\downarrow_s\downarrow_t}-
			N_{\downarrow_s\uparrow_t\downarrow_s\uparrow_t\downarrow_s\uparrow_t\uparrow_s\downarrow_t}}
		^{N_{\downarrow_s\uparrow_t\downarrow_s\uparrow_t\downarrow_s\uparrow_t\downarrow_s\downarrow_t}} \nonumber\\
		&C_{L/4-N_{\downarrow_s\uparrow_t\downarrow_s\uparrow_t\uparrow_s\uparrow_t\uparrow_s\uparrow_t}-
			N_{\downarrow_s\downarrow_t\downarrow_s\uparrow_t\uparrow_s\uparrow_t\uparrow_s\uparrow_t}-
			N_{\downarrow_s\uparrow_t\downarrow_s\uparrow_t\downarrow_s\uparrow_t\uparrow_s\uparrow_t}-
			N_{\downarrow_s\uparrow_t\downarrow_s\uparrow_t\uparrow_s\uparrow_t\uparrow_s\downarrow_t}-
			N_{\downarrow_s\uparrow_t\downarrow_s\uparrow_t\downarrow_s\downarrow_t\uparrow_s\uparrow_t}-
			N_{\downarrow_s\uparrow_t\downarrow_s\uparrow_t\uparrow_s\uparrow_t\downarrow_s\downarrow_t}-
			N_{\downarrow_s\downarrow_t\downarrow_s\uparrow_t\downarrow_s\uparrow_t\uparrow_s\uparrow_t}-
			\atop
			N_{\downarrow_s\downarrow_t\downarrow_s\uparrow_t\uparrow_s\uparrow_t\uparrow_s\downarrow_t}-
			N_{\downarrow_s\downarrow_t\downarrow_s\uparrow_t\downarrow_s\downarrow_t\uparrow_s\uparrow_t}-
			N_{\downarrow_s\downarrow_t\downarrow_s\uparrow_t\uparrow_s\uparrow_t\downarrow_s\downarrow_t}-
			N_{\downarrow_s\uparrow_t\downarrow_s\uparrow_t\downarrow_s\uparrow_t\uparrow_s\downarrow_t}-
			N_{\downarrow_s\uparrow_t\downarrow_s\uparrow_t\downarrow_s\uparrow_t\downarrow_s\downarrow_t}}
		^{N_{\downarrow_s\uparrow_t\downarrow_s\uparrow_t\downarrow_s\downarrow_t\uparrow_s\downarrow_t}} \nonumber\\
		&C_{L/4-N_{\downarrow_s\uparrow_t\downarrow_s\uparrow_t\uparrow_s\uparrow_t\uparrow_s\uparrow_t}-
			N_{\downarrow_s\downarrow_t\downarrow_s\uparrow_t\uparrow_s\uparrow_t\uparrow_s\uparrow_t}-
			N_{\downarrow_s\uparrow_t\downarrow_s\uparrow_t\downarrow_s\uparrow_t\uparrow_s\uparrow_t}-
			N_{\downarrow_s\uparrow_t\downarrow_s\uparrow_t\uparrow_s\uparrow_t\uparrow_s\downarrow_t}-
			N_{\downarrow_s\uparrow_t\downarrow_s\uparrow_t\downarrow_s\downarrow_t\uparrow_s\uparrow_t}-
			N_{\downarrow_s\uparrow_t\downarrow_s\uparrow_t\uparrow_s\uparrow_t\downarrow_s\downarrow_t}-
			N_{\downarrow_s\downarrow_t\downarrow_s\uparrow_t\downarrow_s\uparrow_t\uparrow_s\uparrow_t}-
			\atop
			N_{\downarrow_s\downarrow_t\downarrow_s\uparrow_t\uparrow_s\uparrow_t\uparrow_s\downarrow_t}-
			N_{\downarrow_s\downarrow_t\downarrow_s\uparrow_t\downarrow_s\downarrow_t\uparrow_s\uparrow_t}-
			N_{\downarrow_s\downarrow_t\downarrow_s\uparrow_t\uparrow_s\uparrow_t\downarrow_s\downarrow_t}-
			N_{\downarrow_s\uparrow_t\downarrow_s\uparrow_t\downarrow_s\uparrow_t\uparrow_s\downarrow_t}-
			N_{\downarrow_s\uparrow_t\downarrow_s\uparrow_t\downarrow_s\uparrow_t\downarrow_s\downarrow_t}-
			N_{\downarrow_s\uparrow_t\downarrow_s\uparrow_t\downarrow_s\downarrow_t\uparrow_s\downarrow_t}}
		^{N_{\downarrow_s\uparrow_t\downarrow_s\uparrow_t\downarrow_s\downarrow_t\downarrow_s\downarrow_t}} \nonumber\\
		&C_{L/4-N_{\downarrow_s\uparrow_t\downarrow_s\uparrow_t\uparrow_s\uparrow_t\uparrow_s\uparrow_t}-
			N_{\downarrow_s\downarrow_t\downarrow_s\uparrow_t\uparrow_s\uparrow_t\uparrow_s\uparrow_t}-
			N_{\downarrow_s\uparrow_t\downarrow_s\uparrow_t\downarrow_s\uparrow_t\uparrow_s\uparrow_t}-
			N_{\downarrow_s\uparrow_t\downarrow_s\uparrow_t\uparrow_s\uparrow_t\uparrow_s\downarrow_t}-
			N_{\downarrow_s\uparrow_t\downarrow_s\uparrow_t\downarrow_s\downarrow_t\uparrow_s\uparrow_t}-
			N_{\downarrow_s\uparrow_t\downarrow_s\uparrow_t\uparrow_s\uparrow_t\downarrow_s\downarrow_t}-
			N_{\downarrow_s\downarrow_t\downarrow_s\uparrow_t\downarrow_s\uparrow_t\uparrow_s\uparrow_t}-
			N_{\downarrow_s\downarrow_t\downarrow_s\uparrow_t\uparrow_s\uparrow_t\uparrow_s\downarrow_t}-
			\atop
			N_{\downarrow_s\downarrow_t\downarrow_s\uparrow_t\downarrow_s\downarrow_t\uparrow_s\uparrow_t}-
			N_{\downarrow_s\downarrow_t\downarrow_s\uparrow_t\uparrow_s\uparrow_t\downarrow_s\downarrow_t}-
			N_{\downarrow_s\uparrow_t\downarrow_s\uparrow_t\downarrow_s\uparrow_t\uparrow_s\downarrow_t}-
			N_{\downarrow_s\uparrow_t\downarrow_s\uparrow_t\downarrow_s\uparrow_t\downarrow_s\downarrow_t}-
			N_{\downarrow_s\uparrow_t\downarrow_s\uparrow_t\downarrow_s\downarrow_t\uparrow_s\downarrow_t}-
			N_{\downarrow_s\uparrow_t\downarrow_s\uparrow_t\downarrow_s\downarrow_t\downarrow_s\downarrow_t}}
		^{N_{\downarrow_s\downarrow_t\downarrow_s\uparrow_t\downarrow_s\uparrow_t\uparrow_s\downarrow_t}} \nonumber\\
		&C_{L/4-N_{\downarrow_s\uparrow_t\downarrow_s\uparrow_t\uparrow_s\uparrow_t\uparrow_s\uparrow_t}-
			N_{\downarrow_s\downarrow_t\downarrow_s\uparrow_t\uparrow_s\uparrow_t\uparrow_s\uparrow_t}-
			N_{\downarrow_s\uparrow_t\downarrow_s\uparrow_t\downarrow_s\uparrow_t\uparrow_s\uparrow_t}-
			N_{\downarrow_s\uparrow_t\downarrow_s\uparrow_t\uparrow_s\uparrow_t\uparrow_s\downarrow_t}-
			N_{\downarrow_s\uparrow_t\downarrow_s\uparrow_t\downarrow_s\downarrow_t\uparrow_s\uparrow_t}-
			N_{\downarrow_s\uparrow_t\downarrow_s\uparrow_t\uparrow_s\uparrow_t\downarrow_s\downarrow_t}-
			N_{\downarrow_s\downarrow_t\downarrow_s\uparrow_t\downarrow_s\uparrow_t\uparrow_s\uparrow_t}-
			N_{\downarrow_s\downarrow_t\downarrow_s\uparrow_t\uparrow_s\uparrow_t\uparrow_s\downarrow_t}-
			\atop
			N_{\downarrow_s\downarrow_t\downarrow_s\uparrow_t\downarrow_s\downarrow_t\uparrow_s\uparrow_t}-
			N_{\downarrow_s\downarrow_t\downarrow_s\uparrow_t\uparrow_s\uparrow_t\downarrow_s\downarrow_t}-
			N_{\downarrow_s\uparrow_t\downarrow_s\uparrow_t\downarrow_s\uparrow_t\uparrow_s\downarrow_t}-
			N_{\downarrow_s\uparrow_t\downarrow_s\uparrow_t\downarrow_s\uparrow_t\downarrow_s\downarrow_t}-
			N_{\downarrow_s\uparrow_t\downarrow_s\uparrow_t\downarrow_s\downarrow_t\uparrow_s\downarrow_t}-
			N_{\downarrow_s\uparrow_t\downarrow_s\uparrow_t\downarrow_s\downarrow_t\downarrow_s\downarrow_t}-
			N_{\downarrow_s\downarrow_t\downarrow_s\uparrow_t\downarrow_s\uparrow_t\uparrow_s\downarrow_t}} ^{N_{\downarrow_s\downarrow_t\downarrow_s\uparrow_t\downarrow_s\uparrow_t\downarrow_s\downarrow_t}} \nonumber\\
		&C_{L/4-N_{\downarrow_s\uparrow_t\downarrow_s\uparrow_t\uparrow_s\uparrow_t\uparrow_s\uparrow_t}-
			N_{\downarrow_s\downarrow_t\downarrow_s\uparrow_t\uparrow_s\uparrow_t\uparrow_s\uparrow_t}-
			N_{\downarrow_s\uparrow_t\downarrow_s\uparrow_t\downarrow_s\uparrow_t\uparrow_s\uparrow_t}-
			N_{\downarrow_s\uparrow_t\downarrow_s\uparrow_t\uparrow_s\uparrow_t\uparrow_s\downarrow_t}-
			N_{\downarrow_s\uparrow_t\downarrow_s\uparrow_t\downarrow_s\downarrow_t\uparrow_s\uparrow_t}-
			N_{\downarrow_s\uparrow_t\downarrow_s\uparrow_t\uparrow_s\uparrow_t\downarrow_s\downarrow_t}-
			N_{\downarrow_s\downarrow_t\downarrow_s\uparrow_t\downarrow_s\uparrow_t\uparrow_s\uparrow_t}-
			N_{\downarrow_s\downarrow_t\downarrow_s\uparrow_t\uparrow_s\uparrow_t\uparrow_s\downarrow_t}-
			\atop
			N_{\downarrow_s\downarrow_t\downarrow_s\uparrow_t\downarrow_s\downarrow_t\uparrow_s\uparrow_t}-
			N_{\downarrow_s\downarrow_t\downarrow_s\uparrow_t\uparrow_s\uparrow_t\downarrow_s\downarrow_t}-
			N_{\downarrow_s\uparrow_t\downarrow_s\uparrow_t\downarrow_s\uparrow_t\uparrow_s\downarrow_t}-
			N_{\downarrow_s\uparrow_t\downarrow_s\uparrow_t\downarrow_s\uparrow_t\downarrow_s\downarrow_t}-
			N_{\downarrow_s\uparrow_t\downarrow_s\uparrow_t\downarrow_s\downarrow_t\uparrow_s\downarrow_t}-
			N_{\downarrow_s\uparrow_t\downarrow_s\uparrow_t\downarrow_s\downarrow_t\downarrow_s\downarrow_t}-
			N_{\downarrow_s\downarrow_t\downarrow_s\uparrow_t\downarrow_s\uparrow_t\uparrow_s\downarrow_t}-
			N_{\downarrow_s\downarrow_t\downarrow_s\uparrow_t\downarrow_s\uparrow_t\downarrow_s\downarrow_t}}
		^{N_{\downarrow_s\downarrow_t\downarrow_s\uparrow_t\downarrow_s\downarrow_t\uparrow_s\downarrow_t}} \nonumber\\
		&C_{L/4-N_{\downarrow_s\uparrow_t\downarrow_s\uparrow_t\uparrow_s\uparrow_t\uparrow_s\uparrow_t}-
			N_{\downarrow_s\downarrow_t\downarrow_s\uparrow_t\uparrow_s\uparrow_t\uparrow_s\uparrow_t}-
			N_{\downarrow_s\uparrow_t\downarrow_s\uparrow_t\downarrow_s\uparrow_t\uparrow_s\uparrow_t}-
			N_{\downarrow_s\uparrow_t\downarrow_s\uparrow_t\uparrow_s\uparrow_t\uparrow_s\downarrow_t}-
			N_{\downarrow_s\uparrow_t\downarrow_s\uparrow_t\downarrow_s\downarrow_t\uparrow_s\uparrow_t}-
			N_{\downarrow_s\uparrow_t\downarrow_s\uparrow_t\uparrow_s\uparrow_t\downarrow_s\downarrow_t}-
			N_{\downarrow_s\downarrow_t\downarrow_s\uparrow_t\downarrow_s\uparrow_t\uparrow_s\uparrow_t}-
			N_{\downarrow_s\downarrow_t\downarrow_s\uparrow_t\uparrow_s\uparrow_t\uparrow_s\downarrow_t}-
			N_{\downarrow_s\downarrow_t\downarrow_s\uparrow_t\downarrow_s\downarrow_t\uparrow_s\uparrow_t}-
			\atop
			N_{\downarrow_s\downarrow_t\downarrow_s\uparrow_t\uparrow_s\uparrow_t\downarrow_s\downarrow_t}-
			N_{\downarrow_s\uparrow_t\downarrow_s\uparrow_t\downarrow_s\uparrow_t\uparrow_s\downarrow_t}-
			N_{\downarrow_s\uparrow_t\downarrow_s\uparrow_t\downarrow_s\uparrow_t\downarrow_s\downarrow_t}-
			N_{\downarrow_s\uparrow_t\downarrow_s\uparrow_t\downarrow_s\downarrow_t\uparrow_s\downarrow_t}-
			N_{\downarrow_s\uparrow_t\downarrow_s\uparrow_t\downarrow_s\downarrow_t\downarrow_s\downarrow_t}-
			N_{\downarrow_s\downarrow_t\downarrow_s\uparrow_t\downarrow_s\uparrow_t\uparrow_s\downarrow_t}-
			N_{\downarrow_s\downarrow_t\downarrow_s\uparrow_t\downarrow_s\uparrow_t\downarrow_s\downarrow_t}-
			N_{\downarrow_s\downarrow_t\downarrow_s\uparrow_t\downarrow_s\downarrow_t\uparrow_s\downarrow_t}}
		^{N_{\downarrow_s\downarrow_t\downarrow_s\uparrow_t\downarrow_s\downarrow_t\downarrow_s\downarrow_t}} .
	\end{align}
	
Actually, $Z_4(L,M_1,M_3,M_6)$ may be simplified as
	\begin{equation}
Z_4(L,M_1,M_3,M_6)=M_1!M_3!M_6!\sqrt{C_{L/4}^{M_1}\sum_{l=0}^{\min(M_6,L/4)}C_{L/4}^{l}C_{L/4}^{M_6-l}C_{L/2-M_1-M_6+l}^{M_3}}.
		\label{zlm1m3m5gsm}
	\end{equation}
In addition, we have $H_1 | \, L,M_1,M_3,M_6\rangle_4=(-2M_1-M_3+M_6)|L,M_1,M_3,M_6\rangle_4$, $H_3 | \, L,M_1,M_3,M_6\rangle_4=(L/2-M_1-2M_3-M_6)|L,M_1,M_3,M_6\rangle_4$ and $(H_3-H_1) | \, L,M_1,M_3,M_6\rangle_4=(L/2-M_1-M_3-2M_6) | \, L,M_1,M_3,M_6\rangle_4$,
as long as $M_1\leq L/4$, $M_1+M_3\leq L/2$, $M_6-M_1\leq L/2$ and $M_3+M_6\leq 3L/4$.

\subsection{ An exact Schmidt decomposition and the entanglement entropy}

Here the details to evaluate the entanglement entropy for (i) $| \, L,M_1,M_2,M_3\rangle_4$ in  Eq.~(6) and  (ii) $| \, L,M_1,M_3,M_6\rangle_4$ in Eq.~(8) are presented.

(i) An exact Schmidt decomposition for the degenerate ground states $| \, L,M_1,M_2,M_3\rangle_4$ given in Eq.~(6) may be performed:
	\begin{align}
		|\, L,M_1,M_2,M_3\rangle_4=& \sum_{k_1=0}^{{\rm min}(n/4,M_1)} \sum_{k_2=0}^{{\rm min}(n/4-k_1,M_2)}\sum_{k_3=0}^{{\rm min}(3n/4-k_1-k_2,M_3)} \nonumber\\
		&\lambda(L,n,k_1,k_2,k_3,M_1,M_2,M_3)
		|n,k_1,k_2,k_3\rangle_4|L-n,M_1-k_1,M_2-k_2,M_3-k_3\rangle_4, 
		\label{svd123gsm}
	\end{align}
	where $n$ is even, and the Schmidt coefficients $\lambda(L,n,k_1,k_2,k_3,M_1,M_2,M_3)$ are
	\begin{equation}
		\lambda(L,n,k_1,k_2,k_3,M_1,M_2,M_3)=C_{M_1}^{k_1}C_{M_2}^{k_2}C_{M_3}^{k_3}\frac{Z_4(n,k_1,k_2,k_3)Z_4(L-n,M_1-k_1,M_2-k_2,M_3-k_3)}{Z_4(L,M_1,M_2,M_3)}.
		\label{m1m2m3gsm}
	\end{equation}
	Here $Z_4(n,k_1,k_2,k_3)$ and $Z_4(L-n,M_1-k_1,M_2-k_2,M_3-k_3)$ take the same form as $Z_4(L,M_1,M_2,M_3)$ in Eq.~(\ref{zlm1m2m3gsm}).
	Therefore one may rewrite $\lambda(L,n,k_1,k_2,k_3,M_1,M_2,M_3)$ as
	\begin{equation}
		\lambda(L,n,k_1,k_2,k_3,M_1,M_2,M_3)=\sqrt{\frac{C_{n/4}^{k_1}C_{n/4-k_1}^{k_2}C_{3n/4-k_1-k_2}^{k_3}C_{(L-n)/4}^{M_1-k_1}C_{(L-n)/4-M_1+k_1}^{M_2-k_2}C_{3(L-n)/4-M_1-M_2+k_1+k_2}^{M_3-k_3}} {C_{L/4}^{M_1}C_{L/4-M_1}^{M_2}C_{3L/4-M_1-M_2}^{M_3}}}.
		\label{lamm1m3sm}
	\end{equation}
	Hence for the degenerate ground states in Eq.~(6), the entanglement entropy $S_{\!L}(n,M_1,M_2,M_3)$ follows from the expression
	\begin{align}
		S_{\!L}(n,M_1,M_2,M_3)=&-\sum_{k_1=0}^{{\rm min}(n/4,M_1)} \sum_{k_2=0}^{{\rm min}(n/4-k_1,M_2)}\sum_{k_3=0}^{{\rm min}(3n/4-k_1-k_2,M_3)}\nonumber \\
		&\Lambda(L,n,k_1,k_2,k_3,M_1,M_2,M_3)
		\log_{2}\Lambda(L,n,k_1,k_2,k_3,M_1,M_2,M_3),
		\label{snkm1m2m3sm}
	\end{align}
	where $\Lambda(L,n,k_1,k_2,k_3,M_1,M_2,M_3)$ are the eigenvalues of the reduced density matrix $\rho_L(n,M_1,M_2,M_3)$: $\Lambda(L,n,k_1,k_2,k_3,M_1,M_2,M_3)=[\lambda(L,n,k_1,k_2,k_3,M_1,M_2,M_3)]^2$.
	
	For fixed fillings $f_1=M_1/L$, $f_2=M_2/L$ and $f_3=M_3/L$ in the thermodynamic limit $L \rightarrow \infty$, the eigenvalues of the reduced density matrix 
	$\Lambda(L,n,k_1,k_2,k_3,M_1,M_2,M_3)$ in Eq.~(\ref{snkm1m2m3sm}) become
	\begin{equation}
		\Lambda_\infty(n,k_1,k_2,k_3,f_1,f_2,f_3)=C_{l}^{k_3}\nu^{k_3}(1-\nu)^{l-k_3}\frac{(n/4)!(4f_1)^{k_1}(4f_2)^{k_2}(1-4f_1-4f_2)^{m}}
		{{k_1!}{k_2!}{m!}},
	\end{equation}
	with $l=3n/4-k_1-k_2$, $\nu=4f_3/(3-4f_1-4f_2)$ and $m=n/4-k_1-k_2$.
	
	Hence the entanglement entropy $S(n,f_1,f_2,f_3)$, in the thermodynamic limit, takes the following form the expression
	\begin{align}
		S(n,f_1,f_2,f_3)=-\sum_{k_1=0}^{{\rm min}(n/4,M_1)} \sum_{k_2=0}^{{\rm min}(n/4-k_1,M_2)}\sum_{k_3=0}^{{\rm min}(3n/4-k_1-k_2,M_3)}
		\Lambda_\infty(n,k_1,k_2,k_3,f_1,f_2,f_3)
		\log_{2}\Lambda_\infty(n,k_1,k_2,k_3,f_1,f_2,f_3).
		\label{snkf3f1sm}
	\end{align}

	(ii) An exact Schmidt decomposition for the degenerate ground states $| \, L,M_1,M_3,M_6\rangle_4$ given in Eq.~(8) may be performed:
	\begin{align}
		| \, L,M_1,M_3,M_6\rangle_4=& \sum_{k_1=0}^{{\rm min}(n/4,M_1)}\sum_{k_3=0}^{{\rm min} (n/2-k_1,M_3)}\sum_{k_6=0}^{{\rm min} (n/2+k_1,3n/4-k_3, M_6)}\nonumber \\
		&\lambda(L,n,k_1,k_3,k_6,M_1,M_3,M_6)
		|n,k_1,k_3,k_6\rangle_4|L-n,M_1-k_1,M_3-k_3,M_6-k_6\rangle_4, 
		\label{svd136sm}
	\end{align}
	where $n$ is a multiple of four, and the Schmidt coefficients $\lambda(L,n,k_1,k_3,k_6,M_1,M_3,M_6)$ are
	\begin{equation}
		\lambda(L,n,k_1,k_3,k_6,M_1,M_3,M_6)=C_{M_1}^{k_1}C_{M_3}^{k_3}C_{M_6}^{k_6}\frac{Z_4(n,k_1,k_3,k_6)Z_4(L-n,M_1-k_1,M_3-k_3,M_6-k_6)}{Z_4(L,M_1,M_3,M_6)}.
		\label{m1m3sm}
	\end{equation}
	Here $Z_4(n,k_1,k_3,k_6)$ and $Z_4(L-n,M_1-k_1,M_3-k_3,M_6-k_6)$ take the same form as $Z_4(L,M_1,M_3,M_6)$ in Eq.~(\ref{zlm1m3m5gsm}).
	Hence  $\lambda(L,n,k_1,k_3,k_6,M_1,M_3,M_6)$ may be rewritten as
	\begin{align}
		&\lambda(L,n,k_1,k_3,k_6,M_1,M_3,M_6)\nonumber \\
		=&\sqrt{\frac{C_{n/4}^{k_1}\sum_{l=0}^{\min(k_6,n/4)}C_{n/4}^{l}C_{n/4}^{k_6-l}C_{n/2-k_1-k_6+l}^{k_3}
				C_{(L-n)/4}^{M_1-k_1}\sum_{m=0}^{\min(M_6-k_6,(L-n)/4)}C_{n/4}^{m}C_{(L-n)/4}^{M_6-k_6-m+l}C_{(L-n)/2-M_1-M_6+l-k_1-k_6-m}^{M_3-k_3}} {C_{L/4}^{M_1}\sum_{l=0}^{\min(M_6,L/4)}C_{L/4}^{l}C_{L/4}^{M_6-l}C_{L/2-M_1-M_6+l}^{M_3}}}.
		\label{lamm1m3m5sm}
	\end{align}
	Then for the degenerate ground states in Eq~(8),  the entanglement entropy $S_{\!L}(n,M_1,M_3,M_6)$ follows from
	\begin{align}
		S_{\!L}(n,M_1,M_3,M_6)=&-\sum_{k_1=0}^{{\rm min}(n/4,M_1)}\sum_{k_3=0}^{{\rm min} (n/2-k_1,M_3)}\sum_{k_6=0}^{{\rm min} (n/2+k_1,3n/4-k_3, M_6)}\nonumber \\
		&\Lambda(L,n,k_1,k_3,k_6,M_1,M_3,M_6)
		\log_{2}\Lambda(L,n,k_1,k_3,k_6,M_1,M_3,M_6),
		\label{snkm1m3m5sm}
	\end{align}
	where $\Lambda(L,n,k_1,k_3,k_6,M_1,M_3,M_6)$ are the eigenvalues of the reduced density matrix $\rho_L(n,M_1,M_3,M_6)$: 
	$\Lambda(L,n,k_1,k_3,k_6,M_1,M_3,M_6)=[\lambda(L,n,k_1,k_3,k_6,M_1,M_3,M_6)]^2$.

	For fixed fillings $f_1=M_1/L$, $f_3=M_3/L$ and $f_6=M_6/L$, the eigenvalues of the reduced density matrix $\Lambda(L,n,k_1,k_3,k_6,M_1,M_3,M_6)$ in Eq.~(\ref{snkm1m3m5sm}), in the thermodynamic limit $L \rightarrow \infty$, are 
	\begin{equation}
		\Lambda_\infty(n,k_1,k_3,k_6,f_1,f_3,f_6)=C_{n/4}^{k_1}(4f_1)^{k_1}(1-4f_1)^{n/4-k_1}\sum_{l}C_{n/4}^{l}C_{n/4}^{k_6-l}(2f_6)^{k_6}(1-2f_6)^{n/2-k_6}
		C_{n/2-k_1-k_6-l}^{k_3}u^{k_3}(1-u)^{v},
	\end{equation}
	where $u=2f_3/(1-2f_1-2f_6+2l/n)$ and $v=n/2-k_1-k_3-k_6+l$. 
    We note that in the main text, $|L,M_1,M_3,M_6\rangle_4$ is denoted as $|L,M_1^*,M_2^*,M_3^*\rangle_4$, with $M_1^*=M_1$, $M_2^*=M_3$ and $M_3^*=M_6$. 
	The filling factors follow, i.e., $f_1^*=M_1^*/L$, $f_2^*=M_2^*/L$ and $f_3^*=M_3^*/L$.
	Therefore, the entanglement entropy $S(n,f_1,f_3,f_6)$, with $S(n,f_1^*,f_2^*,f_3^*)\equiv S(n,f_1,f_3,f_6)$, in the thermodynamic limit, follows from
	\begin{align}
		S(n,f_1,f_3,f_6)=-\sum_{k_1=0}^{{\rm min}(n/4,M_1)}\sum_{k_3=0}^{{\rm min} (n/2-k_1,M_3)}\sum_{k_6=0}^{{\rm min} (n/2+k_1,3n/4-k_3, M_6)}
		\Lambda_\infty(n,k_1,k_3,k_6,f_1,f_3,f_6)
		\log_{2}\Lambda_\infty(n,k_1,k_3,k_6,f_1,f_3,f_6).
		\label{snkf3f1sm}
	\end{align}

\subsection{Scaling of the entanglement entropy with the block size in the thermodynamic limit}

The logarithmic scaling behaviour of the entanglement entropy $S_{\!\!f}(n)$ in the thermodynamic limit $L \rightarrow \infty$ for the highly degenerate ground states 
$| \, L,M_1,M_2,M_3\rangle_2$ in Eq.~(4) is revealed by performing an asymptotic analysis, following from Stirling's approximation. 
For large $L$, we have $L! \approx\sqrt{2\pi L} \, L^L \exp(-L)$ in general. 
Denoting $f_\alpha=M_\alpha/L$ with $\alpha=1$, $2$, $3$, the eigenvalue $\Lambda_\infty(k_1,k_2,k_3,f_1,f_2,f_3)$ of the reduced density matrix, in Eq.~(13), labelled by $f_\alpha$, when $L \rightarrow \infty$ and $M_\alpha \rightarrow \infty$ ($\alpha=1$, $2$, $3$), simplifies to
	\begin{equation}
		\Lambda_\infty(k_1,k_2,k_3,f_1,f_2,f_3)=C_{l}^{k_3}\nu^{k_3}(1-\nu)^{l-k_3}\frac{(n/2)!(2f_1)^{k_1}(2f_2)^{k_2}(1-2f_1-2f_2)^{m}}
		{{k_1!}{k_2!}{m!}},
	\end{equation}
with $l=n-k_1-k_2$, $\nu=f_3/(1-f_1-f_2)$ and $m=n/2-k_1-k_2$.
	
In this way, in the thermodynamic limit, the entanglement entropy $S(n,f_1,f_2,f_3)$ as computed follows from
	\begin{align}
		S(n,f_1,f_2,f_3)=-\sum_{k_1=0}^{\min(M_1,n/2)}\sum_{k_2=0}^{\min(M_2,n/2)}\sum_{k_3=0}^{\min(M_3,n-k_1-k_2)}
		\Lambda_\infty(k_1,k_2,k_3,f_1,f_2,f_3)
		\log_{2}\Lambda_\infty(k_1,k_2,k_3,f_1,f_2,f_3).
		\label{sf1f2f3}
	\end{align}

The entanglement entropy $S_{\!\!f}(n)\equiv S(n,f_1,f_2,f_3)$ or $S_{\!\!f}(n)\equiv S(n,f_1,f_3,f_6)$, in the thermodynamic limit, takes the same form as Eq.~(\ref{sf1f2f3}) for each of the three choices among the degenerate ground states:
$| \, L,M_1,M_2,M_3\rangle_2$, $| \, L,M_1,M_2,M_3\rangle_4$ and $| \, L,M_1,M_3,M_6\rangle_4$ in Eq.~(4), Eq.~(6) and Eq.~(8), with $n$ being a multiple of two, a multiple of four and a multiple of four, respectively.

\begin{figure}[ht]
	\centering
	\includegraphics[width=0.85\textwidth]{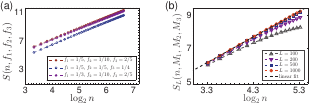}
	\caption{ (a) The entanglement entropy $S(n,f_1,f_2,f_3)$ vs $\log_2 n$ for the staggered $\rm{SU(4)}$ ferromagnetic spin-orbital model in the thermodynamic limit.
    Here $f_1$, $f_2$, and $f_3$ are varied: $f_1=1/5$, $f_2=1/10$, and $f_3=2/5$,  $f_1=1/5$, $f_2=1/5$, and $f_3=1/4$, and $f_1=1/3$, $f_2=1/10$, and $f_3=2/5$.
		(b) The entanglement entropy $S_{\!L}(n,M_1,M_2,M_3)$ vs $\log_2 n$ for the staggered $\rm{SU(4)}$ ferromagnetic spin-orbital model, with $M_1=L/10$, $M_2=L/5$, and $M_3=L/4$, when $L$ is varied: $L=$ 100, 200, 500 and 1000.
		Here $n$  is a multiple of two, and ranges from $10$ to $40$. }
	\label{Su4lm1m2m3sm}
\end{figure}

\begin{figure}[ht]
	\centering
	\includegraphics[width=0.85\textwidth]{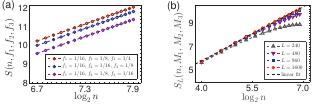}
	\caption{ 	(a) The entanglement entropy $S(n,f_1,f_2,f_3)$ vs $\log_2 n$ for the staggered $\rm{SU(4)}$ ferromagnetic 
		spin-orbital model in the thermodynamic limit.
		Here $f_1$, $f_2$ and $f_3$ are varied:
		$f_1=1/16$, $f_2=1/8$ and  $f_3=1/4$,
		$f_1=1/16$, $f_2=1/16$ and $f_3=1/8$,
		and $f_1=1/16$, $f_2=1/8$ and  $f_3=1/16$.
		(b) The entanglement entropy $S_{\!L}(n,M_1,M_2,M_3)$ vs $\log_2 n$ for the  staggered $\rm{SU(4)}$ ferromagnetic spin-orbital model, with $M_1=L/16$, $M_2=L/16$ and $M_3=L/8$, when $L$ is varied: $L=$ 240, 480, 960 and 1600. Here a generalized highest weight state is chosen to be $\vert\otimes_{k=1}^{L/4}  \{\downarrow_s\uparrow_t\uparrow_s\uparrow_t\uparrow_s\uparrow_t\uparrow_s\uparrow_t\}_{\;k}\rangle$, and $n$  is a multiple of four, and ranges from $16$ to $120$.}
	\label{Su4lm1m2m3gsm}
\end{figure}

\begin{figure}[ht]
	\centering
	\includegraphics[width=0.85\textwidth]{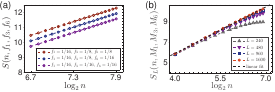}
	\caption{(a) The entanglement entropy $S(n,f_1,f_3,f_6)$ vs $\log_2 n$ for the staggered $\rm{SU(4)}$ ferromagnetic spin-orbital 
		model in the thermodynamic limit.
		Here, $f_1$, $f_3$ and $f_6$ are varied:
		$f_1=1/16$,  $f_3=1/8$ and $f_6=1/8$,
		$f_1=1/16$, $f_3=1/8$ and $f_6=1/16$,
		and $f_1=1/16$, $f_3=1/16$ and  $f_6=1/16$.
		(b) The entanglement entropy $S_{\!L}(n,M_1,M_3,M_6)$ vs $\log_2 n$ for the staggered $\rm{SU(4)}$ ferromagnetic spin-orbital model, with $M_1=L/16$, $M_3=L/8$  and $M_6=L/16$, when $L$ is varied: $L=$ 240, 480, 960 and 1600.   Here a generalized highest weight state is chosen to be $\vert\otimes_{k=1}^{L/4}  \{\downarrow_s\uparrow_t\downarrow_s\uparrow_t\uparrow_s\uparrow_t\uparrow_s\uparrow_t\}_{\;k}\rangle$, and  
		$n$  is a multiple of four, and ranges from $16$ to $120$. }
	\label{Su4lm1m3m5gsm}
\end{figure}

We plot $S(n,f_1,f_2,f_3)$ vs $\log_2 n$ in Fig.~\ref{Su4lm1m2m3sm} (a) and in Fig.~\ref{Su4lm1m2m3gsm} (a) for different fillings $f_1$, $f_2$ and $f_3$.  
The prefactor is close to $3/2$, within an error less than 1$\%$.
In addition, we plot $S(n,f_1,f_3,f_6)$ vs $\log_2 n$ in Fig.~\ref{Su4lm1m3m5gsm} (a) for different fillings $f_1$, $f_3$ and $f_6$. 
The prefactor is close to $3/2$, within an error less than 1$\%$.
That is, the contribution from the fillings $f_1$, $f_2$, and $f_3$ or $f_1$, $f_3$ and $f_6$ is a non-universal additive constant.

In order to observe how the logarithmic scaling behavior emerges in the thermodynamic limit from a numerical perspective, we plot $S_{\!L}(n,M_1,M_2,M_3)$ vs $\log_2 n$ for $| \, L,M_1,M_2,M_3\rangle_2$, as presented in Eq.~(4), in Fig.~\ref{Su4lm1m2m3sm} (b) and for $| \, L,M_1,M_2,M_3\rangle_4$, as presented in Eq.~(6), in Fig.~\ref{Su4lm1m2m3gsm} (b). 
In addition, the entanglement entropy $S_{\!L}(n,M_1,M_3,M_6)$ vs $\log_2 n$ for $| \, L,M_1,M_3,M_6\rangle_4$, as presented in Eq.~(8), is plotted in Fig.~\ref{Su4lm1m3m5gsm} (b).
The plots show a significant deviation from the logarithmic scaling behavior when $L$ is relatively small, which tends to vanish as $L$ increases. 
The prefactor is close to the exact value $3/2$, with the error being less than 2$\%$, when $L=1000$: $S_{\!L}(n,M_1,M_2,M_3)= 1.528\log_2n + 1.100$  in Fig.~\ref{Su4lm1m2m3sm} (b), with an error being less than 1$\%$,  when $L=1600$: $S_L(n,M_1,M_2,M_3)=1.511\log_2n-0.231$ in Fig.~\ref{Su4lm1m2m3gsm} (b), and  with an error less than 1$\%$, when $L=1600$: $S_{\!L}(n,M_1,M_3,M_6)=1.515\log_2n-0.159$ in Fig.~\ref{Su4lm1m3m5gsm} (b).

\end{widetext}